\documentclass[12pt]{article}

\usepackage{newtxtext,newtxmath}

\usepackage{graphicx}

\usepackage[letterpaper,margin=1in]{geometry}

\usepackage{float}

\renewenvironment{abstract}
    {\quotation}
    {\endquotation}

\date{}

\usepackage{scicite}

\usepackage{url}

\usepackage[version=3]{mhchem}
\usepackage{gensymb}
\usepackage{xcolor}
\usepackage{placeins}
\usepackage[bookmarks=false]{hyperref}

\newcommand{\AuPbP}{Au$_2$Pb$_{0.914}$P$_2$}
\newcommand{\AuPbPfull}{Au$_2$PbP$_2$}

\def\scititle{
    Stripping Symmetry: Electrochemical Oxidation to a Superconducting Polar Metal in Au$_2$Pb$_{0.914}$P$_2$
}

\title{\bfseries \boldmath \scititle}

\author{
    Scott B. Lee$^{1}$,
    Stephanie R. Dulovic$^{1}$,
    Joseph W. Stiles$^{1}$,
    Xin Zhang$^{1}$,\and
    Fatmag\"ul Katmer$^{1}$,
    Sudipta Chatterjee$^{1}$,
    Jaime Moya$^{1}$,\and
    Allana G. Iwanicki$^{2,3}$,
    Abby N. Neill$^{2}$,
    Chris Lygouras$^{3}$,
    Tieyan Chang$^{4}$,\and
    Tyrel M. McQueen$^{2,3,5}$,
    Yu-Sheng Chen$^{4}$,
    Leslie M. Schoop$^{1}$\small$^*$
\\
\\
    \small$^{1}$Department of Chemistry, Princeton University, Princeton, NJ 08544\and
    \small$^{2}$Department of Chemistry, The Johns Hopkins University, Baltimore, MD 21218\and
    \small$^{3}$Institute for Quantum Matter, William H. Miller III Department of Physics and Astronomy, Baltimore, MD 21218\and
    \small$^{4}$NSF's ChemMatCARS, The University of Chicago, Lemont, IL 60439, USA\\
    \small$^{5}$Department of Materials Science and Engineering, The Johns Hopkins University, Baltimore, MD 21218\and
    \\
    \small$^*$Corresponding author. Email: lschoop@princeton.edu
}

\begin{document}

\maketitle

\begin{abstract} \bfseries \boldmath
Polar metals and noncentrosymmetric superconductors are exceptionally rare, yet their broken inversion symmetry can give rise to emergent electronic phenomena including mixed singlet--triplet superconducting pairing. As only a few such materials have been found among known compounds, accessing new examples requires synthetic strategies that go beyond conventional crystal growth. Here, we use electrochemical topotactic deintercalation to remove Pb from the centrosymmetric parent compound \AuPbPfull, producing the polar metal \AuPbP. Unlike conventional chemical doping, this transformation actively drives structural symmetry-breaking: the partial removal of Pb triggers a cooperative electronic and geometric rearrangement, mediated by a second-order Jahn-Teller effect and stereochemically active lone pairs, that locks the product into a polar, noncentrosymmetric superspace group $Ama2(01\gamma)ss0$. We solve the complete (3+1)D modulated structure by synchrotron single-crystal X-ray diffraction and confirm the polar assignment through nonlinear electronic transport. Below $T_c = 1.52$~K, \AuPbP\ becomes a type-II superconductor whose heat capacity and AC susceptibility both exhibit power-law behavior, suggestive of a gap structure governed by the broken inversion symmetry of the host lattice. This work establishes electrochemical oxidation as a rational route to metastable noncentrosymmetric superconductors through chemically directed symmetry-breaking.
\end{abstract}

\noindent

Polar metals, in which polarity and metallic conductivity coexist despite the expectation that itinerant electrons screen internal electric fields, have been a long focus of study~\cite{Wang_2022, Legg_2022, SuarezRodriguez_2025, Bhowal_2023}. Polar metallicity enables nonlinear transport phenomena~\cite{Wang_2022, Legg_2022}, the magnetoelectric effect~\cite{Ye_2024, Johansson_2024}, and radiofrequency radiation harvesting~\cite{Kumar_2021, Kumar_2024, SuarezRodriguez_2025}. Because polar space groups lack inversion symmetry, any superconductor in such a space group is inherently a noncentrosymmetric superconductor~\cite{Smidman_2017, Yip_2014}. In these systems, antisymmetric spin-orbit coupling can mix singlet and triplet pairing channels, potentially stabilizing exotic superconducting states with mixed parity. Still, despite this symmetry-permitted mixing, most known noncentrosymmetric superconductors exhibit conventional multi-gap $s$-wave character rather than nodal behavior.~\cite{Smidman_2017} 

Although noncentrosymmetric superconductors are well established across a broad range of material families,~\cite{Bauer_2004, Smidman_2017, Joshi_2011, Shang2022, Shresta_2026} the likelihood of discovering new examples among thermodynamically stable compounds is small, calling for new strategies for materials discovery. Indeed, significant advances in computational strategies have enabled the prediction of new stable compounds,\cite{Tran_2024, Bai_2024} yet substantial challenges remain, especially for large-scale efforts \cite{Cheetham_2024, Leeman_2024, Zhang_2026_Arxiv}. A powerful alternative approach to access some new materials is manipulation of known materials through post-synthetic chemical transformations \cite{Guo_2017, Song_2023, Takada_2003, Rajapakse_2021, Cui_2019, Li_2019}. Such materials are challenging to predict computationally, as they often exhibit large unit cells and may show significant degrees of disorder such as vacancies or mixed-site occupancy. Therefore, post-synthetic processes, such as topochemical modifications represent an underexplored paradigm that experimentalists can leverage to access new, unpredictable, metastable phases.

Topotactic reactions are redox-mediated transformations in which one or more crystallographic axes remain preserved from reactant to product.\cite{Li_2019, Meng_2023, Eder_2025, Song_2023, Villalpando_2024, Aharon_2025,Xie_2025} The general scheme to achieve topotactic transformations requires a combination of redox-active elements and high ionic mobility. As high-mobility ions move through the crystal, many topotactically transformed materials break the symmetries inherent to the parent structure,\cite{Wernert_2026, Martinez_2024, jeen_2013, Varela-Dominguez2024}, offering a powerful tool for controlling physical properties \cite{Wright_2025, Zheng_2021}. Importantly, topotactic reactions provide access to metastable phases that are unreachable in conventional high-temperature synthesis, where thermodynamic stability governs accessible compositions. However, while topotactic reactions have been studied in many semiconducting or insulating materials, they have rarely been utilized for metallic systems. This is particularly relevant for discovery of new superconductors, as emergent superconductivity frequently appears in the vicinity of structural or electronic instabilities. \cite{Taddei_2018, Pang_2015, Adroja_2015}

\begin{figure}[H]
    \centering
    \includegraphics[width=0.5\textwidth]{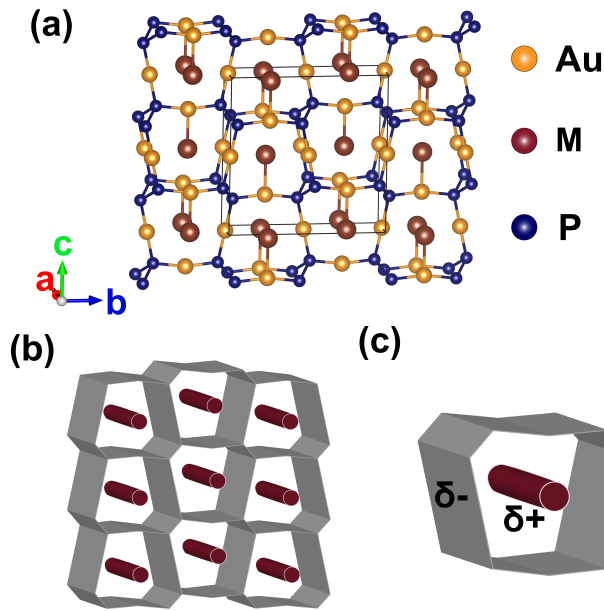}
    \caption{\textbf{Crystal structure and electronic structure of the Au$_2M$P$_2$ family.} \textbf{(a)} The structure consists of a covalently bonded [Au$_2$P$_2$] tunnel framework enclosing a linear chain of $M$ atoms, where $M$ = Hg, Tl, Pb, Pb/Bi, or Bi. The structure is quasi-one-dimensional with $M$ atoms residing in the tunnels along the crystallographic $a$-axis. The framework accommodates varying electron counts, maintaining the $M$ atom in a near-neutral charge state. \textbf{(b)} The structure viewed along the chain axis, highlighting the tunnel geometry. \textbf{(c)} Schematic electronic structure showing the role of the [Au$_2$P$_2$] framework as an electron reservoir. The $\delta$+/$\delta$- partitioning shown schematically reflects a previous Bader charge analysis,~\protect\cite{Lee_2024} which reveals that $M$ atoms carry increasing cationic character across the series while remaining near-neutral overall.}
    \label{Intro_Figure}
\end{figure}

Here, we target the Au$_2M$P$_2$ ($M$ = Hg, Tl, Pb, Pb/Bi, or Bi) family of quasi-one-dimensional compounds, which adopt an isotypic, highly anisotropic structure type across a range spanning three valence electrons per formula unit~\cite{Eschen_2002, Lee_2024}. The structure consists of a covalently bonded [Au$_2$P$_2$] tunnel framework in which a linear chain of $M$ atoms resides (Figure~\ref{Intro_Figure}\textbf{(a)} and \textbf{(b)}). The [Au$_2$P$_2$] framework functions as an electron reservoir, accommodating varying electron counts through charge transfer from $M$ while maintaining the $M$ atom in a near-neutral charge state (Figure~\ref{Intro_Figure}\textbf{(c)})~\cite{Lee_2024}. The structural instability of the linear $M$ chain creates an inherent driving force toward symmetry-breaking distortions~\cite{Wen_2009}. The combination of a flexible electron-accepting framework and an energetic driving force toward distortion makes the Au$_2M$P$_2$ family a compelling candidate for topotactic exploration.

We show that electrochemical oxidation of \AuPbPfull\ partially removes Pb from the chain, producing the metastable polar metal \AuPbP. The reaction is highly precise: only \AuPbP\ is accessible; no intermediate or further-deintercalated compositions can be stabilized under our experimental parameters, as incomplete deintercalation results in a mixture of \AuPbPfull\ and \AuPbP. Through a combination of XPS, $\mu$CT, synchrotron single-crystal X-ray diffraction, DFT calculations, nonlinear transport measurements, and low-temperature thermodynamic and magnetic characterization, we establish three central results. First, the electrochemical method produces homogeneous, high-quality single crystals suitable for a complete structural solution. Second, the topotactic transformation breaks inversion symmetry through a cooperative mechanism involving electronic depopulation, a second-order Jahn-Teller (SOJT) distortion, and stereochemically active lone pairs. Third, the resulting polar metal superconducts below 1.52~K with thermodynamic signatures consistent with a nodal gap. We thus introduce electrochemical topotactic deintercalation as a post-synthetic route to high-spin-orbit-coupling noncentrosymmetric superconductors. 

\subsection*{Synthesis and sample optimization}

Single crystals of orthorhombic \ce{Au2PbP2} were grown by the self-flux method (see Materials and Methods). To induce topotactic deintercalation of Pb, we employed electrochemical oxidation using a three-electrode potentiostat with single crystals of \ce{Au2PbP2} as the working electrode. Initial attempts using acid treatment (\ce{HNO3}) produced inhomogeneous samples, as revealed by $\mu$CT imaging showing two distinct phases within individual crystals and depth-dependent XPS profiles, indicating incomplete penetration of the chemical transformation (Figure~S\ref{XPS}\textbf{(a)--(f)}).

Electrochemical treatment of parent crystals resolves this issue. Repeating the depth-profile XPS experiments on electrochemically treated crystals reveals uniform binding energies from surface to bulk, with no residual parent-phase signatures even after extended sputtering (Figure~\ref{XPS}\textbf{(g)--(i)}). Along with XPS, $\mu$CT scans confirm a single-phase product throughout the crystal volume\cite{Pressley_2022}. These results demonstrate that the electrochemical route produces phase-pure, homogeneous Au$_2$Pb$_{1-x}$P$_2$ crystals, where only one $x$ value is accessible. To the best of our knowledge, the crystallinity retained through this process is unusual for topotactic reactions, which typically introduce significant disorder. The phase-pure crystals, which are stable in air for at least several months, obtained here proved essential for the synchrotron diffraction studies described below.

XPS further indicated a change in oxidation state accompanying the topotactic transformation. In \AuPbPfull\, Pb resides in a near-neutral oxidation state, in agreement with previous reports.~\cite{Lee_2024, Wen_2009} Upon electrochemical oxidation the Pb 4$f$ binding energies shift to values characteristic of Pb$^{2+}$. The Au 4$f$ region shows a modest shift in binding energy, while a shift in the P 2$p$ region to higher binding energy reflects redistribution of electron density in the [Au$_2$P$_2$] framework. The transition of Pb from near-neutral to 2+ is crucial for the symmetry-breaking mechanism discussed below. 

\subsection*{Structural characterization}

Synchrotron diffraction studies on \AuPbP\ crystals reveal characteristic satellite reflections along the reciprocal axis corresponding to the linear chain of Pb atoms, indicating a commensurately modulated structure resulting from ordered Pb deintercalation rather than random vacancy formation (Figure~S\ref{APP_Prec}). The systematic arrangement of these satellite reflections demonstrates that the structural transformation proceeds with a high degree of order, consistent with a topotactic mechanism in which the fundamental framework of the parent structure remains intact.

Following crystallographic conventions, the C-centered parent cell is transformed into an A-centered cell with satellite peaks along the \textbf{c}-axis. The satellite peaks, now indexed at integer multiples of 0.07166\textbf{c*}, require a (3+1)D superspace crystallography approach where the additional dimension represents the phase of modulation~\cite{DeWolff_1974, VanSmaalen_1995}. This framework enables precise refinement of the aperiodic electron density perturbations induced by partial Pb deintercalation.

The observed reflection conditions constrain the possible superspace groups. The combined systematic absence conditions yield optimal refinement statistics with an agreement factor of $3.18\%$ for all observed reflections in superspace group $Ama2(01\gamma)ss0$ (SSG 40.1.15.6), with a goodness of fit of 3.36. The integrated data closely match the theoretical value for noncentrosymmetric crystals ($\langle |E^2-1| \rangle = 0.736$), consistent with the superspace group assignment. The refined composition, Au$_2$Pb$_{0.914}$P$_2$, closely approximates the theoretical value of 0.929 corresponding to the removal of one Pb atom from a 14-fold supercell, confirming the precise stoichiometry of the topotactic transformation.

\begin{figure}[H]
    \resizebox{\textwidth}{!}{\includegraphics{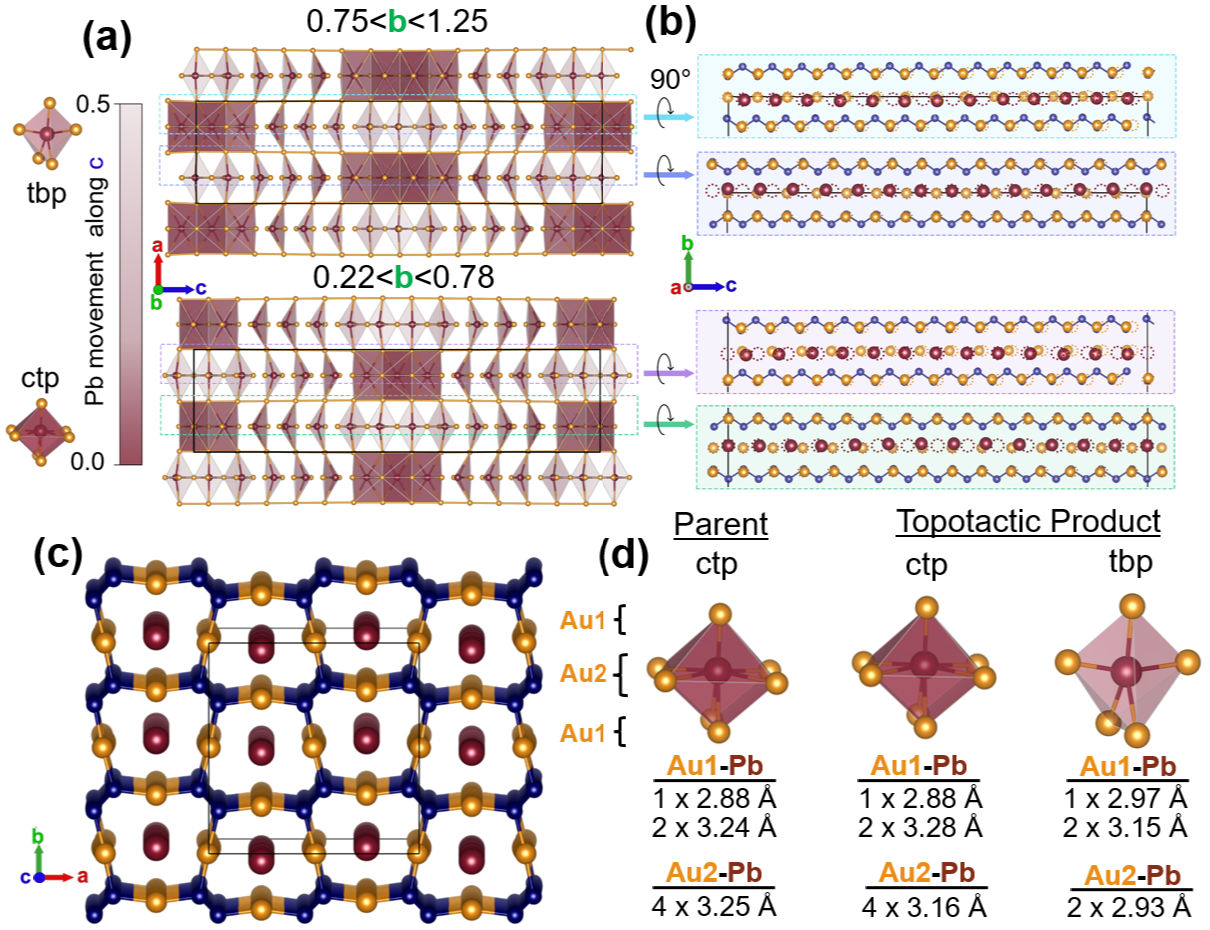}}
    \caption{\textbf{Approximate modulated structural solution for \AuPbP.} \textbf{(a)} Cross-sectional view of tunnel-like structure in the \textit{a}--\textit{c} plane. P atoms are removed and select contacts are drawn at a cutoff of 3.4~\AA\ to place particular emphasis on the Pb--Au coordination environments evolving from capped trigonal prismatic (ctp, opaque) to trigonal bipyramidal (tbp, transparent) along \textit{c} and Au2 positional modulations along \textit{a}. \textbf{(b)} Cross-sectional view of the tunnel. Au2 atoms out of the plane of the page are removed for clarity. Open circles represent Pb atom placement in a 14-fold supercell of the undistorted parent structure-type to emphasize Pb positional modulations along \textit{b} and environments and bond distances between the parent and modulated phase. \textbf{(c)} View of the modulated superstructure parallel to the chain axis, emphasizing the breathing-like motion of Au2 atoms. \textbf{(d)} Select Pb--Au coordination environments and bond distances between the parent and modulated phase.}
    \label{Struct_Figure}
\end{figure}

Figure~\ref{Struct_Figure} illustrates the modulated structure. The remaining Pb atoms adopt different coordination environments through their redistribution along the chain. In regions of minimal displacement, Pb atoms retain their 7-coordinated capped trigonal prismatic (ctp) coordination environment characteristic of the parent structure (Figure~\ref{Struct_Figure}\textbf{(d)}). At positions of maximal displacement (approximately 0.5 average unit cells along the chain axis), the Pb coordination transforms to a 5-coordinated trigonal bipyramidal (tbp) geometry. This coordination dichotomy is consistent with the theoretical predictions of Wen and Hoffmann~\cite{Wen_2009}, who remarkably identified the tbp coordination as the energetically preferred environment during structural relaxation, when the initial position of Pb atoms is sufficiently translated off the inversion center.

During the displacement of Pb, the coordinated Au2 atoms in the \textbf{(a)}--\textbf{(c)} plane undergo substantial modulation along the \textit{a} axis, with an amplitude of approximately 1.26~\AA. This modulation is directly coupled with the positional changes of Pb atoms, creating a coordinated breathing-like motion of the [Au$_2$P$_2$] framework. In the 7-coordinated ctp geometries, the surrounding Au2 atoms contract inward, creating shorter Au--Pb contacts compared to the parent structure (Figure~\ref{Struct_Figure}\textbf{(d)}). This contraction is balanced by expansion in adjacent tunnels, where Pb atoms in the tbp geometry have shifted to positions nearly linearly coordinated by Au atoms with increased Au--Pb distances.

In summary, electrochemical oxidation transforms \AuPbPfull\ from centrosymmetric to polar: approximately 1 in 14 Pb atoms are removed, and the remaining Pb atoms adopt different coordinate environments through their redistribution along the chain. With this modulated structure now established, we can elucidate the mechanism by which the partial deintercalation breaks centrosymmetry.

\subsection*{Mechanisms behind the symmetry-breaking}

Our observations raise the question of why the topotactic reaction produces an ordered polar structure rather than a disordered vacancy arrangement that retains centrosymmetry.
We suggest that the answer lies in three complementary mechanisms previously used to describe noncentrosymmetric displacements. The first explains how Pb ions begin to move, while the latter two explain why the transition ends in a metastable polar structure.

\begin{figure}[H]
    \centering
\includegraphics[width=1\linewidth,height=0.8\textheight,keepaspectratio]{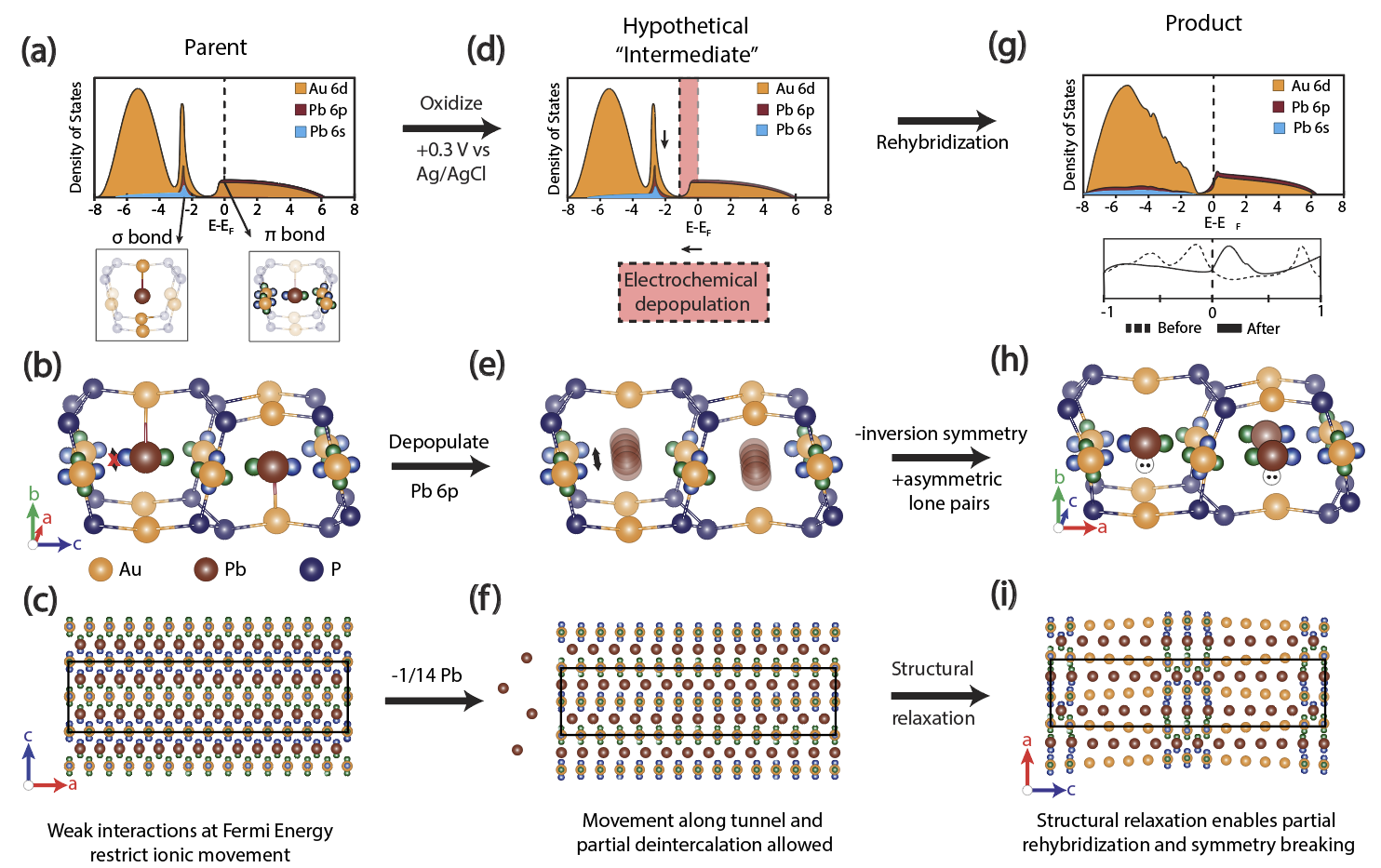}
    \caption{\textbf{Schematic of proposed topotactic oxidation mechanism in Au$_2$Pb$_{1-x}$P$_2$.} \textbf{(a)} Partial density of states plot of \ce{Au2PbP2}, depicting strong Au--Pb $\sigma$ bonding (along \textbf{(b)}) located $\sim$2.25~eV below the Fermi level and weaker Au--Pb $\pi$ bonds (in the \textit{ac} plane) located near the Fermi level. \textbf{(b)} Combined, these two bonding interactions restrict ionic movement of Pb atoms, visualized in two adjacent tunnels. \textbf{(c)} A cross-section in the \textit{ac} plane of a 14-fold supercell of the parent structure. The weak $\pi$ interactions hold the Pb atom in a position such that centrosymmetry is retained in the parent. \textbf{(d)} Upon oxidation, the Fermi Energy is lowered (red shaded area), depopulating the weaker Pb $p$ states. \textbf{(e)} This depopulation allows Pb atoms to move along the chain axis of the tunnel. \textbf{(f)} The deintercalation of Pb continues until $\frac{1}{14}$ of the Pb atoms are removed from the structure. The remaining Pb atoms redistribute along the chain axis creating varying local coordination environments. \textbf{(g)} Partial density of states plot for a 14-fold supercell reconstruction of \AuPbP. In addition to the peak at $\sim$-2.25 no longer being observed, non-negligible amounts of Au--Pb interactions are still apparent at the Fermi level, indicating an electronic reconstruction. \textbf{(h)} This electronic reconstruction is consistent with rehybridization of Au--Pb interactions, reflecting the SOJT effect and asymmetric lone-pairs. \textbf{(i)} Structural relaxation of Au atoms in \AuPbP, visualized in the \textit{ac} plane show that bonding interactions are observed only along certain points in the supercell.}
    \label{Mechanism_Figure}
\end{figure}

\textit{Electronic depopulation and ionic mobility.} The first mechanism is an electronic decoupling principle proposed as a design strategy for noncentrosymmetric metals~\cite{Puggioni_2014}. In this framework, ionic displacements that break inversion symmetry become accessible when the relevant electronic degrees of freedom of the displacing atoms are decoupled from the Fermi level. In \AuPbPfull, the Pb atoms are held in place by two bonding interactions: a strong Au--Pb $\sigma$ bond ($\sim$2.25~eV below the Fermi level) and a weaker Au--Pb $\pi$ interaction near the Fermi level, as reflected in the density of states (Figure~\ref{Mechanism_Figure}\textbf{(a)} detailed further in Figures S\ref{fig:pDOS_atom} and \ref{fig:pDOS_orbital}). Electrochemical oxidation selectively depopulates the weaker frontier Pb~6$p$ states (Figure~\ref{Mechanism_Figure}\textbf{(d)}), initiating Pb displacement along the chain axis (Figure~\ref{Mechanism_Figure}\textbf{(e)} and \textbf{(f)}).

Critically, electrochemical depopulation alone does not account for the formal oxidation state of Pb$^{2+}$ evidenced by XPS. As the Pb oxidation state of the parent compound is close to neutral, achieving this state would require the removal of far more electrons than is experimentally feasible  (see SI section 3.3). Instead, we attribute the change in the oxidation state of Pb to a charge transfer that accompanies the structural transformation: As the Pb ions displace, the electrons corresponding to the $\sigma$ bonding states (previously localized at $\sim$-2.25~eV) are redistributed lower into the Au~5$d$ band, transferring electrons from Pb to the surrounding framework and completing the transition of Pb$^0 \rightarrow$ Pb$^{2+}$ (Figure~\ref{Mechanism_Figure}\textbf{(g)}). Thus, the process is a two-step sequence: electrochemical depopulation of Pb~6$p$ states triggers ionic motion, followed by charge transfer producing Pb$^{2+}$, which is the electronic configuration necessary for the symmetry-breaking mechanisms that follow.

\textit{Second-order Jahn-Teller effect.} The SOJT distortion provides the energetic driving force for the noncentrosymmetric displacement of Pb atoms. A SOJT distortion can occur when two atoms have filled and empty shells of opposite parity; when these conditions are met, the total energy of the product is lowered, stabilizing a noncentrosymmetric, symmetry-broken configuration~\cite{Bersuker_1966, Bersuker_2013, Mallick_2021}. In \AuPbP, the XPS binding energies indicate formal oxidation states of Au$^{0}$ and Pb$^{2+}$, suggesting filled 5$d$ ($\ell = 3$) shells and empty 6$p$ ($\ell = 2$) orbitals --- precisely the parity-alternating configuration required for a SOJT distortion. This oxidation state assignment is supported by cyclic voltammetry (Figure~S\ref{CV2}), which shows an oxidation peak at +0.16~V vs Ag/AgCl consistent with the oxidation of Pb(0) to Pb(II) from the parent structure. Although both those experiments formally suggest closed-shell configurations, partial density of states calculations on the modulated structure solution reveal that states near the Fermi level remain dominated by Au--Pb interactions, indicating that hybridization between filled Au~5$d$ and empty Pb~6$p$ states persists and stabilizes the distorted geometry (Figure~\ref{Mechanism_Figure}\textbf{(g)}--\textbf{(i)}). It is this interaction that selects for the polar, noncentrosymmetric configuration: the SOJT lowers the energy when Pb is displaced off its inversion center, making the ordered polar superstructure thermodynamically preferred over a centrosymmetric vacancy arrangement.

\textit{Stereochemically active lone pair.} Third, the formal Pb$^{2+}$ oxidation state introduces a complementary driving force. The 6$s^2$ lone pair of Pb$^{2+}$ is well known to adopt an asymmetric spatial distribution through mixing with empty 6$p$ states, producing a directed, off-center displacement that disfavors high-symmetry coordination environments~\cite{Yan_2021, Walsh_2011}. This mechanism is intimately related to the SOJT effect~\cite{Fabini_2016} and has been demonstrated to drive polar distortions across a wide range of Pb-based compounds.\cite{Li_2024, Straus_2021} In \AuPbP, the remaining Pb$^{2+}$ ions adopt a continuous distribution of coordination geometries along the modulation, ranging between the seven-coordinate ctp geometry of the parent structure and a five-coordinate tbp geometry at maximal displacement.  While both the idealized ctp and tbp endpoints create a locally centrosymmetric environment, the intermediate geometries are individually asymmetric, where the stereoactive lone pair stabilizes these off-center configurations. More fundamentally, it is the collective ordered sequence of these coordination environments strung along the chain axis that breaks global inversion symmetry. The SOJT distortion and stereoactive lone-pair activity thus act cooperatively and can be thought of as a thermodynamic lock. Once electrochemical decoupling enables the Pb atoms to modulate their position, these mechanisms stabilize the continuously modulated, globally polar configuration over any centrosymmetric alternative.

Together, these three mechanisms act cooperatively: electrochemical depopulation of Pb~6$p$ states triggers ionic mobility, the SOJT effect provides the energetic preference for noncentrosymmetric displacement, and the stereoactive lone pair stabilizes the polar end state. The combination of a flexible electron-accepting framework and these cooperative driving forces is what makes this system produce an ordered polar structure rather than a disordered vacancy phase.

\subsection*{Generalization of the chemical reaction}

To test how general this noncentrosymmetric displacement is across the Au$_2$$M$$P_2$ ($M$ = Hg, Tl, Pb, Bi) series is, we also synthesized the pseudo-isomorphic Au$_2$Tl$_{1-x}$P$_2$. Diffraction studies suggest a general principle for symmetry-breaking in this family upon topotactic oxidation. Analogous satellite peaks can be distinguished in chemically treated \ce{Au2TlP2} (Figure~S\ref{fig:ATP_precession}). In this series, satellite peaks for Au$_2$Tl$_{1-x}$P$_2$ can be indexed to a q$_{1, Tl}$ $=$ -0.144\textbf{c'*}. While a full structural solution remains elusive, $<|E^2-1|>$ for data from Au$_2$Tl$_{1-x}$P$_2$ results in a value of 0.767, which closely matches the theoretical value of 0.736 expected for noncentrosymmetric crystals.

The relationship between the satellite peaks, q$_{1, Tl}$ $\approx$ 2$\times$q$_{1, Pb}$, hints at a possible relationship between the element in the tunnel $M$ and the distortion length. For Au$_2$Pb$_{0.914}$P$_2$, the reciprocal length of q$_{1, Pb}$ (0.07166 \textit{c}*) dictates the real space 14-fold ($\frac{1}{0.07166}$-fold) supercell. In addition, our structural analysis suggests that the value of q$_{1, Pb}$ reflects the percentage of Pb atoms that leave the structure, that is for Au$_2$Pb$_{1-x}$P$_2$, $x$ $\approx$ q$_{1, Pb}$. If we extend these results to Au$_2$Tl$_{1-x}$P$_2$, where q$_{1, Tl}$$=$-0.144\textbf{c'*}, we suspect that the resulting structural solution will be a 7-fold supercell with formula $\sim$Au$_2$Tl$_{0.85}$P$_2$. If we also assume that this structural transition is accompanied by an oxidation of Tl$^0$ $\rightarrow$ Tl$^{1+}$, the total amount of charge that leaves the parent structure is roughly equivalent between Pb and Tl($i.e.$ 0.144 $\times$ 1 $\approx$ 0.07166 $\times$ 2). Subsequently, the now cationic $M$ atoms of the distorted structures, would have the same formal valence electron configuration ([Xe] 4$f$$^{14}$ 5$d$$^{10}$ 6$s$$^2$). This suggests that this closed-shell $d^{10}s^2$ configuration, with its filled $d$ and empty $p$ shells perfectly positioned for SOJT activity and its stereoactive $s^2$ lone pair is the electronic end point that determines both the extent of deintercalation and the noncentrosymmetric outcome across this series.

\subsection*{Nonlinear transport properties}

The structural solution predicts that \AuPbP\ is a polar metal in the $mm2$ point group. We test this assignment through symmetry-sensitive nonlinear transport measurements. 

The $mm2$ point group of \AuPbP\ restricts nonlinear transport to very specific symmetry-allowed configurations. A second-harmonic voltage ($V_{ij}^{2\omega}$) in response to an AC current ($I^{\omega}$) manifests exclusively along the polar $c$-axis, appearing in both longitudinal ($V_{zz}^{2\omega}$) and transverse ($V_{zy}^{2\omega}$ and $V_{zx}^{2\omega}$) geometries.
Here, we define our subscripts $i,j$ as the crystallographic directions aligned with the voltage and current leads, respectively. All other configurations are forbidden by symmetry, and the corresponding $V^{2\omega}$ should vanish~\cite{SuarezRodriguez_2025, Aroyo_2006, Aroyo_2006b, Gallego_2019}. We performed current-dependent lock-in measurements to isolate $V^{2\omega}$ and confirm the polar nature of our crystals.

\begin{figure}[H]
    \resizebox{1\textwidth}{!}{\includegraphics{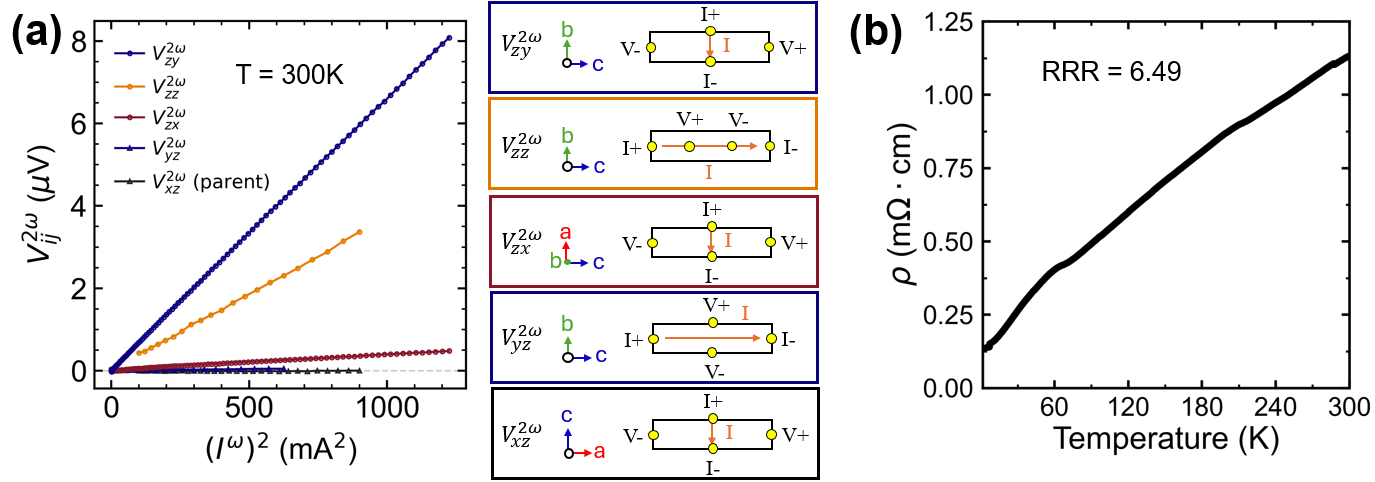}}
    \caption{\textbf{Electrical transport behavior in \AuPbP.} \textbf{(a)} Nonlinear transport measurements at 300~K confirm the $V^{2\omega} \propto (I^\omega)^2$ scaling expected for point group $mm2$. Each color represents a different measurement geometry; the two blue datasets ($V_{zy}^{2\omega}$ and $V_{yz}^{2\omega}$) are measured on the same crystal face with current and voltage leads interchanged. Circles indicate geometries where second-harmonic voltage response is expected, while triangles denote symmetry-forbidden geometries. The parent compound (black) shows no measurable signal, consistent with the absence of broken inversion symmetry. Diagrams to the right illustrate the contact configuration for the devices for each measurement. \textbf{(b)} Temperature-dependent resistivity of \AuPbP, confirming metallic behavior and therefore characterizing it as a polar metal.}
    \label{Nonlinear}
\end{figure}

We indeed observe significant nonlinear components in longitudinal measurements along the polar axis ($V_{zz}^{2\omega}$) and both symmetry-allowed transverse ($V_{zy}^{2\omega}$, $V_{zx}^{2\omega}$) configurations. Figure~\ref{Nonlinear}\textbf{(a)} shows the second-harmonic voltage as a function of squared applied current at 300~K. The data reveal a pronounced quadratic dependence of $V^{2\omega}$ as predicted for a second-order effect ($V^{2\omega} \propto (I^{\omega})^2$), and it is observed in the configurations allowed by the $mm2$ point group symmetry (for further details, see Supplementary Section 4).

To confirm the intrinsic, polar origin of these signals and rule out Joule heating or other possible artifacts, we performed two additional tests detailed in the Supporting Information: (i) simultaneous contact-swap measurements, which demonstrate that $V^{2\omega}$ reverses sign under full current and voltage lead inversion while $V^{\omega}$ does not, consistent with the expected antisymmetry of a genuine nonlinear polar response; and (ii) a symmetry-disallowed geometry ($V_{yz}^{2\omega}$), which yields a second harmonic signal approximately two orders of magnitude smaller than the allowed $V_{zy}^{2\omega}$ on the same crystal face (Figure~\ref{Nonlinear}\textbf{(a)}, blue triangles). We further confirmed the complete absence of any $V^{2\omega}$ signal in the crystals of the centrosymmetric parent compound \AuPbPfull\ (Figure~\ref{Nonlinear}\textbf{(a)}, black triangles). Together, the contact-swap antisymmetry, geometric null in a disallowed direction, and the absence of signal in the centrosymmetric parent compound constitute a robust symmetry argument for intrinsic polar nonlinear conductivity in \AuPbP, directly confirming the topotactic induction of polarity and demonstrating how post-synthetic symmetry-breaking can generate novel electronic functionality absent in the parent structure.

The temperature-dependent longitudinal resistivity (Figure~\ref{Nonlinear}\textbf{(b)}) confirms metallic behavior across the full measured range, with a residual resistivity ratio of $RRR = 6.49$, establishing \AuPbP\ as a polar metal. Having established the polar metallic ground state, we turn to the low-temperature properties of this material.

\subsection*{Superconductivity characterization}

\begin{figure}[H]
    \includegraphics[width=1\textwidth]{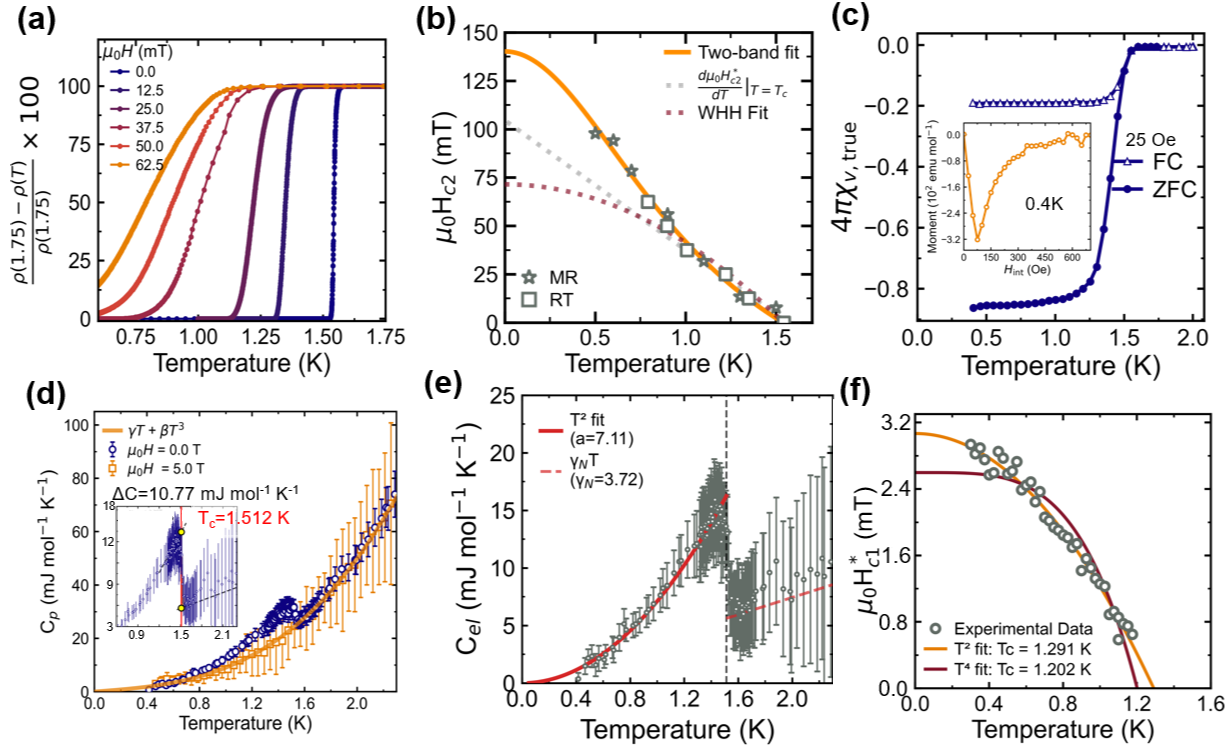}
    \caption{\textbf{Transport and magnetic characterization of the superconducting transition in \AuPbP.} \textbf{(a)} Normalized resistivity measurements at various applied magnetic fields and \textbf{(b)} Temperature dependence of the upper critical field fit to a simplified Werthamer-Helfand-Hohenberg (WHH) fit (red) and a two-band Ginzburg-Landau fit (gold). Experimental data shows an upward curvature close to $T_c$, and departure from WHH behavior, suggestive of multi-gap behavior. \textbf{(c)} DC magnetic susceptibility temperature dependence of the superconducting transition, where ZFC-FC irreversibility indicates type-II behavior. Inset shows a M-H sweep measured at 0.4~K. The sharp dip, and longer tail further indicate type-II behavior. \textbf{(d)} Zero-field (blue) and Normal state (gold) fitting of Heat capacity data showing the observed heat capacity jump of the superconducting transition. Inset shows a zoomed in portion of the normalized jump, with an equal area constructing fit with a critical temperature of $T_c = 1.521$~K and jump of $\Delta C = 10.77$~meV. \textbf{(e)} Below $T_c$, the electronic heat capacity fits well to a $T^2$ dependence. \textbf{(f)} Results of the low-temperature AC susceptibility experiments, finding a better fit to $1-(T/T_c)^2$, suggesting power-law behavior, compared to the nearly-exponential (T$^4$, red) fit expected for an isotropic $s$-wave superconducting gap.}
    \label{Superconductor}
\end{figure}

We find that \AuPbP\ is a bulk superconductor with a resistivity drop at 1.54~K (Figure~\ref{Superconductor}\textbf{(a)}). In zero field, the transition exhibits remarkable sharpness with a transition width of only $\Delta T = 0.016$~K, suggesting that the high sample quality and crystallinity is retained through the electrochemical deintercalation. This finding is notable for two reasons: First, noncentrosymmetric conductors with strong spin-orbit coupling can enable unique pairing mechanisms in a superconducting state. Second, the parent compound \AuPbPfull\ is not superconducting above 0.22~K (see Fig. S\ref{Cp_Parent}), implying that superconductivity is induced via the topotactic transformation. The superconducting transition remains robust in these crystals even after weeks of exposure to air.

We characterize the superconducting state through complementary measurements. Normalized resistivity measurements at various applied magnetic fields (Figure~\ref{Superconductor}\textbf{(a)}) show systematic suppression of $T_c$. The temperature dependence of the upper critical field fits to a simplified Werthamer-Helfand-Hohenberg (WHH) model (red) and a phenomenological two-band Ginzburg-Landau fit (gold), with experimental data showing an upward curvature close to $T_c$ and departure from WHH behavior, suggestive of multi-gap behavior (Figure~\ref{Superconductor}\textbf{(b)}). DC magnetic susceptibility shows type-II behavior, with M-H sweeps confirming the superconducting transition (Figure~\ref{Superconductor}\textbf{(c)}).

Heat capacity measurements performed across the superconducting transition show a clear jump at $T_c = 1.52$~K (Figure~\ref{Superconductor}\textbf{(d)}), with $\Delta C/(\gamma_N T_c) = 1.92 \pm 0.49$. While this may imply strong coupling behavior, the temperature dependence of the electronic contribution $C_{el}$, obtained after subtracting the normal-state phonon contribution $\beta T^3$, points to an alternative interpretation. We modeled $C_{el}$ using five approaches described in Section~5.1 of the Supporting Information: a single isotropic $\alpha$-model, a two-gap $\alpha$-model, both models augmented by a residual electronic term $\gamma_\mathrm{res}T$ to account for an impurity phase, and a $T^2$ power law. Of these, only the $T^2$ power law provides a physically consistent description of the data (Figure~\ref{Superconductor}\textbf{(e)}).

A $T^2$ dependence of $C_{el}$ suggests the presence of nodes in the superconducting gaps, where quasiparticle excitations at low temperatures arise from regions where the superconducting gap vanishes.\cite{Sigrist_1991,Smidman_2017} To support this, we employed AC susceptibility measurements which reveal the phase diagram of the relative lower critical field ($\mu_0 H_{c1}^{*}$) with respect to temperature, shown in Figure~\ref{Superconductor}\textbf{(f)}. These data points are then fit to equations representing behavior expected from a fully-gapped (Figure~\ref{Superconductor}\textbf{(f)} T$^4$, red) and nodal (T$^2$, gold) superconducting behavior. There is a strong deviation from the fully gapped behavior in the temperature dependence of the lower critical field. Instead, the data agree better with a $n=2$ power law, which extrapolates to a lower critical field of $\mu_0 H_{c1}^{*} = 3.050$~mT.

In the dirty limit, the quadratic temperature dependence of both $C_{el}$ and $\chi'$ cannot rigorously distinguish between point and line nodes in the superconducting gap~\cite{Gross_1986}. It can, however, differentiate a nodal gap structure from a single fully-gapped isotropic one. The mutual consistency of the $T^2$ behavior observed in both the heat capacity and the lower critical field, combined with structural and transport evidence for broken inversion symmetry, thus provides circumstantial evidence that \AuPbP\ hosts nodes in its superconducting gap.

We note, however, that $T^2$ behavior in $C_{el}$ is not uniquely diagnostic of nodal superconductivity. A fully-gapped multiband superconductor, consistent with the two-band character suggested by $H_{c2}(T)$, can exhibit apparent power-law behavior at intermediate temperatures as the smaller gap is thermally depopulated. Models incorporating a residual Sommerfeld term $\gamma_{res} T$ to account for a non-superconducting impurity fraction were explicitly considered, however, the impurity contribution required to describe the low-temperature $C_{el}$ produces a superconducting jump ratio $\Delta C / (\gamma_{SC} T_c) \approx 3.7$ inconsistent with any physically reasonable coupling scenario (see SI Section~5.1). Furthermore, the $T^2$ temperature dependence of $H_{c1}^*$ provides a line of evidence largely independent of non-superconducting impurity contributions, which would affect the electronic heat capacity but not the penetration depth of the superconducting condensate. Taken together, the nodal interpretation remains the most self-consistent description of all available data, though a fully-gapped multiband scenario cannot be rigorously excluded.

\subsection*{Discussion}

The results presented here establish electrochemical topotactic oxidation as a versatile route to polar metals through chemically directed symmetry-breaking. In the specific case presented here, the reaction produces a polar superconductor with thermodynamic signatures suggestive of a gap structure governed 
by the broken inversion symmetry of the host lattice. The transition from the centrosymmetric parent structure, \AuPbPfull, into the polar deintercalated structure, \AuPbP, is verified through nonlinear transport. Both diffraction and transport experiments point to high crystal quality even after the post-synthetic electrochemical treatment, a result that distinguishes this approach from other topotactic methods.

This synthetic approach offers several advantages over traditional solid-state synthesis. First, post-synthetic electrochemical treatment provides very selective control over the thermodynamics of the system, allowing access to metastable compositions otherwise outside thermodynamic stability. Second, the methodology allows one to grow bulk single-crystal phases that are retained through treatment, enabling definitive single-crystal X-ray diffraction and transport characterization. Third, electrochemistry allows real-time monitoring and precise control of reaction progress.

To fully capture the advancements provided by this method, it is helpful to compare to other systems where (de)intercalation induces novel electronic states. Cu$_x$Bi$_2$Se$_3$ represents the closest methodological analogy, where electrochemical intercalation of Cu into pristine Bi$_2$Se$_3$ has been 
shown to introduce superconductivity with $T_c \approx 3.8$~K~\cite{Hor_2010, Kriener_2011}. However, despite over a decade of study, the position of the Cu dopant has never been established by crystallography. Single-crystal neutron diffraction experiments testing all proposed intercalation and interstitial sites find no statistically significant Cu occupation at any position, and are concluded to be positionally disordered beyond crystallographic detection~\cite{Frohlich_2020}. 

Cu$_x$TiSe$_2$ offers an instructive comparison from a different synthetic perspective: here Cu is incorporated during conventional solid-state synthesis rather than by post-synthetic electrochemical treatment, and suppresses a CDW transition to produce superconductivity near $T_c \sim 4.15$~K~\cite{Morosan_2006}. Synchrotron single-crystal diffraction has located Cu in the high-temperature phase at an interstitial site between TiSe$_2$ layers~\cite{Kitou_2019}, and with increasing Cu content the commensurate CDW of pristine TiSe$_2$ gives way to an incommensurate modulation coinciding with the onset of superconductivity~\cite{Kogar_2017, Kitou_2019}. The structural and mechanistic origin of this behavior remains an open question~\cite{Kitou_2019, Budko_2007}, and critically, the modulated structure has not been solved. Similarly, partial deintercalation combined with water intercalation has been shown to unveil superconductivity in Na$_x$CoO$_2 \cdot y$H$_2$O~\cite{Chou_2004}, though here too a precise structural description of the active phase remains challenging due to the disorder introduced by the guest species. In all of these cases, the weak or poorly characterized dopant-host interactions limit the degree of structural control available through (de)intercalation, and no complete ordered crystallographic description of the transformed phase has been achieved.

Our approach differs fundamentally in both strategy and mechanism. In Au$_2$Pb$_{0.914}$P$_2$, the remaining Pb atoms are not passive guests but active agents of the symmetry-breaking transformation. Partial deintercalation changes the oxidation state of the remaining Pb from near-neutral to Pb$^{2+}$, directly triggering the SOJT distortion and stereoactive lone pair activity that lock the structure into the polar superspace group $Ama2(01\gamma)ss0$. Deintercalation and symmetry-breaking are not separable events; they are inseparable consequences of the same electronic reconstruction. This is categorically different from the systems discussed above, where the inserted or removed species modifies the electronic structure of the host and the structural response either cannot be located crystallographically or remains incompletely characterized. The structural precision we achieve, a complete (3+1)D superspace solution with unambiguous polar symmetry assignment, is a direct consequence of the cooperative mechanism in which Pb deintercalation and symmetry-breaking are one and the same event, naturally selecting for an ordered polar superstructure over a disordered vacancy arrangement.

\section*{Conclusion}

We have demonstrated that electrochemical oxidation provides a controllable route to polar metallicity starting from the non-polar parent structure of \AuPbPfull, enabling a phase-pure superconducting \AuPbP. The topotactic reaction provides single crystals suitable for comprehensive structural and physical characterization. Three main achievements distinguish this work:

First, methodologically, through electrochemical routes, we solved sample inhomogeneity issues that were inherent in soft-chemical approaches. The resulting uniformity, validated through a combination of XPS depth profiling and $\mu$CT measurements, proved essential for all following structural and physical properties characterization. This optimized post-synthetic method, specifically targeted to perturb ions that could be heavily involved in chemically directed symmetry-breaking, represents a significant advancement in the modification of symmetry elements in quantum materials.

Second, structurally, we achieved a precise model of a (3+1)D commensurately modulated superstructure through synchrotron X-ray diffraction, establishing the symmetry reduction of the topotactically oxidized phase to superspace group $Ama2(01\gamma)ss0$. To the best of our knowledge, this represents the first complete structural solution of a post-synthetic electrochemically processed polar metal. The solved structure reveals insight into the exact ordering of Pb ions and its coupling to the breathing-like motions in the distorted [Au$_2$P$_2$] framework.

Third, physically, we validate the predicted polar metal behavior through symmetry-allowed nonlinear transport measurements. Moreover, we found that this symmetry-breaking leads to observation of a new bulk type-II superconductor with thermodynamic and magnetic measurements suggestive of a gap structure governed by the broken inversion symmetry of the host lattice. The $T^2$ power-law behavior observed in both heat capacity and AC susceptibility is consistent with a gap structure shaped by antisymmetric spin-orbit coupling, a consequence of the broken inversion symmetry established by structural and transport characterization.

Beyond the specific case of \AuPbP, our results suggest a generalizable strategy for retro-synthetic design of polar metals through topotactic postprocessing. This is a binary system --- only \AuPbPfull\ and \AuPbP\ are accessible --- but close reexamination of partially deintercalated structures across the broader Au$_2M$P$_2$ family may offer a broadly applicable route to finding new polar metals through breaking symmetries of parent structures. Looking forward, the combination of controlled topotactic reactions, superspace crystallography, and symmetry-directed property prediction provides a template for rational design of quantum materials with broken symmetries, an area where electrochemistry, structural science, and condensed matter physics meet to enable the discovery of new, computationally unpredictable, quantum materials.

\clearpage
\bibliography{science_template}
\bibliographystyle{sciencemag}

\section*{Acknowledgments}

\paragraph*{Funding:}
This material is based upon work supported by the Air Force Office of Scientific Research under award number FA9550-25-1-0177. Any opinions, findings, and recommendations expressed in this material are those of the author(s) and do not necessarily reflect the views of the United States Air Force. Further support was provided by the NSF through the Princeton Center for Complex Materials DMR-2011750 an NSF-funded MRSEC, as well as by the Gordon and Betty Moore Foundation's EPiQS initiative through Grant No. GBMF9064 and the David and Lucile Packard foundation. SBL is supported by the National Science Foundation Graduate Research Fellowship Program under Grant No. DGE-2039656. X.Z. was supported by the National Science Foundation Graduate Research Fellowship Program under grant number DGE-2146755. Any opinions, findings, or conclusions or recommendations expressed in this material are those of the author(s) and do not necessarily reflect the views of the National Science Foundation. The authors acknowledge the use of Princeton's Imaging and Analysis Center (IAC), which is partially supported by the Princeton Center for Complex Materials (PCCM), a National Science Foundation (NSF) Materials Research Science and Engineering Center (MRSEC; DMR-2011750). ChemMatCARS, Sector 15 at the Advanced Photon Source (APS), Argonne National Laboratory (ANL), is supported by the Divisions of Chemistry (CHE) and Materials Research (DMR), National Science Foundation, under grant number NSF/CHE-2335833. This research was performed on APS beam time award(s) (DOI: <beam time award DOI hyperlinks>) from the Advanced Photon Source, a U.S. Department of Energy (DOE) Office of Science user facility operated for the DOE Office of Science by Argonne National Laboratory under Contract No. DE-AC02-06CH11357. The 3He MPMS was funded by the National Science Foundation, Division of Materials Research, Major Research Instrumentation Program, under Award number 1828490. SBL would like to thank Manuel Suarez-Rodriguez for insights into nonlinear transport, Vaclav Petricek for discussion of (3+1)D crystallography, Andrew B. Bocarsly for discussion on electrochemical measurements, as well as Fang Yuan, Suchismita Sarker, and Connor Pollak for insightful discussion of the project.

\paragraph*{Author contributions:}
S.B.L. and L.M.S. conceived and designed the study. S.B.L., S.R.D., and X.Z. synthesized and electrochemically processed the crystals. S.B.L. S.R.D. and J.W.S. performed the XPS and UPS measurements. S.B.L. performed nonlinear and electronic transport measurements as well as AC susceptibility measurements. S.B.L. X.Z. S.C. and J.M. performed heat capacity measurements. F.K.B. performed DFT calculations. S.C., X.Z., C.L., A.G.I., A.N.N., and T.M.M. assisted with magnetic susceptibility measurements. S.B.L., T.C., and Y.-S.C. collected synchrotron diffraction data. S.B.L. performed the modulated structure refinement. X.Z. performed $\mu$CT measurements. S.B.L. and L.M.S. wrote the manuscript with input from all authors.

\paragraph*{Competing interests:}
The authors declare no competing interests.

\paragraph*{Data and materials availability:}
Crystallographic data have been deposited at The Cambridge Crystallographic Data Centre (CCDC) with Deposition Number 2547689. All other data needed to evaluate the conclusions in the paper are present in the paper or the Supplementary Materials.

\paragraph*{Supplementary materials}
Materials and methods, sample optimizations, details of (3+1)D refinement, crystallographic tables, density of states and band structure calculations, ultraviolet photoemission spectroscopy, nonlinear resistivity, heat capacity, electronic transport, and magnetic susceptibility.

\newpage

\section*{Supplementary Materials for\\ \scititle}
\setcounter{figure}{0}
\subsection*{This PDF file includes:}

Figures S1 to S41\\
Tables S1 to S19\\

\subsection*{Other Supplementary Material for this manuscript includes the following:}

Crystallographic Information Files (CIF)

\newpage

\section*{Materials and Methods}

\subsection*{Materials and Synthesis}
\textbf{Safety Warning:} P has a low solubility in a Pb metal flux. Attempts to scale up this reaction may lead to an explosion of the ampule containing toxic Pb fumes. Additionally, reacted samples occasionally contain small amounts of excess white phosphorus. Any future attempts of these syntheses, and the subsequent opening of ampules, should be carried out in a fumehood or properly ventilated environment.

Samples of orthorhombic \ce{Au2PbP2} were synthesized using the self-flux method utilizing a Canfield crucible. In the bottom alumina crucible, Au (Thermo Scientific, 99.999\%), red P (Alfa Aesar, 99.999+\%), and Pb (Alfa Aesar, 99.999\%) were loaded in a 1:12:1 stoichiometric ratio of Au:Pb:P. The crucible was placed in a fused-quartz tube and sealed under dynamic vacuum at $\sim$70~mTorr after being back-filled three times with Ar. The sealed tube was then placed in a muffle furnace, ramped to 450~$^\circ$C over 4 hours, and held there for 24 hours to pre-react phosphorus. The furnace was then ramped to 950~$^\circ$C over 5 hours, kept at this temperature for 48 hours, and then quickly cooled to 700~$^\circ$C. At 700~$^\circ$C, the cooling rate was decreased to 2~$^\circ$C per hour to a target set point of 375~$^\circ$C, at which point the crucible was centrifuged. All crystals were metallic grey in color.

\subsection*{Electrochemical Postprocessing}
Electrochemical experiments were performed using a CH Instruments 1140 potentiostat with a three-electrode cell that employed an Ag/AgCl reference electrode and a Pb counter electrode. Single crystals of the undistorted \ce{Au2PbP2} served as the working electrode. The working electrode was assembled by attaching a 0.25mm diameter gold wire (Thermo Scientific, 99.9\%) with silver paste (Dupont 4929N) to provide ohmic contact. The contacts are positioned on the face with smallest area, perpendicular to the longest crystal axis. The contact area was then surrounded by insulating epoxy as to not interfere with the following electrochemical experiments. All electrolytes were degassed through bubbling Ar for 30 minutes prior to each experiment.

\subsection*{X-ray Photoelectron Spectroscopy}
X-ray photoelectron spectroscopy (XPS) was performed on a Thermo Fisher Scientific K-Alpha+ XPS system equipped with an Al K$\alpha$ X-ray source. Small (800 $\times$ 300 $\times$ 200 $\mu$m) single crystals of undistorted \ce{Au2PbP2}, acid etched Au$_2$Pb$_{1-x}$P$_2$ and electrochemically treated Au$_2$Pb$_{0.914}$P$_2$ were mounted on Cu tape with the largest surface area mounted parallel to the tape for XPS. Ar ion etching was used to ablate the surface of the crystal. We found that the intermediate values of Ar1000 cluster size and 5~keV total energy effectively remove material without significant drift in observed binding energies. Samples were ablated in 1 hour intervals between each scan. All measurements were taken under ultrahigh vacuum.

\subsection*{Micro-Computed X-ray Tomography}
Micro-Computed X-ray Tomography ($\mu$CT) data were collected using a Zeiss Xradia Versa 630 and Xradia Versa 520. The Versa 630 was operated at a voltage of 160 kV and a power of 25W. The Versa 520 was operated at a voltage of 140 kV and a power of 10W. The samples were mounted on a 27mm nickel-plated steel extra-fine satin pin using GE 7031. Filters were selected such that the transmission percentage is between 20-35\%. For all samples measured, the voxel size was between 2 to 5 microns. The variation was due to the use of different magnifications to keep the entire measured sample within the instrument's field of view. Two-dimensional images were collected by rotating the sample 360$^\circ$ in increments of at least 0.15$^\circ$. Exposure times were chosen such that the average counts in the measured sample were greater than 5000. Reconstruction of the two-dimensional exposures was performed with Zeiss Scout-and-Scan Control System Reconstructor software (Version 16.8.19905.48502), where beam hardening correction algorithms were applied.  The reconstructed images were visualized with the Avizo software (Version 9.0.0).

\subsection*{Ultraviolet Photoelectron Spectroscopy}
Ultraviolet photoelectron spectroscopy (XPS) was performed on a Thermo Fisher Scientific K-Alpha+ XPS system using a He I source ($hv$ = 21.218~eV) with an applied sample bias of -6~V to access the secondary electron cutoff (SECO). Work functions were determined by linear extrapolation of the SECO rising edge to zero intensity, with the bias voltage subtracted so that the Fermi edge falls at $hv$ = 21.218~eV.

\subsection*{Single Crystal Diffraction Experiments}
Small single crystals of the resulting samples were screened by single-crystal X-ray diffraction (SCXRD) analysis using an APEX2 CCD diffractometer equipped with a Mo K$\alpha$ ($\lambda$=0.71073 \AA) sealed-tube X-ray source and graphite monochromator at 100 K. Screened crystals were then taken for full data collection at NSF's ChemMatCARS, Sector 15 at the Advanced Photon Source, Argonne National Laboratory on a Pilatus3 X 2M detector with a photon energy of 30keV ($\lambda = 0.41328$~\AA). Indexation and integration were completed for a full sphere collection out to a resolution of 0.7 \AA.  Run list generation APEX 5 and frame data processing were done in APEX 4. An analytical absorption correction was used to scale the data before importing the peak list into JANA2020 \cite{Petricek_2023}. 

\subsection*{DFT Calculations}
All density functional theory calculations were performed using the Vienna \textit{ab initio} Simulation Package (VASP) with the Perdew-Burke-Ernzerhof (PBE) exchange-correlation functional. Structures were taken directly from the single-crystal solutions without further geometric optimization. These calculations do account for spin-orbit coupling, except for those explicitly mentioned. Self-consistent calculations employed the generalized gradient approximation and the projector augmented wave potentials provided with the package \cite{Blochl_1994, Kresse_1999} with Pb semi-core $d$ orbitals treated as part of the valence set. Band unfolding calculations were done with easyunfold package\cite{easyunfold}. In single point and density of states calculations, a convergence of 0.1~meV was achieved by using a $\Gamma$-centered $34 \times 34 \times 17$ k-point mesh and energy cutoff of 520~eV.

\subsection*{Electronic transport}
Alternating current electronic transport measurements were performed using a Quantum Design Physical Property Measurement System with a dilution fridge attachment. Gold wires were connected in a four probe geometry using conducting silver paste (Dupont 4929N) to small needle-like single crystals of the electrochemically treated phase. The crystallographic axes for the transport measurements were determined from the morphology of the crystals and alignment in SCXRD.

Data for nonlinear resistivity measurements utilized a LakeShore Cryotronics lock-in amplifier with frequencies adjusted such that the measured second harmonic generator ensures a phase angle of 90 degrees, the expected phase shift from second harmonic generation. Measurement probes for simultaneous first and second harmonics were split off of the same pins. I-V curves for second harmonic generation were taken while sweeping current from 5~$\mu$A to 35~mA in 100 linear steps, followed by sweeping back to 5~$\mu$A for each temperature. Data points at the same current and temperature were then averaged. Measurements were performed on two separate sample, one sample for V$_{zy}$, V$_{yz}$, V$_{zz}$, and the contact-swapped geometries discussed in the SI. The second sample was used to measure V$_{zx}$ and its contact-swapped geometry V$_{-z-x}$.

Data for low temperature measurements utilized a Quantum Design dilution refrigerator with $^3$He/$^4$He mixture. For all measurements that were not frequency dependent, the frequency of the AC current was set at 24.4~Hz.

\subsection*{Heat Capacity}
Specific heat measurements were performed using a Quantum Design Physical Property Measurement System with dilution refrigerator and DRHC attachments. Different addendas were needed for both the zero-field and normal state measurement. After the zero-field addenda, zero-field measurements were done with a crystal with mass $0.56 \pm 0.05$~mg. The crystal was then carefully removed and washed 5 times with toluene to dissolve any remaining N grease on the surface of the crystal. After another addendum at 5T, this crystal was measured again.

\subsection*{Magnetic Susceptibility Measurements}
Data for low temperature DC magnetic susceptibility measurements utilized a Quantum Design Magnetic Properties Measurement system equipped with a $^3$He dilution refrigerator. Electrochemically treated samples were mounted on Kapton straws with GE varnish, with their axes aligned on the basis of the orientation of previous single-crystal X-ray diffraction experiments.

\subsection*{AC Susceptibility Measurements}
Data for low temperature AC magnetic susceptibility measurements utilized a Quantum Design Physical Property Measurement System with dilution refrigerator and ACDR attachments. Samples were mounted using a tiny amount of GE varnish on a sapphire substrate. AC excitations were  applied with 1 Oe amplitude and 400 Hz frequency to minimize Eddy currents from forming. DC applied field was swept from 0 to 100 Oe in the temperature range of 0.3-1.6 K with an interval of 0.025 K.

\section{Sample Optimization}
\subsection{Electrochemical Postprocessing} 

Cyclic voltammetric experiments were performed in aqueous 0.5 M Pb(NO$_3$)$_2$ in the potential window of -0.4 V to +0.6 V vs Ag/AgCl at a scan rate of 25 mV/s. A representative cyclic voltammogram can be shown below in Figure S\ref{CV1}. After the first cycle, subsequent CVs showed a peak attributed to Pb oxidation (from Pb(0) to Pb(II)) at +0.18 V vs Ag/AgCl that diminishes with increasing cycles. A reduction event occurs starting at -0.18 V vs Ag/AgCl. Following single-crystal X-ray diffraction experiments, it is determined that this reducing event is attributed to Pb plating on the surface of the crystal, rather than intercalation of Pb into this topotactically transformed structure. This irreversibility is likely due to the atomic relaxation of Au positions in the \textbf{a}-\textbf{c} plane, elaborated further in the main text and supplemental information.

\begin{figure}[H]
\renewcommand{\figurename}{Figure S}
    \centering
    \resizebox{!}{!}{\includegraphics{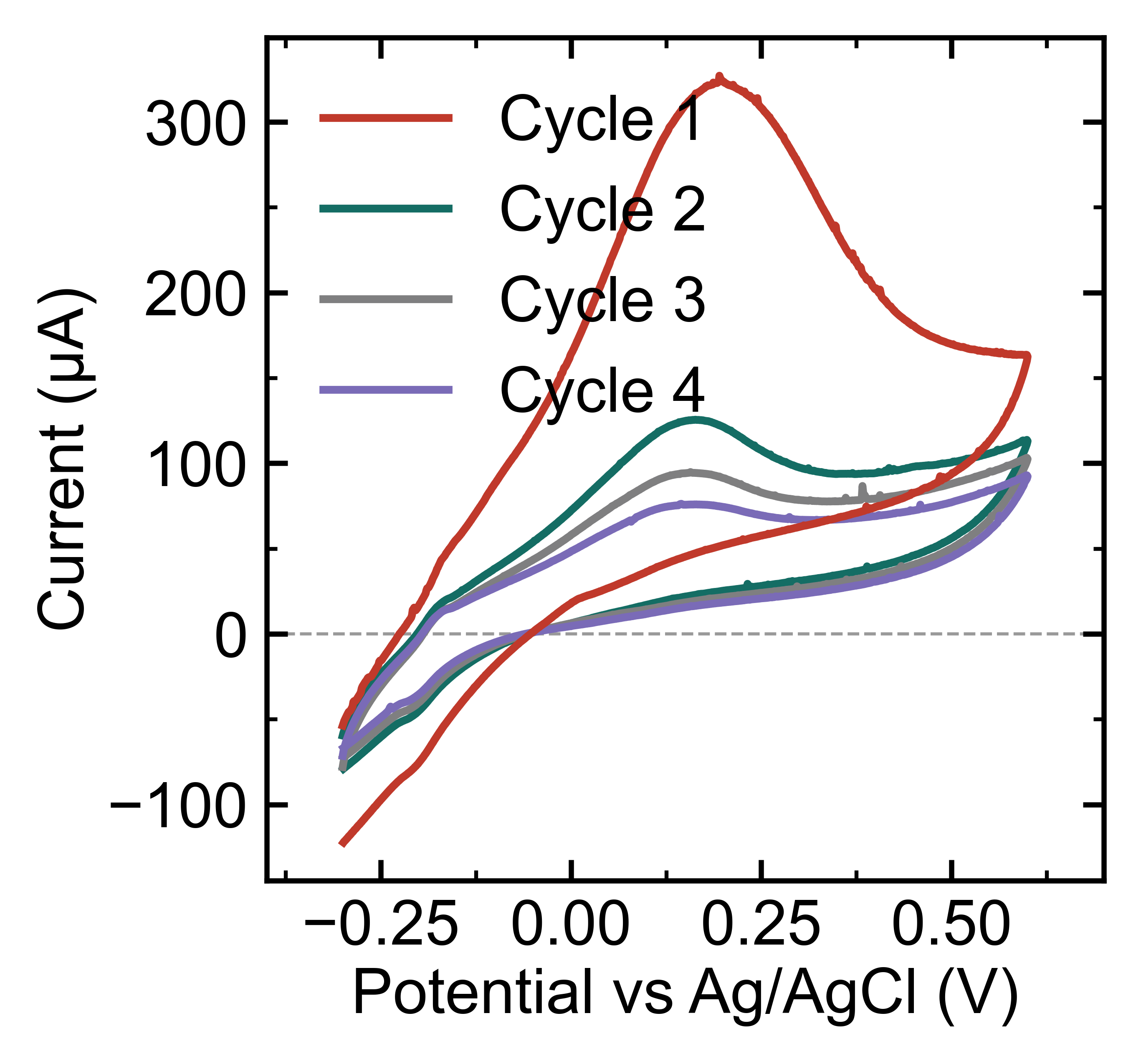}}
    \caption{Cyclic Voltammogram of \ce{Au2PbP2} versus Ag/AgCl as a reference electrode in the potential window of -0.4 V to +0.6 V. The reduction peak at $\sim-0.18$~V corresponds to Pb plating on the surface of our crystal.}
    \label{CV1}
\end{figure}

\begin{figure}[H]
\renewcommand{\figurename}{Figure S}
    \centering
    \resizebox{!}{!}{\includegraphics{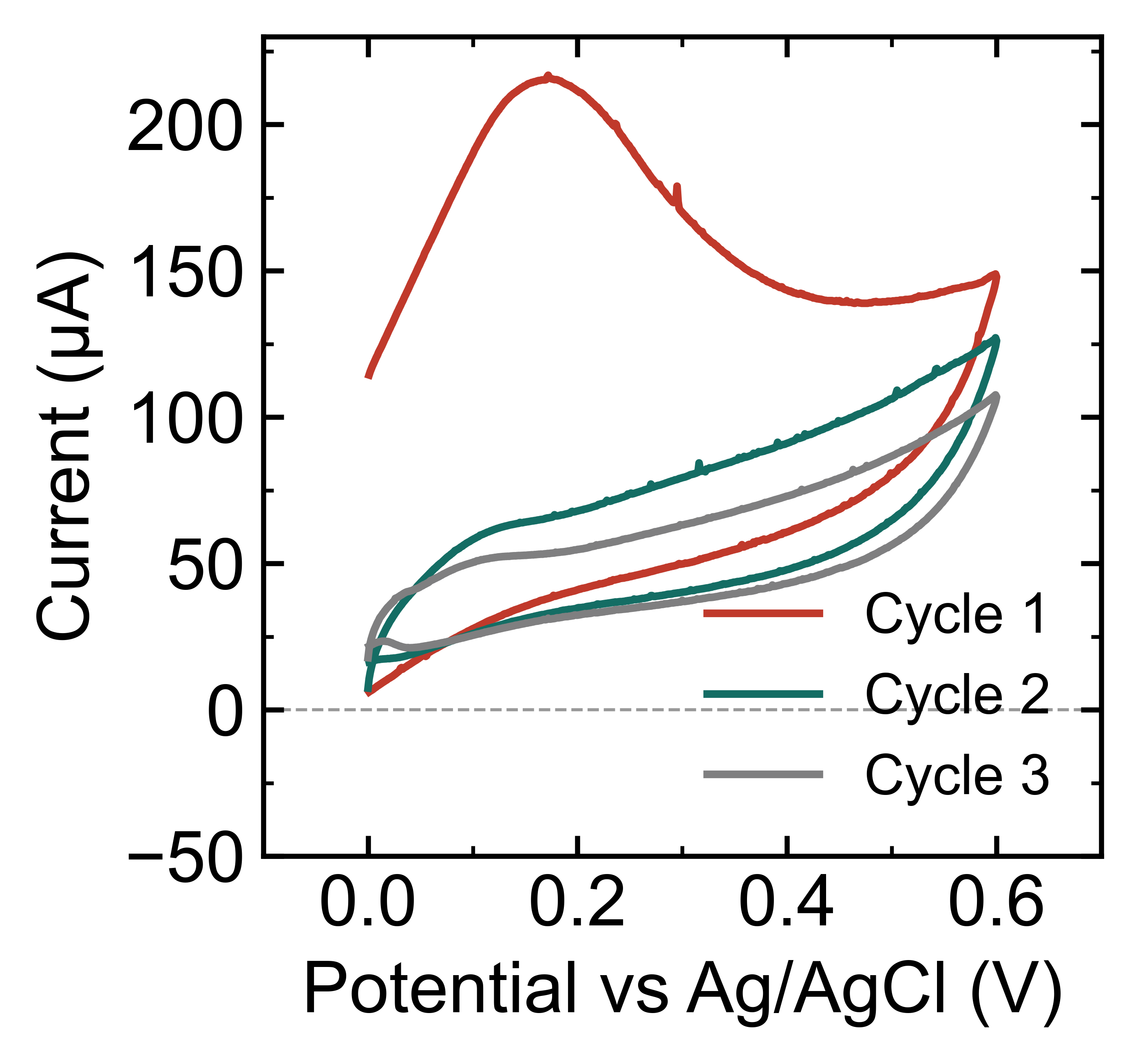}}
    \caption{Cyclic Voltammogram of \ce{Au2PbP2} versus Ag/AgCl as a reference electrode in the reduced potential window of 0 V to +0.6 V. The diminished oxidation peak at $\sim0.18$ in subsequent cycle indicates that in this potential window no more Pb can be deintercalated.}
    \label{CV2}
\end{figure}

Decreasing the potential window to 0 V to +0.6 V vs Ag/AgCl avoids any reduction event shown in Figure S\ref{CV2}. Here, cycle 1 has an oxidation event occurring at +0.18 V vs Ag/AgCl, with subsequent cycles having no such peak, indicative of a complete deintercalation after one cycle. Thus, the peaks in cycles 3 and 4 in Figure S\ref{CV1} are believed to arise from re-oxidation of Pb plated on the crystal.

Controlled potential coulometry was performed at +0.3 V vs Ag/AgCl at various lengths of time, ranging from 210 to 10000 seconds. Samples at shorter times result in crystals to be less brittle, but could leave more Pb flux inclusions compared to those held longer. We found a time of 600 seconds to be a good trade off between these two considerations. A representative potential coulometry can be shown below in Figure S\ref{coulometry}.

\begin{figure}[H]
\renewcommand{\figurename}{Figure S}
    \centering
    \resizebox{0.5\textwidth}{!}{\includegraphics{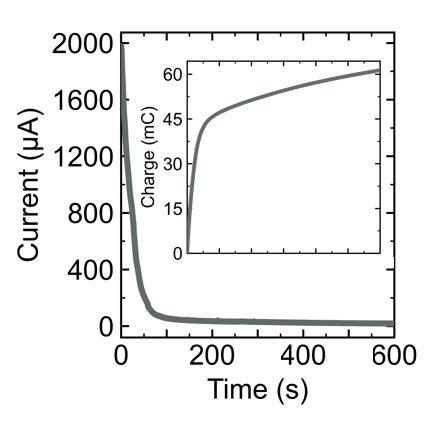}}
    \caption{\textbf{Controlled potential coulometry }performed at +0.3 V vs Ag/AgCl with Pb counter electrode for 600 seconds. Inset shows the total charge passed over the time period.}
    \label{coulometry}
\end{figure}

\FloatBarrier

\subsection{X-ray photoelectron spectroscopy}
\begin{figure}[ht!]
\renewcommand{\figurename}{Figure S}
    \centering
    \includegraphics[width=\textwidth]{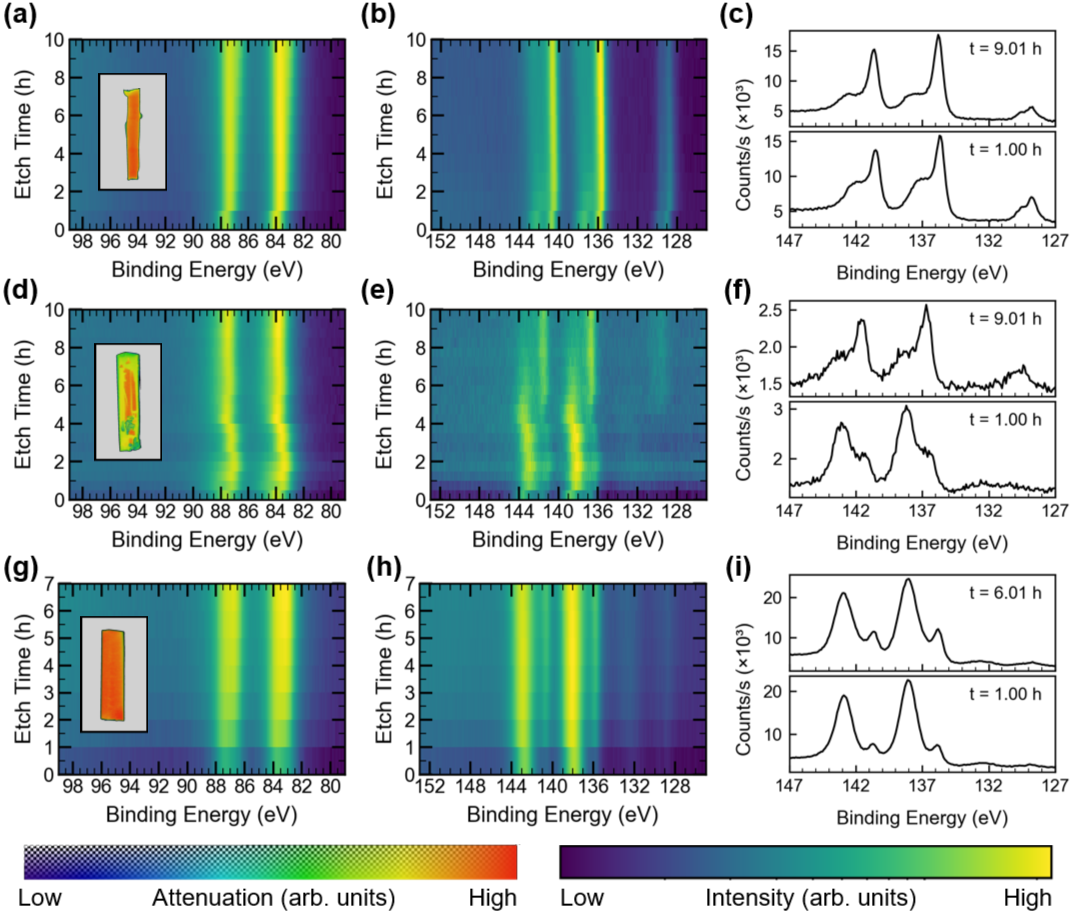}
    \caption{\textbf{X-ray photoelectron spectroscopy depth profiles and micro-computed tomography images for topotactic transformations of \ce{Au2PbP2}.} \textbf{(a)} XPS spectra for the untreated parent compound, \ce{Au2PbP2}, around Au~4$f_{7/2}$ and 4$f_{5/2}$ peaks. The inset shows a $\mu$CT image taken of the crystal demonstrating sample homogeneity. \textbf{(b)} The corresponding XPS depth profile around Pb~4$f$ and P~2$p$ binding energies. \textbf{(c)} Individual scans after Ar ion ablation of the surface of crystals for 1 hour (bottom) and 9 hours (top). \textbf{(d)--\textbf{(f)}} show the same profiles for a \ce{Au2PbP2} crystal soaked in concentrated HNO$_3$ for 15 minutes, demonstrating an inhomogeneous transition. \textbf{(g)--\textbf{(i)}} are the corresponding profiles for an electrochemically treated \ce{Au2PbP2}, demonstrating a more homogeneous sample throughout.}
    \label{XPS}
\end{figure}

To track the topotactic transformation and confirm phase purity, we used X-ray photoelectron spectroscopy (XPS) to probe the oxidation states in the pristine, soft-chemical processed, and electrochemically oxidized structures. The spatial distribution of this electronic transformation was mapped through depth-profile XPS experiments, in which sequential surface ablation at one-hour intervals revealed depth-dependent shifts in oxidation state, or their absence, throughout the crystal (Figure S\ref{XPS}).

\begin{figure}[H]
\renewcommand{\figurename}{Figure S}
    \centering
    \resizebox{\textwidth}{!}{\includegraphics{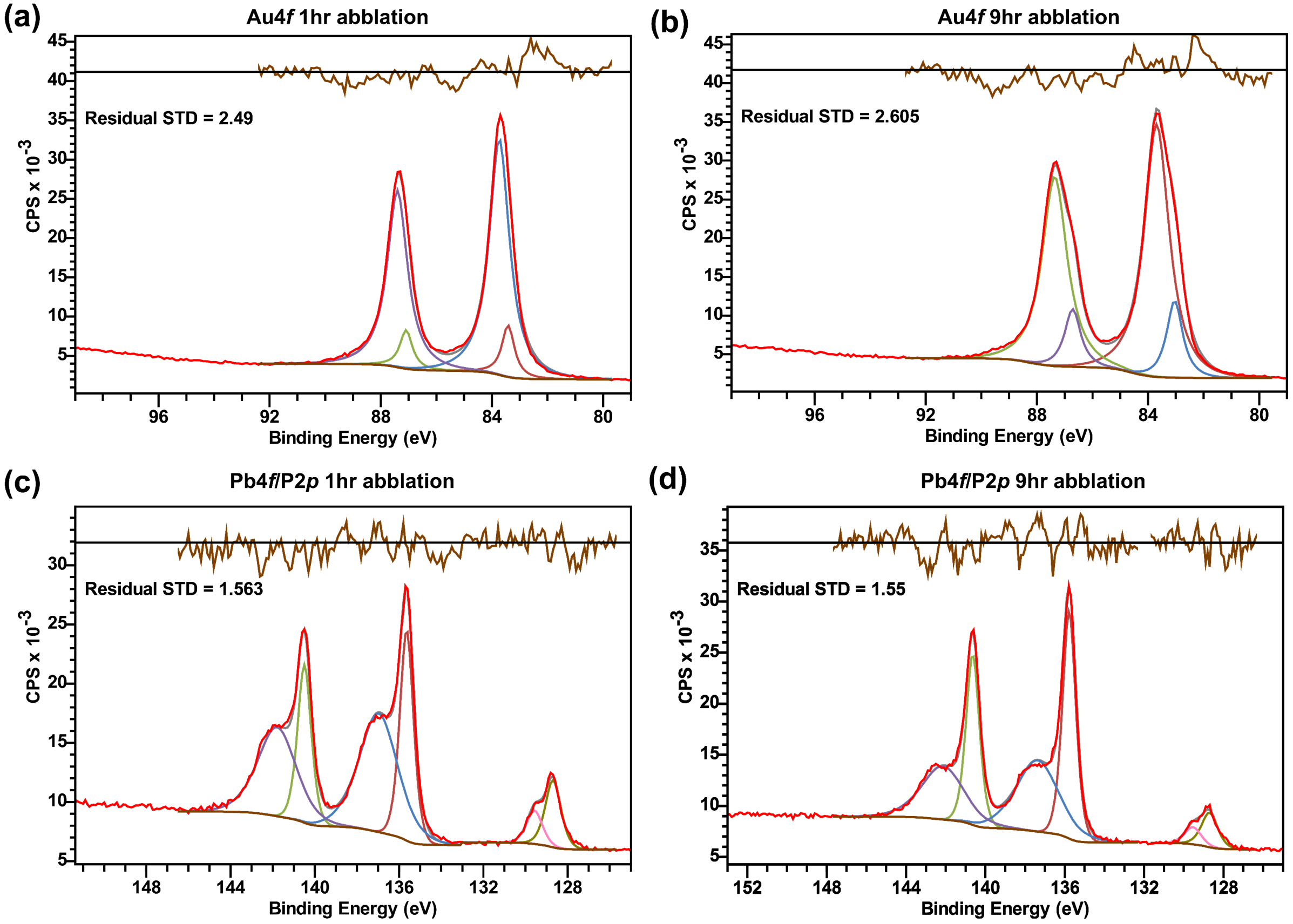}}
    \caption{\textbf{Fits to XPS spectra for the parent compound, \ce{Au2PbP2}}. \textbf{(a)} is the Au spectra taken after T=1 hour of ablation, as opposed to \textbf{(b)} taken after T=9 hours of ablation. \textbf{(c)} and \textbf{(d)} are Pb4f and P2p fits taken at the same respective time intervals.}
    \label{Parent_XPS}
\end{figure}

XPS analysis of untreated \ce{Au2PbP2} benchmarks our parent phases for direct comparison between chemical and electrochemically treated phases. In XPS, uniform oxidation states within the parent structure are seen for Au and Pb 4$f$ and P2$p$ peaks (Figure S\ref{XPS}\textbf{(a)}-\textbf{(c)}).  Fits to selected XPS spectra of the parent structure are shown in Figure S\ref{Parent_XPS}\textbf{(a)}-\textbf{(d)} and the data are tabulated in Table S\ref{ParentXPSFits}. 

\begin{table}[H]
\renewcommand{\tablename}{Table S}
\caption{\textbf{Tabulated \ce{Au2PbP2} XPS data.}}
\centering
\begin{tabular}{c c c c | c c c c}
\hline
Peak & Position (eV) & FWHM & Area & Peak & Position (eV) & FWHM & Area \\
\hline \hline
\multicolumn{4}{ c }{T = 1 hr} & \multicolumn{4}{ c }{T = 9 hr}\\
\hline
Au1 4f$_{7/2}$& 83.7  & 0.87 & 35669.7 & Au1 4f$_{7/2}$ & 83.7  & 1.07 & 45116.2\\
Au1 4f$_{5/2}$& 87.4  & 0.87 & 26752.3 & Au1 4f$_{5/2}$ & 87.4  & 1.07 & 33837.2 \\
Au2 4f$_{7/2}$& 83.4  & 0.54 & 4919.4  & Au2 4f$_{7/2}$ & 83.04 & 0.68 & 8813.3 \\
Au2 4f$_{5/2}$& 87.1  & 0.54 & 3689.5  & Au2 4f$_{5/2}$ & 86.7  & 0.68 & 6610.0  \\
Pb1 4f$_{7/2}$& 135.6 & 0.76 & 15344.6 & Pb1 4f$_{7/2}$ & 135.8 & 0.80 & 19994.4 \\
Pb1 4f$_{5/2}$& 140.5 & 0.76 & 11508.4 & Pb1 4f$_{5/2}$ & 140.6 & 0.80 & 14995.8\\
Pb2 4f$_{7/2}$& 137.0 & 2.16 & 24461.0 & Pb2 4f$_{7/2}$ & 137.4 & 2.51 & 19788.5 \\
Pb2 4f$_{5/2}$& 141.8 & 2.16 & 18345.7 & Pb2 4f$_{5/2}$ & 142.2 & 2.51 & 14841.4\\
P1 2p$_{3/2}$ & 128.7 & 0.87 & 5577.7  & P1 2p$_{3/2}$  & 128.7 & 0.96 & 3663.8 \\
P1 2p$_{1/2}$ & 129.6 & 0.87 & 2788.8  & P1 2p$_{1/2}$  & 129.5 & 0.96 & 1831.9 \\
P2 2p$_{3/2}$ & -     & -    & -       & P2 2p$_{3/2}$  & - & - & - \\
P2 2p$_{1/2}$ & -     & -    & -       & P2 2p$_{1/2}$  & - & - & - \\
\hline
\end{tabular}    
\label{ParentXPSFits}
\end{table}

\begin{figure}[H]
\renewcommand{\figurename}{Figure S}
    \centering
    \resizebox{\textwidth}{!}{\includegraphics{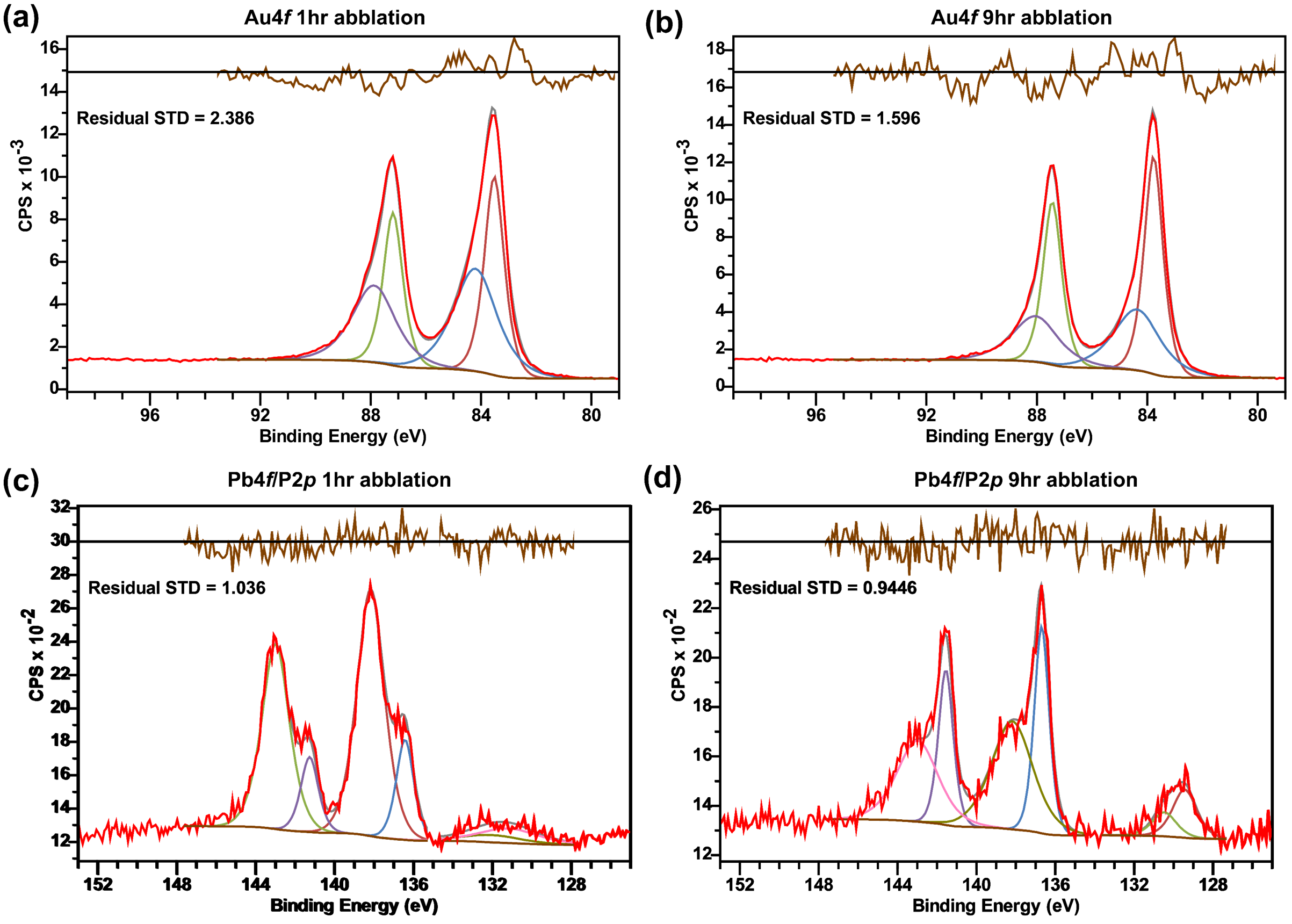}}
    \caption{\textbf{Fits to XPS spectra for a crystal soaked for 15 min in concentrated HNO$_3$.} \textbf{(a)} is the Au spectra taken after T=1 hour of ablation, as opposed to \textbf{(b)} taken after T=9 hours of ablation. \textbf{(c)} and \textbf{(d)} are Pb4f and P2p fits taken at the same respective time intervals.}
    \label{Acid_XPS}
\end{figure}

Upon chemical treatment with \ce{HNO3}, significant modifications in the XPS spectra emerge (Figure \ref{XPS}\textbf{(d)}-\textbf{(f)}), indicative of substantial electronic restructuring accompanying a topotactic transformation. Near the surface (etch time = 1 h, Figure S\ref{Acid_XPS}\textbf{(a)}), the Au oxidation states shift to more reducing potentials. This electronic reconfiguration is partially charge balanced by the concurrent oxidation of both Pb and P, as evidenced by the shifts in their respective binding energies in Figure S\ref{Acid_XPS}\textbf{(b)}. However, after extended sputtering (t = 9 h, Figure S\ref{Acid_XPS}\textbf{(c)} and \textbf{(d)}), the binding energies of Pb revert back to those seen in our parent structure, indicating penetration into regions less affected by the chemical treatment. Binding energies are tabulated in Table S\ref{AcidXPSFits}. This electronic redistribution provides compelling evidence that the structural transformation is fundamentally a redox-mediated process, in which electron transfer between framework and guest atoms drives the structural change. Still, it is evident that etching in concentrated nitric acid will not produce homogeneous samples. For further discussion on potential soft-chemical processing techniques, see SI Section 3.3.  

\begin{table}[H]
\renewcommand{\tablename}{Table S}
\caption{\textbf{Tabulated XPS data for acid etched crystal.}}
\centering
\begin{tabular}{c c c c | c c c c}
\hline 
\multicolumn{4}{ c }{T = 1 hr} & \multicolumn{4}{ c }{T = 9 hr}\\
\hline
Peak & Position (eV) & FWHM & Area & Peak & Position (eV) & FWHM & Area \\
\hline
Au1 4f$_{7/2}$& 83.5 & 0.85 & 9227.3 & Au1 4f$_{7/2}$& 83.8& 0.80 & 10809.9\\
Au1 4f$_{5/2}$& 87.2 & 0.85 & 6920.5 & Au1 4f$_{5/2}$& 87.4& 0.80 & 8107.4 \\
Au2 4f$_{7/2}$& 84.2 & 1.86  & 10361.1  &  Au2 4f$_{7/2}$& 84.4 & 1.99 & 7509.8 \\
Au2 4f$_{5/2}$& 87.9 & 1.86  & 7770.8  & Au2 4f$_{5/2}$& 88.0 & 1.99 & 5632.4  \\
Pb1 4f$_{7/2}$& 138.2 & 1.64 & 2740.0 & Pb1 4f$_{7/2}$& 138.2 & 2.47 & 1218.3 \\
Pb1 4f$_{5/2}$& 143.0 & 1.64 & 2055.0 & Pb1 4f$_{5/2}$& 143.0 & 2.47 & 913.7\\
Pb2 4f$_{7/2}$& 136.5 & 1.05 & 713.3 & Pb2 4f$_{7/2}$& 136.7 & 0.88 & 839.2\\
Pb2 4f$_{5/2}$& 141.3 & 1.05 & 535.0 & Pb2 4f$_{5/2}$& 141.5 & 0.88 & 629.4\\
P1 2p$_{3/2}$& - & - & - & P1 2p$_{3/2}$& 129.5 & 1.43 & 314.6\\
P1 2p$_{1/2}$& - & - & - & P1 2p$_{1/2}$& 130.5 & 1.43 & 157.3 \\
P2 2p$_{3/2}$& 131.2 & 3.56 & 351.6 & P2 2p$_{3/2}$& - & - & -\\
P2 2p$_{1/2}$& 132.4 & 3.56 & 175.8 & P2 2p$_{1/2}$& - & - & -\\
\hline
\end{tabular}    
\label{AcidXPSFits}
\end{table}
\FloatBarrier

\begin{figure}[H]
\renewcommand{\figurename}{Figure S}
    \centering
    \resizebox{\textwidth}{!}{\includegraphics{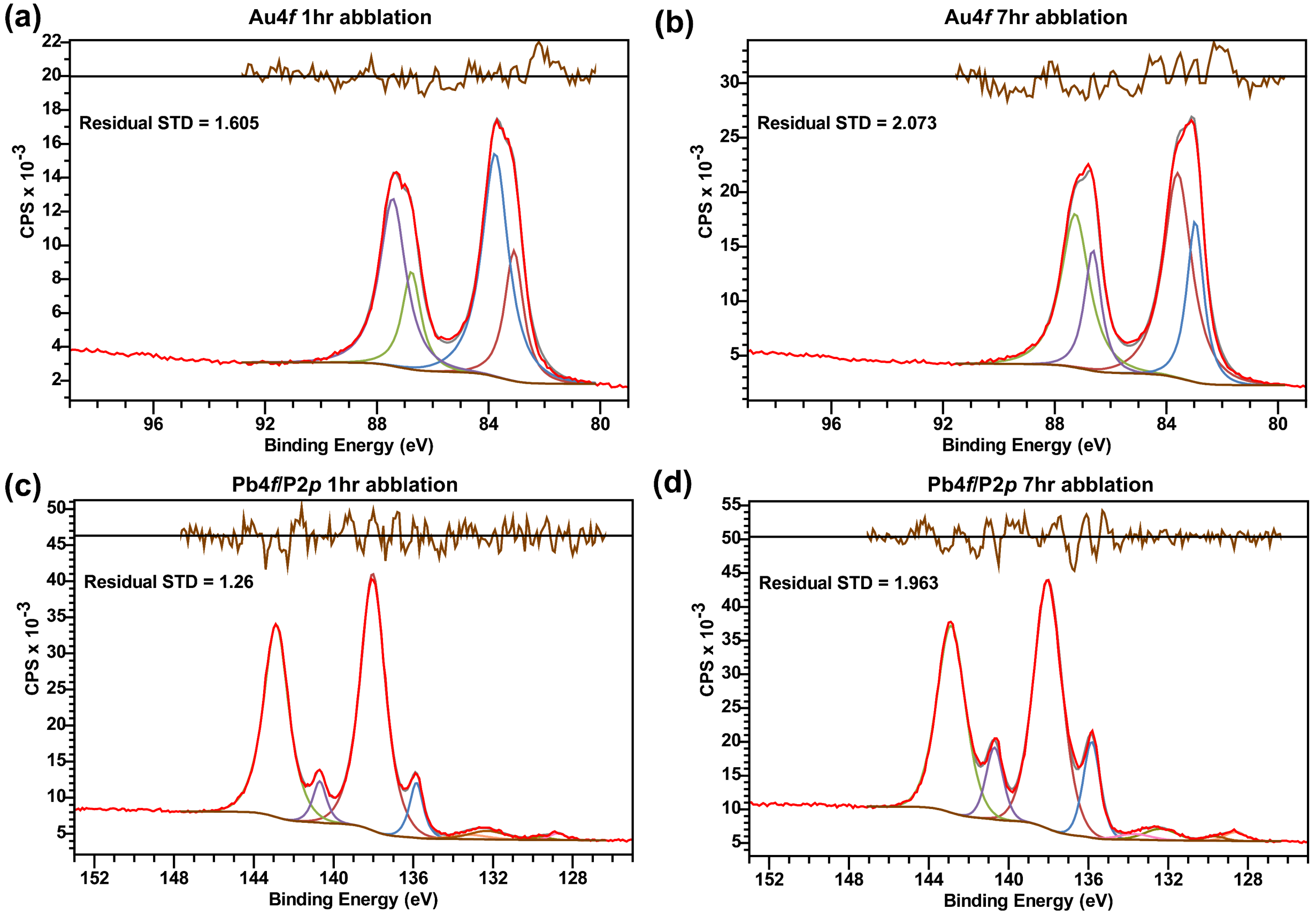}}
    \caption{\textbf{Fits to XPS spectra treated electrochemically for 300 seconds.} \textbf{(a)} is the Au spectra taken after T=1 hour of ablation, as opposed to \textbf{(b)} taken after T=7 hours of ablation. \textbf{(c)} and \textbf{(d)} are Pb4f and P2p fits taken at the same respective time intervals.}
    \label{Echem_XPS}
\end{figure}

Electrochemical treatment of parent crystals apparently remedies the issue of inhomogeneity. Repeating the depth-profile XPS experiments with an electrochemically treated crystal shows the same binding energies as the surface of the acid-treated crystal (Figure \ref{XPS}\textbf{(g)}-\textbf{(i)}, Figure S\ref{Echem_XPS}, and Table S\ref{EchemXPSFits}). However, unlike the acid-treated crystal, no peaks that align with the parent compound are seen, even after etching deep into the crystal (t=7.0~hrs).

\begin{table}[H]
\renewcommand{\tablename}{Table S}
\caption{\textbf{Tabulated XPS data for electrochemically-treated crystal}}.
\centering
\begin{tabular}{c c c c | c c c c}
\hline
Peak & Pos. & FWHM  & Area & Peak & Pos.  & FWHM  & Area  \\
 & (eV) & (eV) & (counts$\cdot$eV) &  &  (eV) &  (eV) & (counts$\cdot$eV) \\
\hline \hline
\multicolumn{4}{ c }{T = 1 hr} & \multicolumn{4}{ c }{T = 7 hr}\\
\hline
Au1 4f$_{7/2}$ & 83.1  & 0.80 & 7966.0  & Au1 4f$_{7/2}$ & 83.0 & 0.72 & 13496.4\\
Au1 4f$_{5/2}$ & 86.8  & 0.80 & 5947.5  & Au1 4f$_{5/2}$ & 86.6 & 0.72 & 10122.3 \\
Au2 4f$_{7/2}$ & 83.8  & 1.12 & 19133.8 & Au2 4f$_{7/2}$ & 83.6 & 1.20 & 28826.2 \\
Au2 4f$_{5/2}$ & 87.4  & 1.12 & 14350.3 & Au2 4f$_{5/2}$ & 87.2 & 1.20 & 21619.6  \\
Pb1 4f$_{7/2}$ & 138.0 & 1.48 & 62557.8 & Pb1 4f$_{7/2}$& 138.0 & 1.68 & 69122.0 \\
Pb1 4f$_{5/2}$ & 142.9 & 1.48 & 46918.3 & Pb1 4f$_{5/2}$& 142.9 & 1.68 & 51841.5 \\
Pb2 4f$_{7/2}$ & 135.9 & 0.81 & 7401.1  & Pb2 4f$_{7/2}$& 135.8 & 1.01 & 16093.1 \\
Pb2 4f$_{5/2}$ & 140.7 & 0.81 & 5550.8  & Pb2 4f$_{5/2}$& 140.7 & 1.01 & 12069.8 \\
P1 2p$_{3/2}$  & 128.7 & 0.93 & 977.6   & P1 2p$_{3/2}$ & 128.6 & 1.18 & 1670.4  \\
P1 2p$_{1/2}$  & 129.6 & 0.93 & 488.8   & P1 2p$_{1/2}$ & 129.7 & 1.18 & 835.2   \\
P2 2p$_{3/2}$  & 132.3 & 2.15 & 2884.8  & P2 2p$_{3/2}$ & 132.4 & 1.90 & 3446.3  \\
P2 2p$_{1/2}$  & 133.1 & 2.15 & 1442.4  & P2 2p$_{1/2}$ & 133.7 & 1.90 & 1723.2  \\
\hline
\end{tabular}    
\label{EchemXPSFits}
\end{table}
\FloatBarrier

Quantitative analysis of binding energies, correlated with previous Bader charge calculations,\cite{Lee_2024} provides compelling evidence that the tunnel-like [Au$_2$P$_2$] framework functions as an electron reservoir during the topotactic transformation. The framework's ability to accommodate variable charge states plays a crucial role in stabilizing the apparent topotactic reaction.

Remarkably, the shift in binding energies are immediate, rather than continuous, suggesting a binary electronic transition rather than a continuous spectrum of oxidation states. This distinct bimodal distribution provides strong evidence that the topotactic transformation propagates from the surface inward and is limited by diffusion kinetics rather than thermodynamic factors.

\FloatBarrier

\subsection{Micro-Computed X-ray Tomography}
 $\mu$CT was used as a non-destructive characterization technique to probe the internal structure of samples to probe sample homogeneity, morphology, and flux incorporation.\cite{Pressley_2022, Withers2021}. The $\mu$CT cross-sections used in the main text are shown below in Figure S\ref{uCT} with a scale bar to reference crystal size. Notably, attenuation values are relative to within the sample and can not be compared between scans for the crystals of the parent (S\ref{uCT}\textbf{(a)}), acid-treated (S\ref{uCT}\textbf{(b)}), or electrochemically-treated (S\ref{uCT}\textbf{(c)} samples since each sample was measured in separate scans. In other words, though panels \textbf{(a)} and \textbf{(c)} appear to have the same relative attenuation, they have different absolute attenuation.

 For the \ce{Au2PbP2} system, cross-sections from $\mu$CT reconstructions were taken to determine sample homogeneity. Shown in Figure S\ref{uCT}\textbf{(a)}, the cross section shows no signs of internal flux inclusions, secondary phases, and voids and is indicative of a homogeneous bulk sample. In contrast, for the acid-treated sample shown in Figure S\ref{uCT}\textbf{(b)}, the cross-section along the rod axis shows two distinct phases with different attenuation contrasts along with voids in the bottom of the crystal. Due to the structural similarity between the parent structure and the partially Pb-deintercalated phase, it is expected that the parent structure would be comparatively more attenuating. Notably, the deintercalated phase appears more prevalent at the front and back of the chain axis, in qualitative agreement with previous predictions for easy Pb ionic movement along the chain axis\cite{Wen_2009}. $\mu$CT scans confirm the homogeneity of electrochemically treated crystals (Figure S\ref{uCT}\textbf{(c)}), where the resulting measurement shows only one phase, which is in agreement with what is seen in the XPS depth profiles and shows that the deintercalation process remains constant across the whole crystal.
 
\begin{figure}[H]
\renewcommand{\figurename}{Figure S}
    \centering
    \resizebox{\textwidth}{!}{\includegraphics{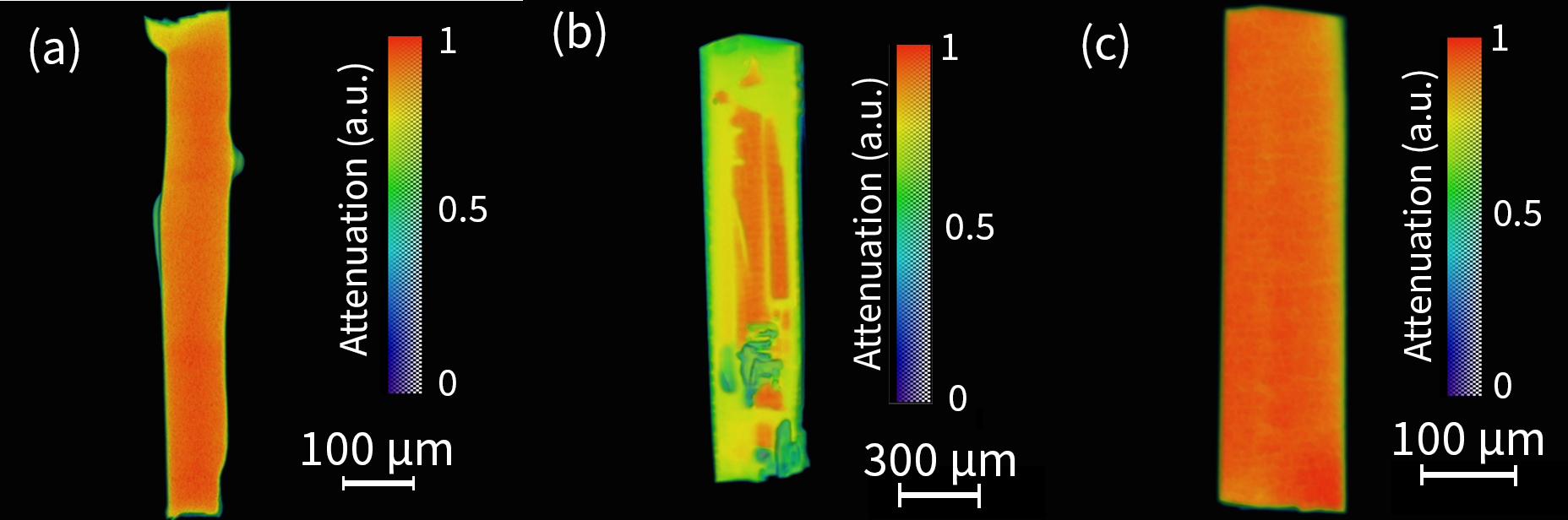}}
    \caption{\textbf{Micro-Computed X-ray Tomography ($\mu$CT) data for various treatments of \ce{Au2PbP2}.} (a) The untreated parent compound, \ce{Au2PbP2}. (b) \ce{Au2PbP2} after being soaked in concentrated HNO$_3$ for 15 minutes. Yellow streaks along the perimeter of the crystal have different attenuation from the core of the crystal, indicative of the partially deintercalated phase. (c) Electrochemically treated product, Au$_2$Pb$_{0.914}$P$_2$. The lack of attenuation contrast suggests a homogeneous deintercalation.}
    \label{uCT}
\end{figure}

\FloatBarrier

\section{Structural Characterization}
\subsection{Single-crystal Diffraction Refinement}

The indexation of the electrochemically treated diffraction patterns requires attention to the relationship between main and satellite reflections. Manual adjustment of lattice parameters was necessary to ensure proper indexing of the main reflections, particularly due to the proximity of intense satellite peaks to systematically absent main reflections. Our refinement strategy prioritized accurate determination of the \textbf{a} parameter, aligning it with the highest intensity reflections while allowing subsequent automated refinement of all parameters. This approach is validated by two key observations: (1) the refined \textbf{a} parameter of 3.1915 \AA\ closely approximates that of the parent structure (3.2351(4) \AA), and (2) the main reflections preserve the systematic absence conditions characteristic of the C-centered orthorhombic parent cell (\textit{hk0}:\textit{h}+\textit{k} = 2\textit{n}, \textit{h0l}:\textit{h}+\textit{l} = 2\textit{n}).

Determination of the appropriate (3+1)D superspace group required a systematic analysis of reflection conditions. Following crystallographic conventions, we transformed the C-centered orthorhombic cell of the parent structure to an A-centered setting through application of a unitary matrix:

\begin{equation}
\begin{bmatrix}
\textbf{a'}\\
\textbf{b'}\\
\textbf{c'}\\
\end{bmatrix} \times
\begin{bmatrix}
0 & 0 & 1\\
0 & -1 & 0 \\
1 & 0 & 0 \\
\end{bmatrix}	
= \begin{bmatrix}
\textbf{a}\\
\textbf{b}\\
\textbf{c}\\
\end{bmatrix}
\end{equation}

This transformation yielded an A-centered cell with parameters \textbf{a} $\sim$ 11.23 \AA, \textbf{b} $\sim$ 11.25 \AA, and \textbf{c} $\sim$ 3.19 \AA.

\begin{figure}[H]
\renewcommand{\figurename}{Figure S}
    \centering
    \resizebox{!}{!}{\includegraphics{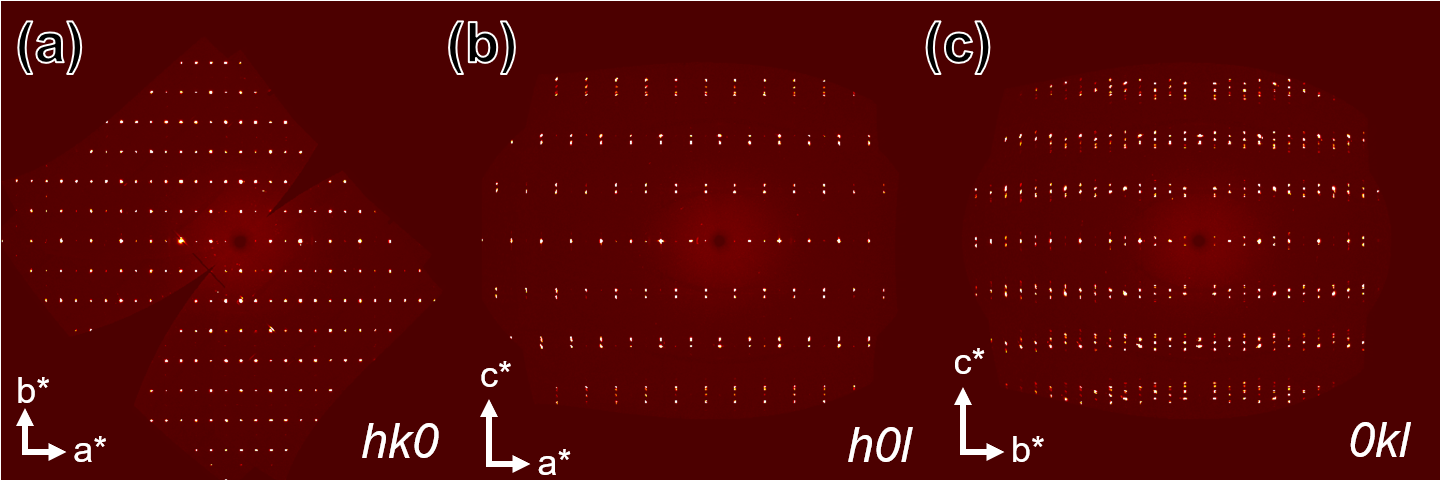}}
    \caption{Reconstructed precession images of Au$_2$Pb$_{0.914}$P$_2$ of the \textbf{(a)} \textit{hk0}, \textbf{(b)} \textit{h0l}, and \textbf{(c)} \textit{0kl} planes.}
    \label{APP_Prec}
\end{figure}

Below, we present reconstructed precession images to determine the superspace group of Au$_2$Pb$_{0.914}$P$_2$. In Figure S\ref{APP_Prec} we show the experimental data, which can be utilized to show the initial indexation of our satellite with q-vector (0, 0, 0.07166) in an A-centered cell. In the \textit{hk0} plane we see systematic absent reflections at larger values of two theta, attributed to satellite reflections bleeding into these planes. 

\begin{figure}[H]
\renewcommand{\figurename}{Figure S}
    \centering
    \resizebox{!}{!}{\includegraphics{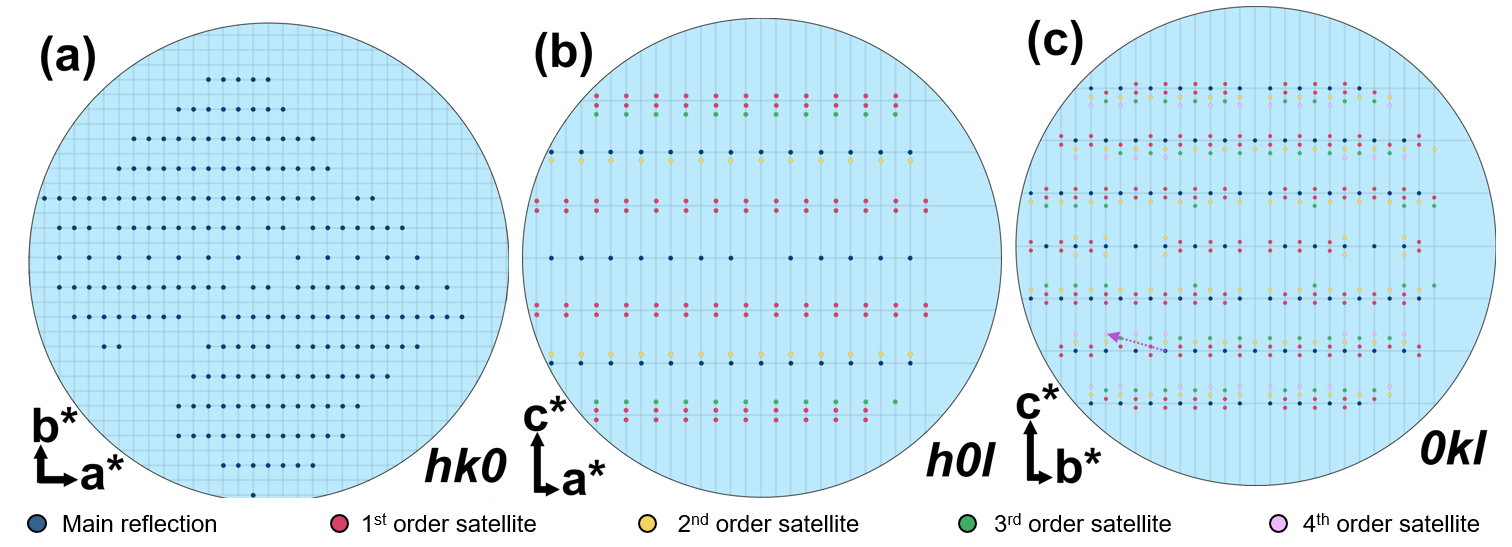}}
    \caption{Schematic of how precession images are interpreted for Au$_2$Pb$_{0.914}$P$_2$. \textbf{(a)} \textit{hk0} slice with \textit{k=2n}. in \textbf{(b)} Strong odd-order satellite peaks are seen adjacent to absent reflections where \textit{l=2n-1}. Similarly, in \textbf{(c)} odd-ordered satellite peaks are seen next to \textit{k+l=2n-1} peaks. Adjusting the q-vector to (0, 1, 0.07166) (purple arrow) now indexes all satellite peaks to A-centering main reflections.}
    \label{APP_Prec_sim}
\end{figure}

 For integration of satellite reflections, we evaluated whether the A-centering extended to the (3+1)D superspace. The conventional extension of A-centering (0, 1/2, 1/2, 0) would impose the condition \textit{hklm}: \textit{k} + \textit{l} = 2\textit{n} for all reflections, including satellites. However, our diffraction data clearly violates this expectation, with observed \textit{m}=$\pm$1\textbf{q$_1$} satellite peaks adjacent to systematically absent main reflections, particularly at positions corresponding to (\textit{h}01).

However, there exists an alternative centering vector (0, 1/2, 1/2, 1/2) that accurately described the observed reflection conditions, imposing \textit{hklm}:\textit{k}+\textit{l}+\textit{m}=2\textit{n}.The observed reflection conditions, following this condition precisely accounts for the distribution of satellite reflections in our reconstruction precession images, where odd (even) order satellites appear exclusively when the associated reciprocal lattice coordinate \textit{l} is odd (even) in the \textit{h}0\textit{l} plane, and similarly when \textit{k}+\textit{l} is odd (even) in the 0\textit{kl} plane. To accommodate this centering condition, we redefined the modulation vector as \textbf{q$_1$}=0\textbf{a*}+1\textbf{b*}-0.07166\textbf{c*}, where the rational component 1\textbf{b*} encodes this identified centering translation.

In Figure S\ref{APP_Prec_sim}, we have created a schematic to unambiguously label what experimental peak is indexed to what satellite peak, resulting in the superspace group symmetry A\textit{ma}2(0,1,$\gamma$)0\textit{ss}.

Computed $\langle|E^2 - 1|\rangle$ values for this data set give a value of 0.750, closely agreeing with the theoretically ideal noncentrosymmetric value of 0.736 compared to the ideal centrosymmetric value of 0.968. Such a discrepancy could be attributed to heavy atoms on special positions, which is expected to lower the calculated score, however, we note a significant deviation of $\langle|E^2 - 1|\rangle$ compared to the parent structure of 0.890.

Table S~\ref{CI_APP} tabulates the main experimental variables for the single-crystal diffraction experiment, as well as the main statistical parameters for the refinement of our structural model. In general, our model agrees very well with the collected data, with $R(I>3\sigma) = 3.18$, goodness of fit $S(I>3\sigma) = 3.36$, and unaccounted for electron density ($\Delta\rho_{\max}$, $\Delta\rho_{\min}$ (e~\AA$^{-3}$)) $=$ (2.9245, -2.7521) all indicative of an accurate refinement.
 {\renewcommand{\arraystretch}{0.70}
\begin{table}[H] 
\renewcommand{\tablename}{Table S}
\caption{\textbf{Crystallographic Data} for the (3+1)D refinement of Au$_2$Pb$_{0.914}$P$_2$.}
 \centering
 \begin{tabular}{l c}
\hline
Refined Composition & Au$_2$Pb$_{0.914}$P$_2$ \\ 
\hline 
Radiation source, $\lambda$ (\AA) & X-ray, 0.41328 \\
Absorption Correction & multi-scan \\
Data Collection Temperature (K)  & 100 \\
(3+1) D superspace Group & A\textit{ma}2(0,1,$\gamma$)0\textit{ss} \\
$a$ (\AA) & 11.2305 \\
$b$ (\AA) & 11.2551 \\
$c$ (\AA) & 3.1915 \\
Cell Volume (\AA$^{3}$)  &  403.4069 \\
$q_1$ & 0, 1, 0.07166 \\
Absorption Coefficient (mm$^{-1}$) & 27.096 \\
$\theta_{min}$ , $\theta_{max}$ (deg) & 1.09, 17.15 \\
Refinement Method & F$^{2}$   \\
R$_{int}$(I\textgreater{}3$\sigma$, all) & 8.87, 8.88\\
$<|E^2-1|>$ & 0.750 \\
\hline 
\multicolumn{2}{c}{\textit{Overall Refinement}} \\
\hline 
Total Reflections (I \textgreater 3$\sigma$, all)  & 15357, 24119  \\
Unique Reflections (I \textgreater 3$\sigma$, all)  & 2375, 3419  \\
Number of Parameters    & 265  \\
R(I\textgreater{}$3\sigma$), R$_{w}$(I\textgreater{}$3\sigma$)  &  3.18,  8.49 \\
R(all), R$_{w}$(all)     &   3.80, 8.70 \\
S(I\textgreater{}$3\sigma$), S(all)   & 3.36, 2.81 \\
$\Delta\rho_{max}$ , $\Delta\rho_{min}$ (e \AA$^{-3}$) & 2.9245, -2.7521 \\
\hline 
\multicolumn{2}{c}{\textit{Main Reflections}} \\
\hline 
Unique Reflections (I \textgreater 3$\sigma$, all) &  487, 487\\
R(I\textgreater{}$3\sigma$), R$_{w}$(I\textgreater{}$3\sigma$)  & 2.86, 7.87\\
R(all), R$_{w}$(all)     & 2.86, 7.87  \\
\hline 
\multicolumn{2}{c}{\textit{Satellites $\pm$1}} \\
\hline 
Unique Reflections (I \textgreater 3$\sigma$, all) & 951, 980 \\
R(I\textgreater{}$3\sigma$), R$_{w}$(I\textgreater{}$3\sigma$)  & 3.37, 9.13\\
R(all), R$_{w}$(all)     & 3.47, 9.17  \\
\hline
\multicolumn{2}{c}{\textit{Satellites $\pm$2}} \\
\hline
Unique Reflections (I \textgreater 3$\sigma$, all) & 595, 981 \\
R(I\textgreater{}$3\sigma$), R$_{w}$(I\textgreater{}$3\sigma$)  &  3.69, 8.14\\
R(all), R$_{w}$(all)     &   5.50, 8.57\\
\hline 
\multicolumn{2}{c}{\textit{Satellites $\pm$3}} \\
\hline
Unique Reflections (I \textgreater 3$\sigma$, all) &  342, 971 \\
R(I\textgreater{}$3\sigma$), R$_{w}$(I\textgreater{}$3\sigma$)  & 3.91, 8.19 \\
R(all), R$_{w}$(all)     &  9.58, 9.45 \\
\hline
\label{CI_APP}
\end{tabular}
\end{table}}

Table S~\ref{Positions_APP} tabulates the atomic positions for this refinement. For the final refinement, Au and Pb atoms were modeled with anharmonic ADPs, while the P atoms were modeled isotropically. Throughout the entire refinement, any additional degrees of freedom afforded to the P ADP parameters caused a non-positive-definite ADP, resulting in a physically impossible model.

\begin{table}[H]
\renewcommand{\tablename}{Table S}
\caption{\textbf{Atomic positions} for the (3+1)D refinement of Au$_2$Pb$_{0.914}$P$_2$.}
 \centering
 \begin{tabular}{c c c c c c c}
\hline
Label  & Element  & $x$ &  $y$  & $z$ &  U$_{ani}$/U$_{iso}$ &  U$_{eq}$ \\
\hline
  Au1  & Au  & 0.5 &  0.5  & 0.6410(14) &  U$_{ani}$ &  0.00577(14) \\
  Au2  & Au  & 0.75  & 0.72749(7) &  0.6294(15) &  U$_{ani}$  & 0.00639(12) \\
  Pb1 &  Pb &  0.75 &  0.47425(12) &  1.2118(14) &  U$_{ani}$  & 0.0228(5) \\
  Pb2 &  Pb  & 0.75 &  0.4697(2) & 0.7542(15) &  U$_{ani}$ &  0.0578(13) \\
  P1 & P  & 0.54493(14) & 0.70213(13) & 0.6318(19) & U$_{iso}$ & 0.0047(3) \\
  \hline
\label{Positions_APP}
\end{tabular}
\end{table}

Table S~\ref{Params_APP} summarizes the parameters used to obtain the final solution. Tables S\ref{U_APP} through S\ref{Legendre_Uanh_Pb2_APP} detail all refined parameters.

\begin{table}[H]
\renewcommand{\tablename}{Table S}
\caption{\textbf{Number of Parameters} for each atom in the (3+1)D refinement of Au$_2$Pb$_{0.914}$P$_2$.}
 \centering
 \begin{tabular}{c c c c c c}
\hline
Label  & Occupancy  & Position & Continuous & ADP 2$^{nd}$  & ADP 3$^{rd}$ \\
\hline
  Au1  & 3  & 2 &  yes  & 0 &  3  \\
  Au2  & 3  & 3  & yes &  0 &  3  \\
  Pb1 &  1 &  2 &  no &  1 & 3 \\
  Pb2 &  1  & 2 &  no & 0 & 3 \\
  P1 & 3  & 3 & yes & 0 & 0 \\
  \hline
\label{Params_APP}
\end{tabular}
\end{table}

\begin{center}
\begin{table} [H]
\renewcommand{\tablename}{Table S}
\caption{\textbf{Components of the anisotropic ADP (U$_{ani}$) parameters} for Au and Pb atoms in the (3+1)D refinement of Au$_2$Pb$_{0.914}$P$_2$.}
 \centering
 \begin{tabular}{c c c c c c c}
\hline
Atom & U$_{11}$ & U$_{22}$ & U$_{33}$ & U$_{12}$ & U$_{13}$ & U$_{23}$ \\
\hline
Au1  & 0.0057(2) &  0.0061(2) &  0.0056(3) &  0.00006(13) &  0 &  0 \\
Au2 &  0.0024(2) &  0.0085(2) & 0.0083(2) &  0  & 0 &  -0.0017(4) \\
Pb1 & 0.0073(4) &  0.0117(5) &  0.0493(14) &  0 &  0 &  0.0036(6) \\
Pb2 & 0.0068(6) &  0.0134(8) & 0.153(4) &  0 &  0 &  0.0225(11) \\
\hline
\label{U_APP}
\end{tabular}
\end{table}
\end{center}

\begin{center}
\begin{table} [H]
\renewcommand{\tablename}{Table S}
\caption{\textbf{Components of the anharmonic ADP (U$_{anhar}$) parameters} for Au and Pb atoms in the (3+1)D refinement of Au$_2$Pb$_{0.914}$P$_2$.}
 \centering
 \begin{tabular}{c c c c c}
\hline
Parameter & Au1 & Au2 & Pb1 & Pb2 \\
\hline
U$_{111}$ & 0 & 0 & 0 & 0 \\
U$_{112}$ & 0 & -0.00019(5) & -0.0001(3) & 0.0023(6) \\
U$_{113}$ & -0.0022(9) & 0.0001(9) & 0.0001(18) & -0.013(3) \\
U$_{122}$ & 0 & 0 & 0 & 0 \\
U$_{123}$ & 0.0002(5) & 0 & 0 & 0 \\
U$_{133}$ & 0 & 0 & 0 & 0 \\
U$_{222}$ & 0 & -0.00023(8) & -0.0011(4) & 0.0096(9) \\
U$_{223}$ & 0.0026(8) & -0.0049(7) & -0.0064(18) & -0.012(3) \\
U$_{233}$ & 0 & -0.0001(5) & -0.060(11) & 0.29(3) \\
U$_{333}$ & 0.13(2) & 0.02(2) & 0.07(10) & 1.4(3) \\
\hline
\label{U_h_APP}
\end{tabular}
\end{table}
\end{center}

\begin{center}
\begin{table} [H]
\renewcommand{\tablename}{Table S}
\caption{\textbf{Trigonometric components of the continuous positional displacement waves} for Au and P atoms in the (3+1)D refinement of Au$_2$Pb$_{0.914}$P$_2$.}
 \centering
 \begin{tabular}{c c c c c}
\hline
Atom & axis & order & cos & sin \\
\hline
 Au1 &  x & 1 &   -0.00621(12)  &   -0.0030(2) \\
 Au1 &  y & 1 &    0.00017(8)   &   -0.00019(13)  \\
 Au1 &  z & 1 &    0            &    0 \\
 Au1 &  x & 2 &    0            &    0 \\
 Au1 &  y & 2 &    0            &    0 \\
 Au1 &  z & 2 &    0.0011(3)    &   -0.0019(3) \\
 Au2 &  x & 1 &    0            &    0 \\
 Au2 &  y & 1 &    0.00173(15)  &   -0.00302(8) \\
 Au2 &  z & 1 &    0.0449(5)    &    0.0163(13) \\
 Au2 &  x & 2 &    0            &    0 \\
 Au2 &  y & 2 &   -0.00091(7)   &   -0.00040(13) \\
 Au2 &  z & 2 &    0.0052(6)    &   -0.0093(4) \\
 Au2 &  x & 3 &    0            &    0 \\
 Au2 &  y & 3 &   -0.00043(12)  &    0.00025(11) \\
 Au2 &  z & 3 &    0.0004(5)    &   -0.0015(5) \\
 P1  &  x & 1 &   -0.0033(3)    &   -0.00094(15) \\
 P1  &  y & 1 &    0.0012(3)    &   -0.00111(12) \\
 P1  &  z & 1 &    0.0272(5)    &    0.0016(17) \\
 P1  &  x & 2 &   -0.00006(14)  &   -0.0003(3) \\
 P1  &  y & 2 &   -0.00016(11)  &    0.0001(2) \\
 P1  &  z & 2 &   -0.0001(11)   &   -0.0026(6) \\
 P1  &  x & 3 &    0.0003(3)    &    0.00018(19) \\
 P1  &  y & 3 &    0.0001(3)    &    0.00002(15) \\
 P1  &  z & 3 &    0.0003(7)    &   -0.0007(8) \\
\hline
\label{Displacement_APP}
\end{tabular}
\end{table}
\end{center}

\begin{center}
\begin{table} [H]
\renewcommand{\tablename}{Table S}
\caption{\textbf{Refined values of the Crenel functions} describing Pb atom displacement in the (3+1)D refinement of Au$_2$Pb$_{0.914}$P$_2$.}
 \centering
 \begin{tabular}{c c c}
\hline
Atom & $x_{4_0}$ & $\delta$ \\
\hline
Pb1 & 0.6501 & 0.5882 \\
Pb2 & 0.1443 & 0.3341 \\
\hline
\label{Crenel_APP}
\end{tabular}
\end{table}
\end{center}

\begin{center}
\begin{table} [H]
\renewcommand{\tablename}{Table S}
\caption{\textbf{Legendre Polynomial parameters} for Au, Pb, and P atom displacements in the (3+1)D refinement of Au$_2$Pb$_{0.914}$P$_2$.}
 \centering
 \begin{tabular}{c c c c}
\hline
Atom & axis & Order & Coefficient \\
\hline
 Pb1 & x  & 1  &    0 \\
 Pb1 & y  & 1  &   -0.0113(3) \\
 Pb1 & z  & 1  &    0.303(2) \\
  \hline
 Pb1 & x  & 2  &    0 \\
 Pb1 & y  & 2  &   -0.0104(5) \\
 Pb1 & z  & 2  &    0.090(2) \\
  \hline
 Pb1 & x  & 3  &    0 \\
 Pb1 & y  & 3  &   -0.0024(4) \\
 Pb1 & z  & 3  &    0.084(2) \\
  \hline
 Pb1 & x  & 4  &    0 \\
 Pb1 & y  & 4  &   -0.0009(6) \\
 Pb1 & z  & 4  &    0.014(3) \\
  \hline
 Pb2 & x  & 1  &    0 \\
 Pb2 & y  & 1  &    0.0020(5) \\
 Pb2 & z  & 1  &    0.291(3) \\
  \hline
 Pb2 & x  & 2  &    0 \\
 Pb2 & y  & 2  &    0.0131(7) \\
 Pb2 & z  & 2  &   -0.121(4) \\
  \hline
 Pb2 & x  & 3  &    0 \\
 Pb2 & y  & 3  &   -0.0116(8) \\
 Pb2 & z  & 3  &    0.077(5) \\
  \hline
 Pb2 & x  & 4  &    0 \\
 Pb2 & y  & 4  &    0.0099(9) \\
 Pb2 & z  & 4  &   -0.050(6) \\
\hline
\label{Legendre_displacement_APP}
\end{tabular}
\end{table}
\end{center}

\begin{center}
\begin{table} [H]
\renewcommand{\tablename}{Table S}
\caption{\textbf{Legendre Polynomial parameters} for Au, Pb, and P ADP parameters in the (3+1)D refinement of Au$_2$Pb$_{0.914}$P$_2$.}
 \centering
 \begin{tabular}{c c c c}
\hline
Atom & U$_{ij}$ & Order & Coefficient \\
\hline
 Pb1 & U11 & 1  &  -0.0003(8) \\
 Pb1 & U22 & 1  &   0.0014(9) \\
 Pb1 & U33 & 1  &  -0.005(4) \\
 Pb1 & U12 & 1  &   0 \\
 Pb1 & U13 & 1  &   0 \\
 Pb1 & U23 & 1  &  -0.0192(14) \\
  \hline
 Pb1 & U11 & 2  &  -0.0008(11) \\
 Pb1 & U22 & 2  &   0.0008(14) \\
 Pb1 & U33 & 2  &   0.080(4) \\
 Pb1 & U12 & 2  &   0 \\
 Pb1 & U13 & 2  &   0 \\
 Pb1 & U23 & 2  &   0.0066(14) \\
\hline
\label{Legendre_Uani_APP}
\end{tabular}
\end{table}
\end{center}

\begin{center}
\begin{table} [H]
\renewcommand{\tablename}{Table S}
\caption{\textbf{Legendre Polynomial parameters} for Au1 ADP parameters in the (3+1)D refinement of Au$_2$Pb$_{0.914}$P$_2$.}
 \centering
 {\renewcommand{\arraystretch}{0.80}
 \begin{tabular}{c c c c c}
\hline
Atom & U$_{ijk}$ & Order & cos & sin \\
\hline
 Au1 & C111  &   1 &    0.00099(17) &  -0.0002(3)  \\
 Au1 & C112  &   1 &    0.00000(8)  &   0.00008(13) \\
 Au1 & C113  &   1 &    0           &   0 \\
 Au1 & C122  &   1 &    0.00012(10) &   0.00052(16) \\
 Au1 & C123  &   1 &    0           &   0 \\
 Au1 & C133  &   1 &   -0.0017(11)  &   0.0105(18) \\
 Au1 & C222  &   1 &    0.00003(12) &   0.0000(2) \\
 Au1 & C223  &   1 &    0           &   0 \\
 Au1 & C233  &   1 &    0.0012(8)   &  -0.0024(12) \\
 Au1 & C333  &   1 &    0           &   0 \\
\hline
 Au1 & C111  &   2 &    0           &   0 \\
 Au1 & C112  &   2 &    0           &   0 \\
 Au1 & C113  &   2 &   -0.0001(3)   &  -0.0006(4) \\
 Au1 & C122  &   2 &    0           &   0 \\
 Au1 & C123  &   2 &   -0.0001(2)   &   0.0001(3) \\
 Au1 & C133  &   2 &    0           &   0 \\
 Au1 & C222  &   2 &    0           &   0 \\
 Au1 & C223  &   2 &   -0.00005(17) &  -0.0002(3) \\
 Au1 & C233  &   2 &    0           &   0 \\
 Au1 & C333  &   2 &   -0.004(6)    &  -0.017(8) \\
\hline
 Au1 & C111  &   3 &    0.00005(17) &  -0.00005(9) \\
 Au1 & C112  &   3 &   -0.00005(16) &  -0.00005(8) \\
 Au1 & C113  &   3 &    0           &   0 \\
 Au1 & C122  &   3 &   -0.00005(13) &  -0.00003(6) \\
 Au1 & C123  &   3 &    0           &   0 \\
 Au1 & C133  &   3 &    0.0007(10)  &  -0.0015(8) \\
 Au1 & C222  &   3 &    0.00000(15) &   0.00000(6) \\
 Au1 & C223  &   3 &    0           &   0 \\
 Au1 & C233  &   3 &    0.0005(10)  &   0.0004(7) \\
 Au1 & C333  &   3 &    0           &   0 \\
\hline
\label{Legendre_Uanh_Au1_APP}
\end{tabular}}
\end{table}
\end{center}

\begin{center}
\begin{table} [H]
\renewcommand{\tablename}{Table S}
\caption{\textbf{Legendre Polynomial parameters} for Au2 anharmonic ADP parameters in the (3+1)D refinement of Au$_2$Pb$_{0.914}$P$_2$.}
 \centering
 {\renewcommand{\arraystretch}{0.8}
 \begin{tabular}{c c c c c}
\hline
Atom & U$_{ijk}$ & Order & cos & sin \\
\hline
 Au2 & C111  &   1 &    0           &   0 \\
 Au2 & C112  &   1 &    0.00041(18) &   0.00010(7) \\
 Au2 & C113  &   1 &   -0.0006(3)   &   0.0025(7) \\
 Au2 & C122  &   1 &    0           &   0 \\
 Au2 & C123  &   1 &    0           &   0 \\
 Au2 & C133  &   1 &    0           &   0 \\
 Au2 & C222  &   1 &    0.0010(3)   &   0.00059(12) \\
 Au2 & C223  &   1 &    0.0002(3)   &  -0.0014(6) \\
 Au2 & C233  &   1 &   -0.0017(17)  &  -0.0011(9) \\
 Au2 & C333  &   1 &    0.017(11)   &  -0.08(2) \\
\hline
 Au2 & C111  &   2 &    0           &   0 \\
 Au2 & C112  &   2 &    0.00009(8)  &  -0.00002(14) \\
 Au2 & C113  &   2 &    0.0003(4)   &  -0.0002(3) \\
 Au2 & C122  &   2 &    0           &   0 \\
 Au2 & C123  &   2 &    0           &   0 \\
 Au2 & C133  &   2 &    0           &   0 \\
 Au2 & C222  &   2 &    0.00044(13) &   0.00002(19) \\
 Au2 & C223  &   2 &   -0.0008(4)   &  -0.0007(4) \\
 Au2 & C233  &   2 &   -0.0022(13)  &   0.0056(14) \\
 Au2 & C333  &   2 &   -0.046(13)   &  -0.038(11) \\
\hline
 Au2 & C111  &   3 &    0           &   0 \\
 Au2 & C112  &   3 &    0.00002(14) &   0.00006(14) \\
 Au2 & C113  &   3 &    0.0001(3)   &  -0.0002(2) \\
 Au2 & C122  &   3 &    0           &   0 \\
 Au2 & C123  &   3 &    0           &   0 \\
 Au2 & C133  &   3 &    0           &   0 \\
 Au2 & C222  &   3 &   -0.00030(17) &   0.0000(2) \\
 Au2 & C223  &   3 &    0.0002(4)   &  -0.0005(3) \\
 Au2 & C233  &   3 &    0.0001(9)   &  -0.0002(15) \\
 Au2 & C333  &   3 &    0.007(15)   &  -0.011(10) \\
\hline
\label{Legendre_Uanh_Au2_APP}
\end{tabular}}
\end{table}
\end{center}

\begin{center}
\begin{table} [H]
\renewcommand{\tablename}{Table S}
\caption{\textbf{Legendre Polynomial parameters} for Pb1 ADP parameters in the (3+1)D refinement of Au$_2$Pb$_{0.914}$P$_2$.}
 \centering
{\renewcommand{\arraystretch}{0.8}
 \begin{tabular}{c c c c c}
\hline
Atom & U$_{ijk}$ & Order & cos & sin \\
\hline
 Pb1 & C111  &   1   &  0           &   0 \\
 Pb1 & C112  &   1   & -0.0004(14)  &   0.0019(9) \\
 Pb1 & C113  &   1   &  0.020(6)    &  -0.006(4) \\
 Pb1 & C122  &   1   &  0           &   0 \\
 Pb1 & C123  &   1   &  0           &   0 \\
 Pb1 & C133  &   1   &  0           &   0 \\
 Pb1 & C222  &   1   & -0.0015(17)  &   0.0056(13) \\
 Pb1 & C223  &   1   &  0.003(6)    &   0.001(3) \\
 Pb1 & C233  &   1   & -0.23(4)     &   0.27(3) \\
 Pb1 & C333  &   1   & -0.8(3)      &   3.7(3) \\
 \hline
 Pb1 & C111  &   2   &  0           &   0 \\
 Pb1 & C112  &   2   &  0.002(3)    &   0.003(2) \\
 Pb1 & C113  &   2   &  0.019(7)    &  -0.018(7) \\
 Pb1 & C122  &   2   &  0           &   0 \\
 Pb1 & C123  &   2   &  0           &   0 \\
 Pb1 & C133  &   2   &  0           &   0 \\
 Pb1 & C222  &   2   & -0.006(3)    &   0.012(3) \\
 Pb1 & C223  &   2   &  0.014(8)    &  -0.016(7) \\
 Pb1 & C233  &   2   & -0.31(4)     &   0.34(4) \\
 Pb1 & C333  &   2   & -1.2(4)      &   4.0(4) \\
\hline
 Pb1 & C111  &   3   &  0           &   0 \\
 Pb1 & C112  &   3   &  0.004(2)    &   0.003(2) \\
 Pb1 & C113  &   3   &  0.011(6)    &  -0.014(7) \\
 Pb1 & C122  &   3   &  0           &   0 \\
 Pb1 & C123  &   3   &  0           &   0 \\
 Pb1 & C133  &   3   &  0           &   0 \\
 Pb1 & C222  &   3   & -0.002(3)    &   0.009(3) \\
 Pb1 & C223  &   3   &  0.022(7)    &  -0.005(7) \\
 Pb1 & C233  &   3   & -0.19(3)     &   0.09(3) \\
 Pb1 & C333  &   3   & -0.9(2)      &   0.7(2) \\
\hline
\label{Legendre_Uanh_Pb1_APP}
\end{tabular}}
\end{table}
\end{center}

\begin{figure}[H]
\renewcommand{\figurename}{Figure S}
    \centering
    \resizebox{!}{!}{\includegraphics{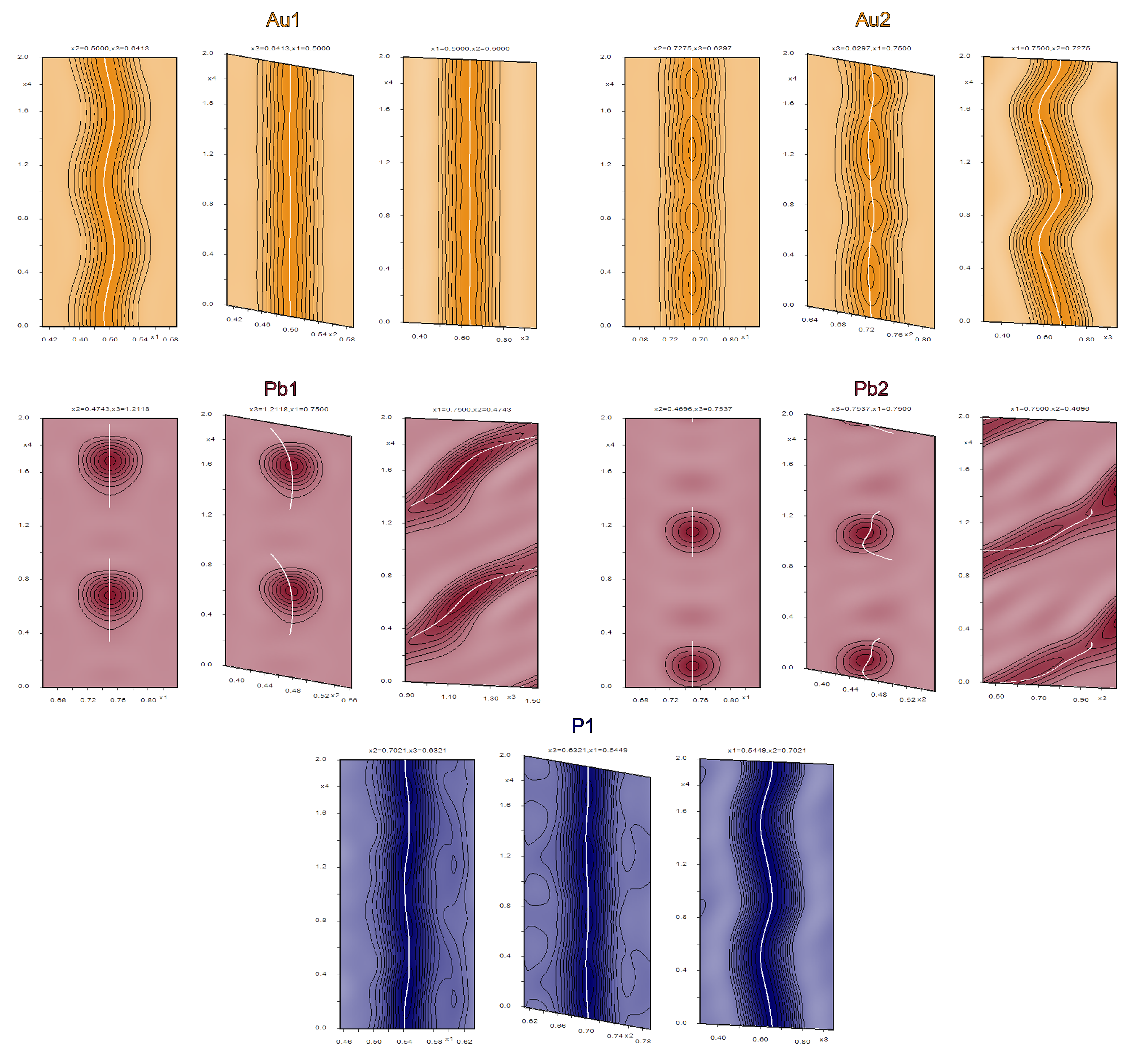}}
    \caption{DeWolff plots for Au$_2$Pb$_{0.914}$P$_2$ observed density. More saturated colors indicate higher electron density, while more white colors represent negative electron density. Black lines are isosurface levels at various step sizes while white lines correspond to the fit electron density of the model.}
    \label{DeWolff Observed}
\end{figure}

\begin{figure}[H]
\renewcommand{\figurename}{Figure S}
    \centering
    \resizebox{!}{!}{\includegraphics{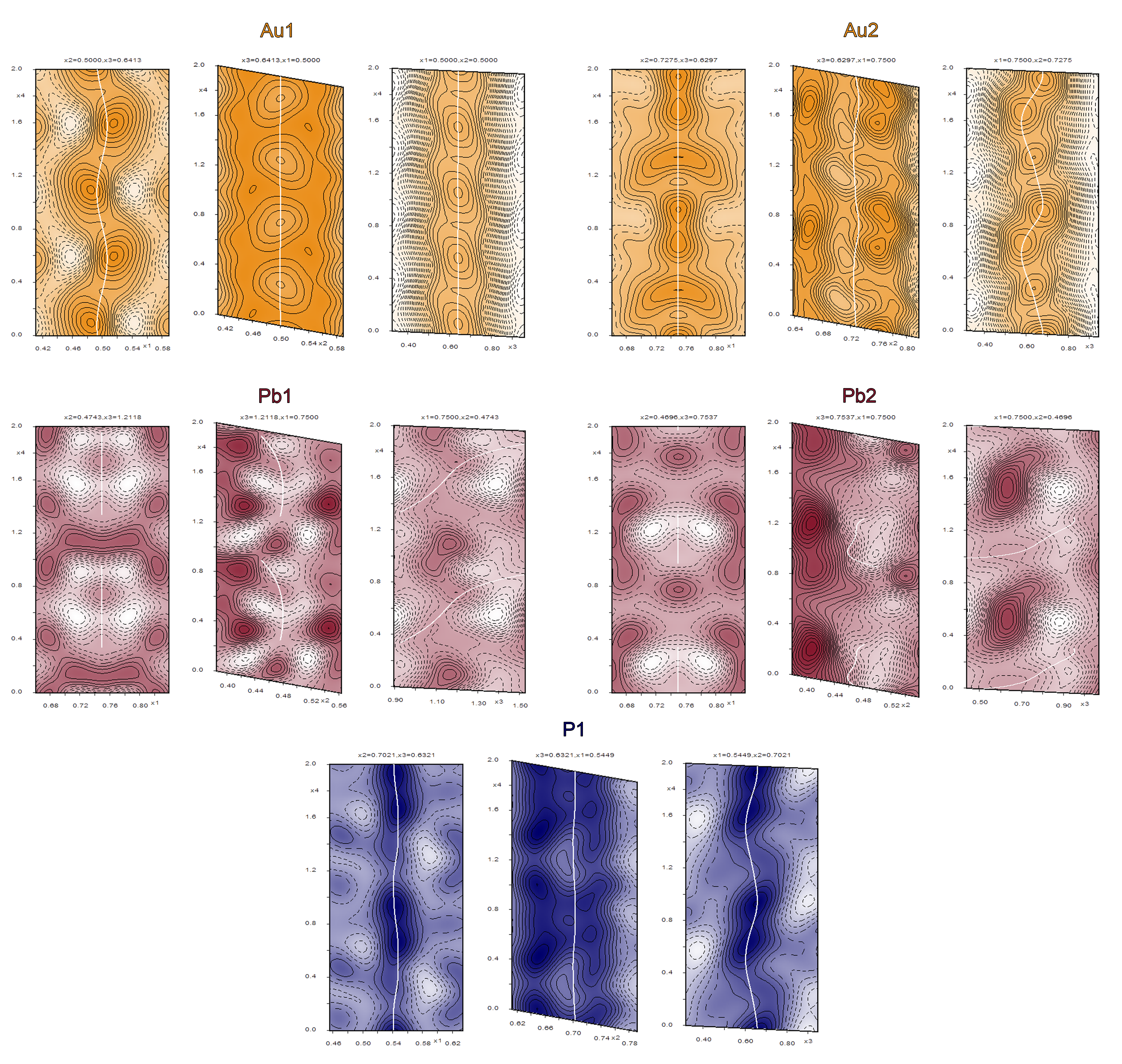}}
    \caption{[\textbf{f(obs)-f(calc)] DeWolff plots of Au$_2$Pb$_{0.914}$P$_2$}. Difference in electronic density between experimentally observed and modeled electron density for the final structural model. More saturated colors surrounded by continuous black lines indicate higher electron density, while more white colors surrounded by dashed lines represent negative electron density. Isosurfaces are drawn at 0.2 e/\AA$^3$ intervals.}
    \label{DeWolff Diff}
\end{figure}

\begin{center}
\begin{table} [H]
\renewcommand{\tablename}{Table S}
\caption{\textbf{Legendre Polynomial parameters} for Pb2 ADP parameters in the (3+1)D refinement of Au$_2$Pb$_{0.914}$P$_2$.}
 \centering
{\renewcommand{\arraystretch}{0.8}
 \begin{tabular}{c c c c c}
\hline
Atom & U$_{ijk}$ & Order & cos & sin \\
\hline
 Pb2 & C111  &   1   &  0           &   0 \\
 Pb2 & C112  &   1   &  0.006(3)    &  -0.0002(17) \\
 Pb2 & C113  &   1   & -0.069(18)   &  -0.012(9) \\
 Pb2 & C122  &   1   &  0           &   0 \\
 Pb2 & C123  &   1   &  0           &   0 \\
 Pb2 & C133  &   1   &  0           &   0 \\
 Pb2 & C222  &   1   &  0.022(4)    &  -0.003(2) \\
 Pb2 & C223  &   1   & -0.083(13)   &   0.006(9) \\
 Pb2 & C233  &   1   &  0.68(11)    &   0.00(6) \\
 Pb2 & C333  &   1   & -2.2(10)     &   3.1(6) \\
  \hline
 Pb2 & C111  &   2   &  0           &   0 \\
 Pb2 & C112  &   2   &  0.013(5)    &   0.005(4) \\
 Pb2 & C113  &   2   & -0.11(4)     &  -0.08(3) \\
 Pb2 & C122  &   2   &  0           &   0 \\
 Pb2 & C123  &   2   &  0           &   0 \\
 Pb2 & C133  &   2   &  0           &   0 \\
 Pb2 & C222  &   2   &  0.011(6)    &   0.007(4) \\
 Pb2 & C223  &   2   &  0.00(3)     &  -0.05(2) \\
 Pb2 & C233  &   2   &  1.6(2)      &   1.15(16) \\
 Pb2 & C333  &   2   &  12(2)       &   8.4(17) \\
  \hline
 Pb2 & C111  &   3   &  0           &   0 \\
 Pb2 & C112  &   3   & -0.010(9)    &   0.005(8) \\
 Pb2 & C113  &   3   & -0.09(4)     &  -0.13(4) \\
 Pb2 & C122  &   3   &  0           &   0 \\
 Pb2 & C123  &   3   &  0           &   0 \\
 Pb2 & C133  &   3   &  0           &   0 \\
 Pb2 & C222  &   3   & -0.022(11)   &  -0.022(10) \\
 Pb2 & C223  &   3   &  0.22(4)     &  -0.03(3) \\
 Pb2 & C233  &   3   &  1.2(2)      &   1.6(3) \\
 Pb2 & C333  &   3   &  11.0(19)    &   3(2) \\
\hline
\label{Legendre_Uanh_Pb2_APP}
\end{tabular}}
\end{table}
\end{center}
The refined structural model shows excellent agreement with experimental data. The DeWolff plots shown in Figure S~\ref{DeWolff Observed} demonstrate the quality of the observed density, while Figure S~\ref{DeWolff Diff} shows the difference density between observed and calculated structures.

This breathing motion of the Au2 atoms in our deintercalated Au$_2$Pb$_{0.914}$P$_2$ can be well understood through the lens of DFT chemical pressure (DFT-CP) analysis.\cite{Lee_2024} This method utilizes \textit{ab initio} calculations to spatially resolve local packing frustrations in solid state systems. DFT-CP analysis of the related orthorhombic \ce{Au2BiP2} reveals that both Au and P atoms exhibit positive chemical pressures along Au-P contacts of [Au$_2$P$_2$] tunnels, indicating the desire for expansion of this framework. For the Au atoms, these positive CPs are partially counterbalanced by the negative CP's found pointing along the Au-Pb, and Au-Au contacts, indicating a desire to shorten these interatomic distances. This dichotomy of positive and negative pressures orthogonally oriented on Au atoms signal the largest source of packing frustration, with their high quadrupole metric values \cite{Kamp_2022} indicating a large frustration conducive to soft phonon modes. Concurrently, negative CP's nearly surrounding the Pb atoms in a triangular fashion in the \textbf{a}-\textbf{b} plane suggest that the tunnel voids remain oversize for single Pb occupancy. The system could hypothetically remedy this by stuffing in more than one Pb atom per unit cell; however, the lack of appreciable CP along Pb-Pb contacts indicate an unwillingness to dimerize or trimerize along the chain axis to realize this experimentally. 

This chemical pressure landscape then provides a compelling framework for understanding the collective breathing motion of Au atoms during deintercalation. When Pb atoms adopt the 5-coordinate tbp geometries, positioning themselves between two Au atoms along the \textbf{a}-axis, they effectively overcompensate for the large negative pressures previously observed in the parent compound. This overcompensation permits Au atoms to expand along the \textbf{a}-direction, which also partially relieves the compressed Au-P contacts. Simultaneously, through this collective breathing motion, Au atoms in regions with 7-coordinated ctp geometries can contract inwards, also mediating the negative pressures found along Au-Pb contacts. The final structure then demonstrates how these collective motions work cooperatively with adjacent tunnels \textemdash in regions of tbp geometries, the elongated Au-Pb bonds is adjacent to regions of ctp geometries, where Au-Pb bonds shorten compared to the parent structure. This demonstrates how chemical pressure analysis not only identifies key regions of structural frustrations but can also qualitatively predict displacement patterns that arise when the system is perturbed.

\FloatBarrier

\section{Mechanisms behind the symmetry-breaking}

With the single-crystal structure solution consistent with our assignment to a polar space group, we return to our hypothesis that the mechanisms driving the symmetry-breaking distortion are a depopulation of frontier orbitals, allowing for the second-order Jahn-Teller (SOJT) and asymmetric lone pair effect. We present three complementary lines of evidence: (1) the generalization of noncentrosymmetric satellite reflections to the \ce{Au$_2$Tl$_{1-x}$P$_2$} analogue; (2) DFT density of states calculations on both the parent and product; and (3) ultraviolet photoelectron spectroscopy (UPS) that places the applied electrochemical potential in direct correspondence with the frontier orbital window responsible for the distortion.

\subsection{Generalization to Au$_2$Tl$_{1-x}$P$_2$}

As discussed in the main text, XPS measurements indicate that the frontier orbitals of both Au and Pb in the product consist of formally filled 5$d$ and empty 6$p$ states, implicating the SOJT effect as a primary symmetry-breaking mechanism. An independent structural test of this picture is afforded by the pseudo-isomorphic compound \ce{Au2TlP2}. If the SOJT mechanism is general to the \ce{Au2$M$P2} series, analogous satellite reflections should appear upon topotactic oxidation of the Tl-containing member.

\begin{figure}[H]
    \centering
    \resizebox{\textwidth}{!}{\includegraphics{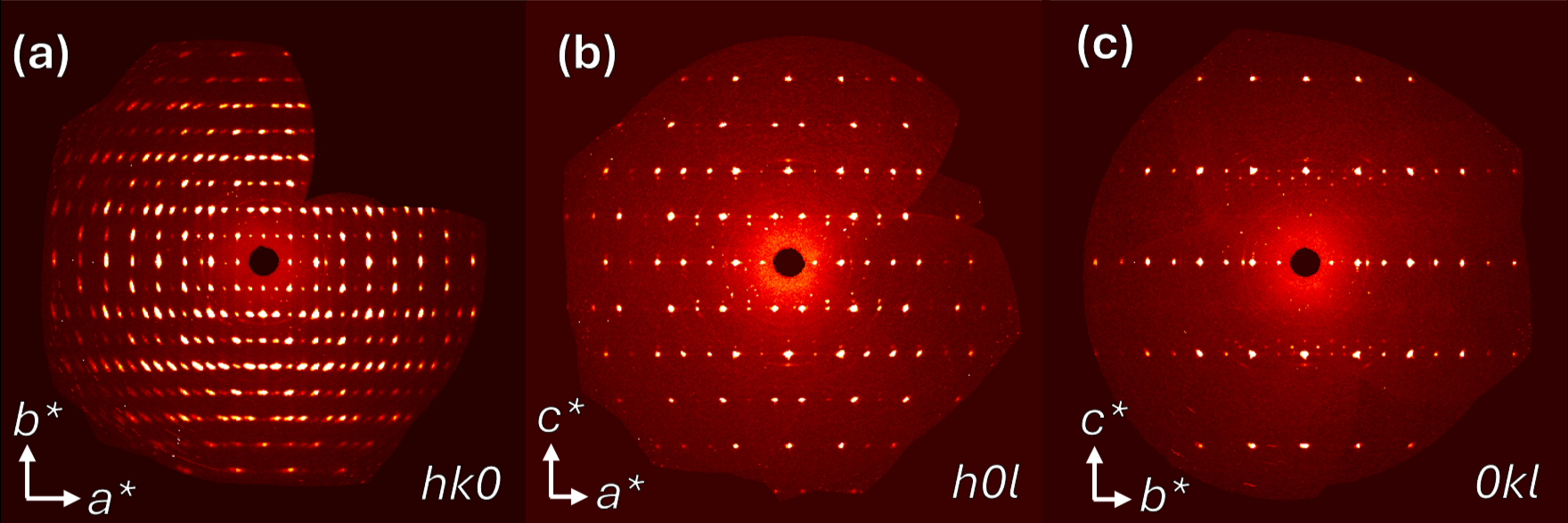}}
    \caption{\textbf{Precession images for \ce{Au2Tl_{1-x}P2}.} Satellite reflections along \textbf{c*} are indexable to $q_{1,\mathrm{Tl}} = -0.144$\,\textbf{c'*}, consistent with a noncentrosymmetric modulated superstructure analogous to \ce{Au2Pb_{0.914}P2}.}
    \label{fig:ATP_precession}
\end{figure}

Diffraction experiments on chemically treated \ce{Au2TlP2} reveal exactly this: satellite reflections can be indexed to $q_{1,\mathrm{Tl}} = -0.144$\,\textbf{c'*} (Figure~S\ref{fig:ATP_precession}, Table~S\ref{tab:ATP_crystal}). While a complete structural solution remains elusive, $\langle|E^2-1|\rangle = 0.767$ for \ce{Au2Tl_{1-x}P2} closely matches the theoretical value of 0.736 expected for a noncentrosymmetric crystal, comparable to the value of 0.750 obtained for \ce{Au2Pb_{0.914}P2}.

The relationship $q_{1,\mathrm{Tl}} \approx 2 \times q_{1,\mathrm{Pb}}$ hints at a systematic dependence of the distortion length on the identity of $M$. For \ce{Au2Pb_{0.914}P2}, the reciprocal length $q_{1,\mathrm{Pb}} = 0.07166$\,c* dictates the 14-fold real-space supercell, and our structural analysis shows that $x \approx q_{1,\mathrm{Pb}}$ for \ce{Au2Pb_{1-x}P2}. Extending this logic to \ce{Au2Tl_{1-x}P2}, where $q_{1,\mathrm{Tl}} = -0.144$\,c'*, we anticipate a 7-fold supercell with composition $\sim$\ce{Au2Tl_{0.85}P2}. Moreover, if this transition is accompanied by a Tl$^0 \to$ Tl$^{1+}$ oxidation, the total charge leaving the structure per formula unit is essentially identical between the two systems ($0.144 \times 1 \approx 0.07166 \times 2$). This charge equivalence implies a common electronic endpoint: in both cases, the $M$-site cation reaches the [Xe]\,4$f^{14}$5$d^{10}$6$s^2$ configuration — filled $d$ and empty $p$ shells ideally positioned for SOJT activity, with a stereoactive 6$s^2$ lone pair.

\begin{table}[H]
\caption{\textbf{Crystallographic data} for the (3+1)D integration of \ce{Au2Tl_{1-x}P2}.}
\centering
\begin{tabular}{lc}
\hline
Parameter & Value \\
\hline
Radiation source, $\lambda$ (\AA) & X-ray, 0.7107 \\
Absorption correction & Multi-scan \\
Data collection temperature (K) & 100 \\
(3+1)D superspace group & $Ama2(0,0,\gamma)0ss$ \\
$a$ (\AA) & 11.21 \\
$b$ (\AA) & 11.28 \\
$c$ (\AA) & 3.20 \\
Cell volume (\AA$^3$) & 404.37 \\
$q_1$ & 0, 0, 0.144 \\
Absorption coefficient (mm$^{-1}$) & 27.096 \\
$\theta_{\min}$, $\theta_{\max}$ (deg) & 1.09, 17.15 \\
$\langle|E^2-1|\rangle$ & 0.767 \\
\hline
\end{tabular}
\label{tab:ATP_crystal}
\end{table}

\subsection{DFT Density of States: Total and Partial}

Density functional theory calculations provide further theoretical support for the SOJT mechanism. Taking the 14-fold approximate structural solution of \ce{Au2Pb_{0.914}P2} as the starting geometry, we computed the density of states (DOS) of the product and compared it to the parent \ce{Au2PbP2}. All DOS values are normalized to the primitive parent cell (2 formula units).

\subsubsection*{Total DOS, Band Structure and the absence of a pseudogap}

\begin{figure}[H]
    \centering
    \resizebox{\textwidth}{!}{\includegraphics{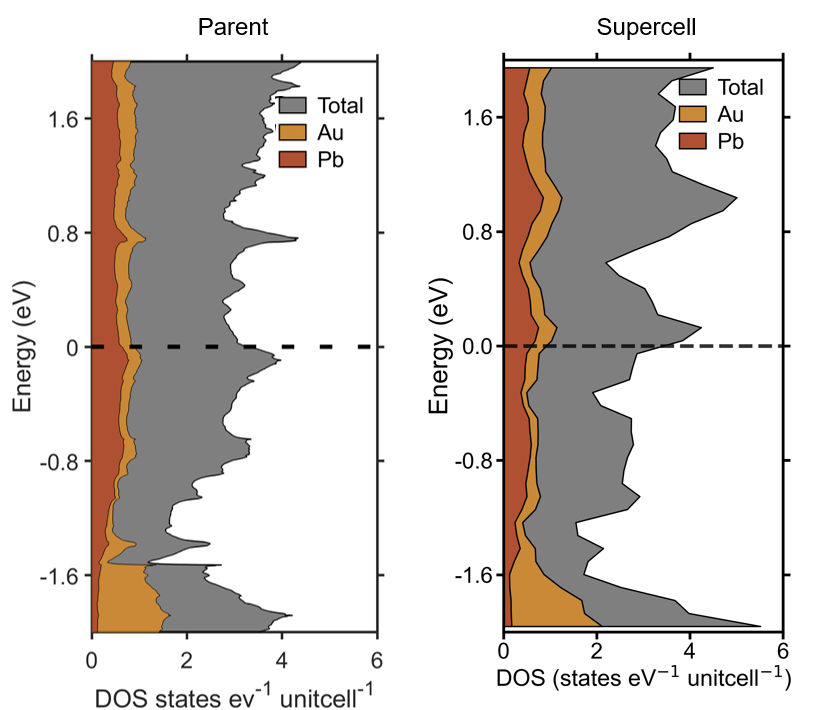}}
    \caption{\textbf{Total DOS comparison of parent \ce{Au2PbP2} and product \ce{Au2Pb_{0.914}P2}.} Both curves are normalized to the primitive cell (2 f.u.). The $\sim$0.3~eV downward shift of $E_F$ in the product reflects the electron depletion associated with Pb oxidation and partial deintercalation. The density of states at $E_F$ is unchanged.}
    \label{fig:DOS_comp}
\end{figure}

The total DOS comparison (Figure~S\ref{fig:DOS_comp}) shows that the overall spectral shape is well preserved upon deintercalation, with a downward shift of $E_F$ by $\sim$0.3~eV in the product, consistent with the loss of electrons during partial Pb removal. Importantly, the DOS at $E_F$ remains essentially unchanged at $\sim$1.5~states~eV$^{-1}$~f.u.$^{-1}$ in both parent and product. This rules out the possibility that the topotactic reaction is terminated by tuning $E_F$ to a pseudogap — a stabilization mechanism common in other intermetallic systems — and instead supports the view that the transformation is terminated by a favorable electron count associated with the SOJT-active $d^{10}s^2$ configuration, as discussed in the main text.

\begin{figure}[H]
    \centering
    \resizebox{\textwidth}{!}{\includegraphics{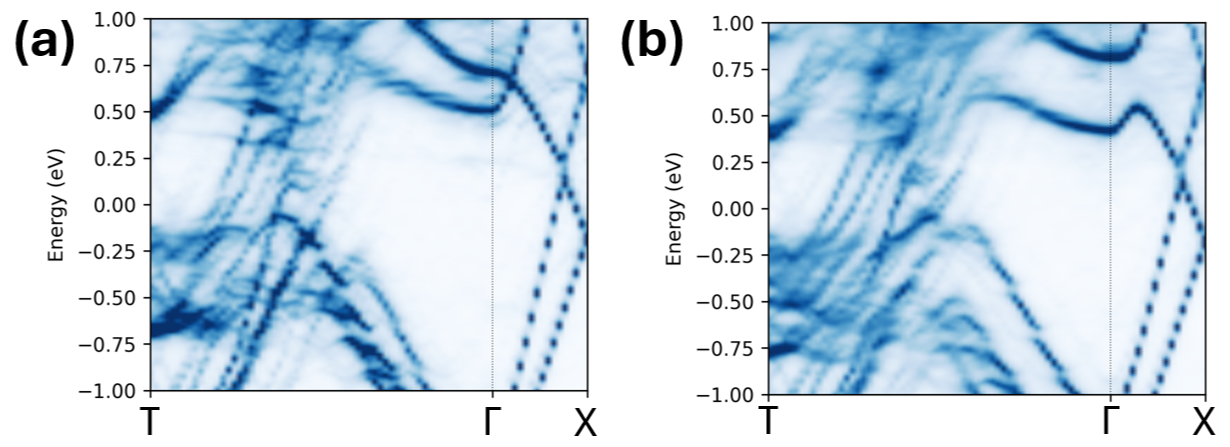}}
    \caption{\textbf{Band Structures of \ce{Au2Pb_{0.914}P2} (a)} without and \textbf{(b)} with spin-orbit coupling. TRIMs were chosen perpendicular to the polar axis with T ([0, 0.5, 0.5]), $\Gamma$ ([0, 0, 0]), and X (0.5, 0, 0).}
    \label{fig:BS}
\end{figure}

In Figure S~\ref{fig:BS}, we plot slices of band structures perpendicular to the polar axis. The band structure further confirms that we expect \ce{Au2Pb_{0.914}P2} to be metallic. Along T $\rightarrow$ $\Gamma$, band splitting appears to be more prevalent than along $\Gamma$ $\rightarrow$ X, suggesting Dresselhaus-type spin-orbit coupling. 

\subsubsection*{Atom-decomposed pDOS}

\begin{figure}[H]
    \centering
    \resizebox{\textwidth}{!}{\includegraphics{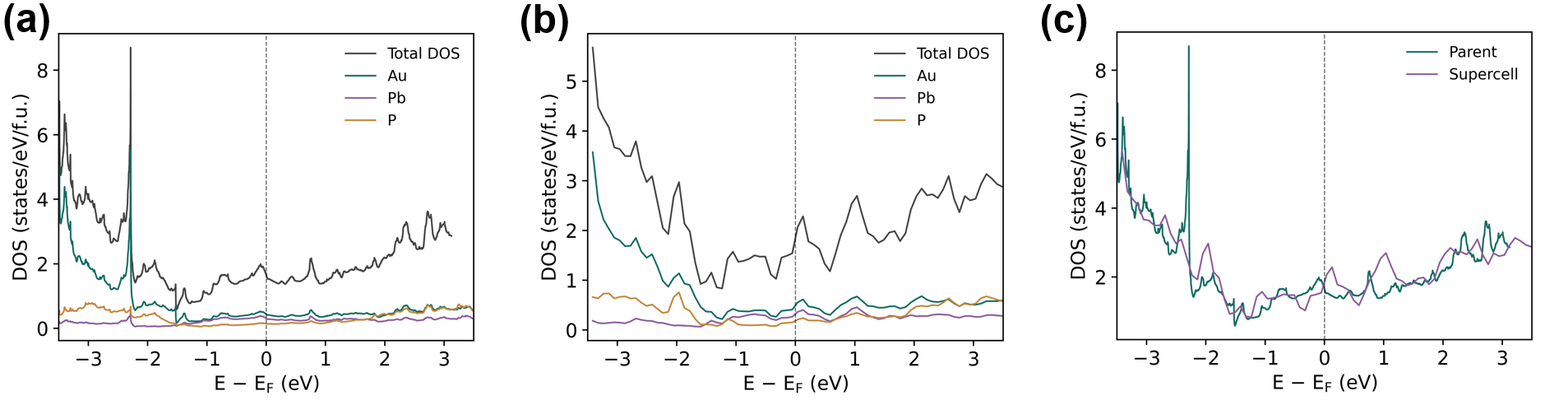}}
    \caption{\textbf{Atom-decomposed pDOS for parent and product.}
    \textbf{(a)} \ce{Au2PbP2}: states near $E_F$ are dominated by Au and Pb, with a local maximum $\sim -0.15$~eV below $E_F$.
    \textbf{(b)} \ce{Au2Pb_{0.914}P2}: the sharp peak at $\sim -2.5$~eV is absent and the Au/Pb local maximum has shifted to $\sim +0.15$~eV above $E_F$.
    \textbf{(c)} Direct overlay of (a) and (b) emphasizing the shift in $E_F$ and the disappearance of the $\sigma$-bonding peak.}
    \label{fig:pDOS_atom}
\end{figure}

The atom-decomposed partial DOS (Figure~S\ref{fig:pDOS_atom}) shows that states near $E_F$ are dominated by Au and Pb contributions in both structures. The most notable change is the disappearance in the product of a sharp peak located $\sim$2.5~eV below $E_F$ in the parent, which is strongly hybridized between Au and Pb states and attributed to the short ($\sim$2.8~\AA) Au1--Pb $\sigma$ contact. Its absence in the supercell calculation is consistent with the Pb$^0 \to$ Pb$^{2+}$ charge transfer: as Pb ions displace from their inversion-symmetric positions, the $\sigma$-bonding states redistribute deeper into the Au~5$d$ band, transferring electrons to the framework.

A second key feature is the shift of the local maximum near $E_F$: in the parent, this maximum lies $\sim$0.15~eV \emph{below} $E_F$ and is attributed to the weaker Au2--Pb $\pi$ bonding states that hold Pb on its inversion center. In the product, this maximum has shifted to $\sim$0.15~eV \emph{above} $E_F$, consistent with electrochemical depopulation of these frontier states and the onset of Pb ionic mobility discussed in the main text.

\subsubsection*{Orbital-decomposed pDOS}

\begin{figure}[H]
    \centering
    \resizebox{\textwidth}{!}{\includegraphics{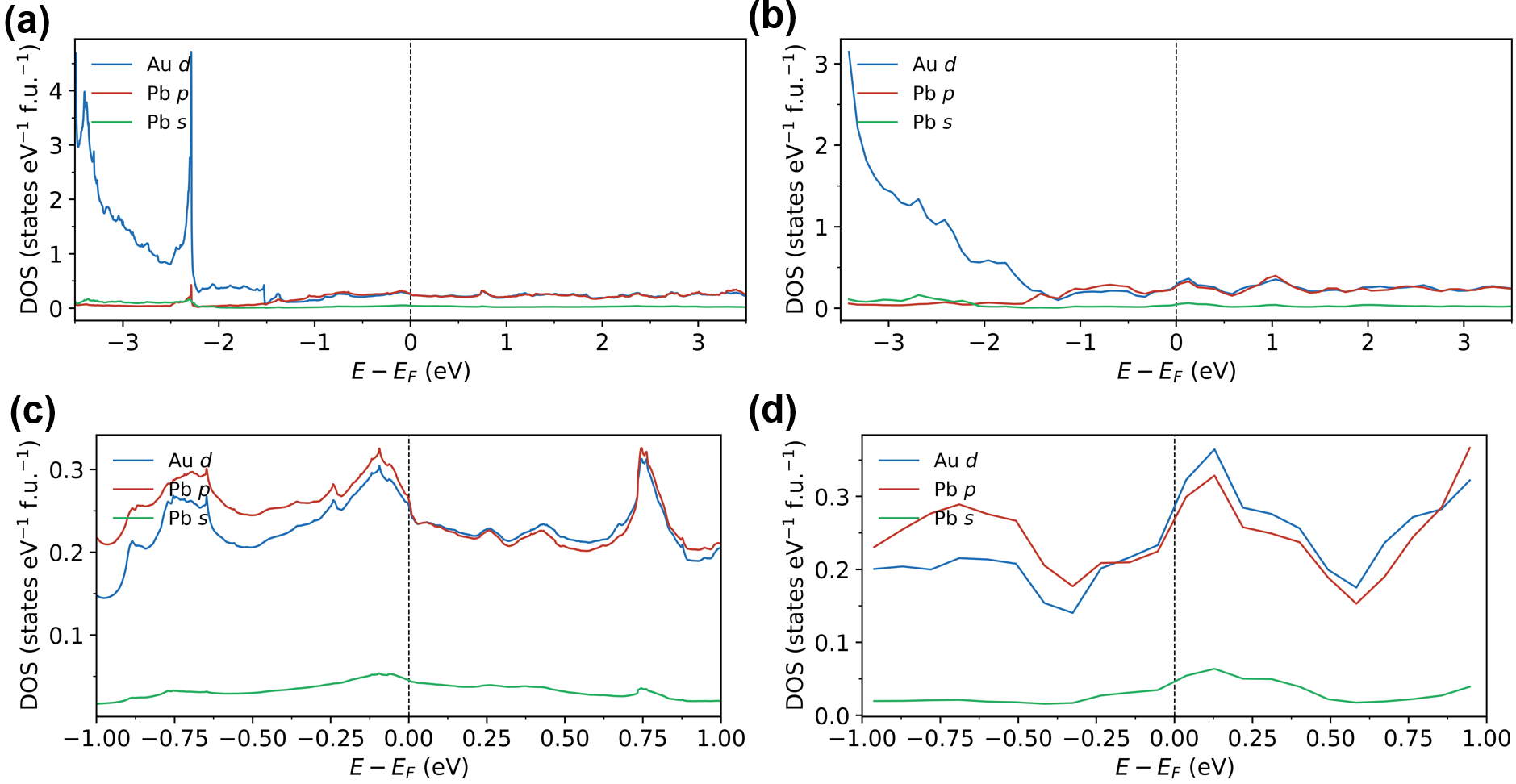}}
    \caption{\textbf{Orbital-decomposed pDOS for parent and product.}
    \textbf{(a)} \ce{Au2PbP2}: a sharp peak at $\sim -2.25$~eV is attributed to the short Au1--Pb contact ($\sim 2.8$~\AA).
    \textbf{(b)} \ce{Au2Pb_{0.914}P2}: the sharp $-2.25$~eV peak is absent, consistent with Pb$^0 \to$ Pb$^{2+}$ charge transfer.
    \textbf{(c)} Enlarged view near $E_F$ for the parent, showing the Au2--Pb $\pi$ local maximum at $\sim -0.15$~eV.
    \textbf{(d)} Enlarged view near $E_F$ for the product. The local maximum has shifted above $E_F$; residual filled Au~$d$--Pb~$p$ states reflect post-distortion SOJT rehybridization.}
    \label{fig:pDOS_orbital}
\end{figure}

The orbital decomposition (Figure~S\ref{fig:pDOS_orbital}) isolates the Au~5$d$ and Pb~6$p$ contributions. In the parent (panels (a) and (c)), the sharp $\sigma$-bonding peak at $\sim -2.25$~eV is clearly visible and the Au2--Pb $\pi$ local maximum at $\sim -0.15$~eV below $E_F$ is attributable to states that hold Pb in its centrosymmetric position. In the product (panels (b) and (d)), the $-2.25$~eV peak is absent, and the local $\pi$ maximum has shifted to $\sim +0.15$~eV above $E_F$. Crucially, non-negligible Au~$d$--Pb~$p$ interactions persist as filled states below $E_F$ in the product, likely corresponding to states rehybridized after the SOJT distortion. This persistence of Au--Pb hybridization at $E_F$ — despite the formally closed-shell Au$^0$/Pb$^{2+}$ configuration — is the defining electronic signature of the SOJT mechanism: the total energy is lowered by mixing filled Au~5$d$ and empty Pb~6$p$ states, and this mixing is maximized only when Pb is displaced off its inversion center.

Taken together, the DFT analysis shows that the states directly implicated in holding Pb on its inversion center — the Au2--Pb $\pi$ frontier states — are precisely those depopulated by the applied electrochemical potential, and that the resulting structural relaxation produces a new hybridized electronic structure consistent with the SOJT-stabilized polar geometry.

\subsection{Ultraviolet Photoelectron Spectroscopy and Work Function Analysis}

The DFT picture can be placed on a quantitative experimental footing using ultraviolet photoelectron spectroscopy (UPS). The work functions ($\phi$) of both \ce{Au2PbP2} and \ce{Au2Pb_{0.914}P2} were determined with a He~I source ($h\nu = 21.218$~eV) and an applied sample bias of $-6$~V. In all spectra the bias is subtracted, placing the Fermi edge at $h\nu = 21.218$~eV.

\begin{figure}[H]
    \centering
    \resizebox{\textwidth}{!}{\includegraphics{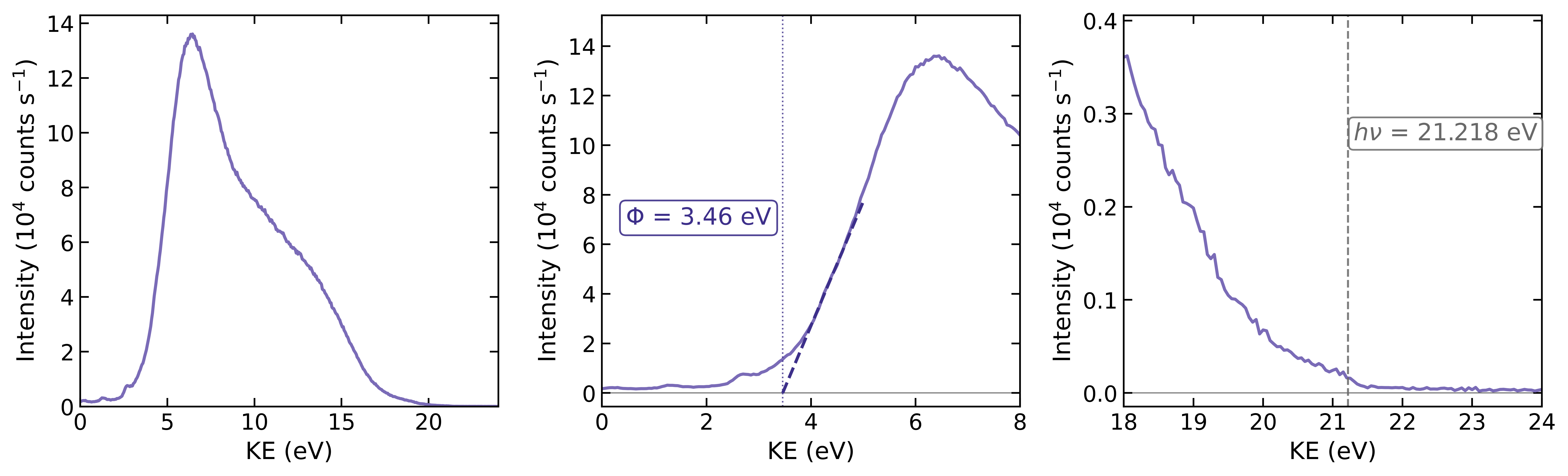}}
    \caption{\textbf{UPS of the parent compound \ce{Au2PbP2}.}
    \textbf{(a)} Full spectrum corrected for the $-6$~V bias.
    \textbf{(b)} Work function determination by linear extrapolation of the SECO: $\phi = 3.46$~eV.
    \textbf{(c)} Fermi edge region confirming correct bias subtraction (small drop at $h\nu = 21.218$~eV).}
    \label{fig:UPS_parent}
\end{figure}

\begin{figure}[H]
    \centering
    \resizebox{\textwidth}{!}{\includegraphics{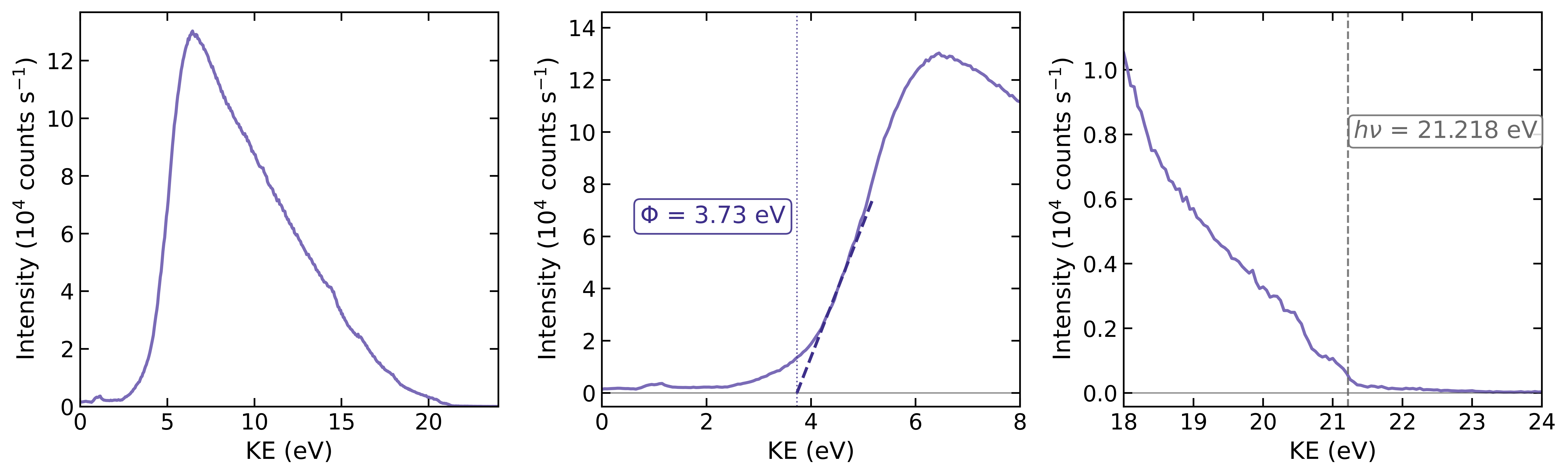}}
    \caption{\textbf{UPS of the electrochemically treated compound \ce{Au2Pb_{0.914}P2}.}
    \textbf{(a)}~Full spectrum corrected for the $-6$~V bias.
    \textbf{(b)} Work function determination: $\phi = 3.73$~eV.
    \textbf{(c)} Fermi edge region.}
    \label{fig:UPS_echem}
\end{figure}

Linear extrapolation of the secondary electron cutoff (SECO) rising edge yields $\phi = 3.46$~eV for the parent (Figure~S\ref{fig:UPS_parent}) and $\phi = 3.73$~eV for the product (Figure~S\ref{fig:UPS_echem}).

\subsubsection*{Connecting the electrochemical potential to the orbital window}

The applied potential of $+0.30$~V vs.\ Ag/AgCl (sat.\ KCl) can be converted to the vacuum scale using the standard hydrogen electrode (SHE) as an intermediate reference.\cite{Bard_2022} Taking the absolute potential of SHE as $+4.5$~eV vs.\ vacuum, and the Ag/AgCl (sat.\ KCl) electrode as $+0.197$~V vs.\ SHE, the applied potential corresponds to a Fermi level position of $-(4.5 + 0.197 + 0.30) \approx -5.0$~eV vs.\ vacuum.

Comparing this to the measured parent work function ($\phi = 3.46$~eV, $E_F = -3.46$~eV vs.\ vacuum), the electrochemical driving force represents a shift of $\sim$1.5~eV below the parent Fermi level. This is precisely the binding energy window (0--1.5~eV) that, from the orbital-decomposed pDOS, contains the hybridized Pb~6$p$--Au~5$d$ states responsible for the SOJT distortion and the stereochemical activation of the Pb$^{2+}$ lone pair. The selective depopulation of these states — and not deeper-lying states — is what initiates Pb displacement rather than wholesale structural decomposition.

The work function of the product ($\phi = 3.73$~eV, $E_F = -3.73$~eV vs.\ vacuum) is measurably larger than that of the parent ($\phi = 3.46$~eV), consistent with the oxidative removal of Pb lowering the E$_F$ with respect to vacuum. However, the product's Fermi level remains substantially higher (closer to vacuum) than the applied electrochemical potential of $-5.0$~eV. Had the crystal simply equilibrated rigidly with the electrode, one would expect $\Delta\phi \approx 1.5$~eV rather than the observed $\Delta\phi = 0.27$~eV.

This incomplete equilibration directly evidences an electronic reconstruction in response to Pb vacancy formation. The SOJT mechanism provides the explanation: rehybridization of the formerly-depleted Pb~6$p$--Au~5$d$ frontier states partially restores electron density near $E_F$, so the Fermi level settles at an intermediate position that reflects the new hybridized electronic structure of the polar product rather than the bare electrochemical driving force applied to the centrosymmetric parent. This self-limiting nature of the transformation is also consistent with the binary character of the reaction: only one product composition, \ce{Au2Pb_{0.914}P2}, is accessible, because the electronic reconstruction terminates further deintercalation once the SOJT-favorable $d^{10}s^2$ configuration is achieved.

This mechanistic insight may also provide unique strategies for improving this topotactic oxidation through soft-chemical processing. Initial attempts using concentrated \ce{HNO3} as oxidant produced phase-inhomogeneous samples (Figure S~\ref{uCT}\textbf{(b)}). This outcome can be understood directly in terms of the electronic structure established above. The effective oxidizing potential of concentrated \ce{HNO3} exceeds $+1.0$~V vs.\ SHE, corresponding to a Fermi level position of approximately $-5.7$~eV or lower on the vacuum scale. Referencing the parent work function ($\phi = 3.46$~eV, $E_\mathrm{F} = -3.46$~eV vs.\ vacuum), this represents a driving force of more than 2.2~eV below $E_\mathrm{F}$ --- well beyond the ${\sim}1.5$~eV window containing the target Pb~$6p$--Au~$5d$ frontier states, and extending into the energy range of the deep Au--Pb $\sigma$ bonding states at $-2.25$~eV below $E_\mathrm{F}$. Depopulation of these $\sigma$ states, as well as Au 6$d$ or P2$p$ orbitals, disrupts the structural framework that makes topotactic preservation of crystallinity possible, producing the disordered, multiphase product observed by $\mu$CT and depth-profile XPS (Figure S~\ref{uCT}\textbf{(b)}). 

By contrast, the electrochemical potential of $+0.3$~V vs.\ Ag/AgCl ($-5.0$~eV vs.\ vacuum) selectively accesses only the frontier $\pi$ states while leaving the lower-lying bonding states populated. In principle, a solution-phase redox couple with $E^\circ$ matched to the target potential and present in both oxidized and reduced forms could function as an electrochemical buffer, pinning the solution potential near the frontier state window and replicating the selectivity of the potentiostat without requiring electrochemical apparatus.

Taken together, the DFT and UPS analyses confirm the creation of a polar metal in \ce{Au2Pb_{0.914}P2}: ample states remain at $E_F$ for metallic conduction, while the crystallographic point group is reduced from $mmm$ to $mm2$. The polar space group enables physical phenomena forbidden in centrosymmetric structures, and the heavy-element composition introduces strong spin-orbit coupling. These observations lead to two specific predictions that are tested experimentally in the following sections: (1) symmetry-allowed nonlinear transport signatures consistent with the $mm2$ point group, and (2) antisymmetric spin-orbit coupling that may influence the superconducting gap structure at low temperature.

\section{Nonlinear Transport Properties}

In materials lacking inversion symmetry, the second-order approximation of a voltage response to an applied current deviates from simple Ohmic behavior and can be expressed as:
\begin{equation}
V = R^{(1,0)}I + R^{(1,2)}IB^2+ R^{(1,1)}IB +R^{(2,1)}I^2B +R^{(2,0)}I^2
\end{equation}

\noindent where the superscript of R indicate the scaling dependence on current and applied field, respectively.

In this expression, the conventional linear resistance term ($R^{(1,0)}I$) is supplemented by higher-order contributions that depend on both current and magnetic field. Notably, in the absence of an external magnetic field ($B = 0$), only the linear ($R^{(1,0)}I$) and quadratic ($R^{(2,0)}I^2$) terms remain, with the latter serving as a direct manifestation of broken inversion symmetry. This quadratic response can be formulated in terms of the second-order nonlinear conductivity tensor ($\chi_{ijk}^{(2)}$), which relates the second-harmonic current density ($J_i^{2\omega}$) to the product of two electric field components at the fundamental frequency (E$^{\omega}_j$):
\begin{equation}
J_i^{2\omega}=\chi_{ijk}^{(2)} E_j^\omega E_k^\omega
\end{equation}

For the mm2 point group of Au$_2$Pb$_{0.914}$P$_2$, group theoretical analysis\cite{SuarezRodriguez_2025, Aroyo_2006, Aroyo_2006b, Gallego_2019} dictates that the $\chi_{ijk}^{(2)}$ tensor adopts the form:
\begin{equation}
\chi_{ijk}^{(2)}=\begin{bmatrix}
\textbf{\;0\;\;\;\; 0\;\;\;\;\; 0\;\;\;\;\; 0\;\;\;  $\chi_{xxz}$\;  0}\\
\textbf{\,\;0\;\;\;\; 0\;\;\;\;\; 0\;\;\;\; $\chi_{yyz}$\; 0\;\; \;0}\\
\textbf{$\chi_{zxx}$ $\chi_{zyy}$ $\chi_{zzz}$\;\;\; 0\;\;\;\; 0\;\;\; 0}\\
\end{bmatrix}
\end{equation}

To confirm this hypothesis, we then made multiple four probe devices in 3 different geometries: (1) a longitudinal configuration with current and voltage contacts aligned along the principle axis ($V_z(I^{\omega}_z)$), (ii) a Hall configuration where the current is fed along the crystallographic $a$-axis ($x$) and voltage measured along the crystallographic $c$-axis ($z$) ($V_z(I^{\omega}_x)$), and (iii) The same Hall configuration, instead with current along y ($V_z(I^{\omega}_y)$). For all measurements, the contacts were split to simultaneously lock-in to the first- and second-harmonic signals individually such that each set of measurements shown below are taken at the same time. 

\begin{figure}
\renewcommand{\figurename}{Figure S}
    \centering
    \includegraphics[width=0.5\linewidth]{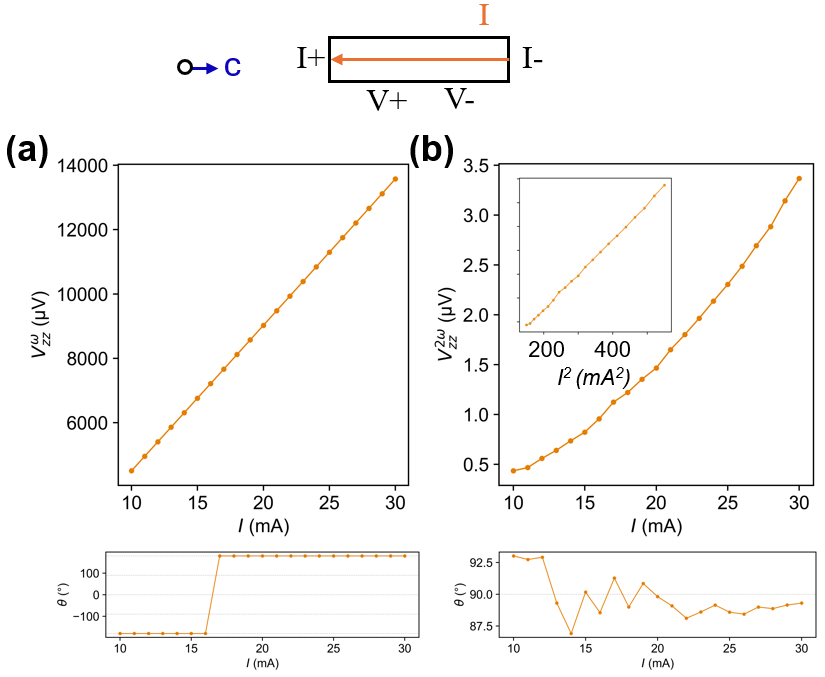}
    \caption{\textbf{Nonlinear longitudinal conductivity of Au$_2$Pb$_{0.914}$P$_2$.} \textbf{(top)} schematic of device geometry for the longitudinal geometry where current and voltage contacts are applied along the polar $c$-axis. \textbf{(a)} at 300k, first-order harmonic signal is linear with a phase angle very close to $\pm180^\circ$. \textbf{(b)} Second-order harmonic signal taken simultaneously with the first-order harmonic. The signal is quadratic, as indicated by the linearity of $I^2-V$ shown in the inset, with a phase angle locking in at 90$^\circ$.}
    \label{Vzz}
\end{figure}

Measurements for $V_z(I^{\omega}_z)$ are consistent with our sample obtaining a polar z-axis. Measured in the current range of 10 to 30 mA, the first-harmonic signal shows a clean, linear $I^\omega-V^\omega$ relation that is consistent with expected behavior (Figure S\ref{Vzz}\textbf{(a)}). The phase angle, expected to be 0 or $\pm$ 180$^\circ$ is also shown below this plot. The second-harmonic voltage, V$^{2\omega}_{zz}$ (Figure \ref{Vzz}\textbf{(b)}) grows quadratically with current, as confirmed by the linear trend in the $V_{zz}^{2\omega}$ versus $I^2$ inset 
(Figure~\ref{Vzz}\textbf{b}). The phase angle of the second-harmonic also stabilizes at expected values of $\pm 90^\circ$, consistent with a genuine second-order response.

\begin{figure}
\renewcommand{\figurename}{Figure S}
    \centering
    \includegraphics[width=1\linewidth]{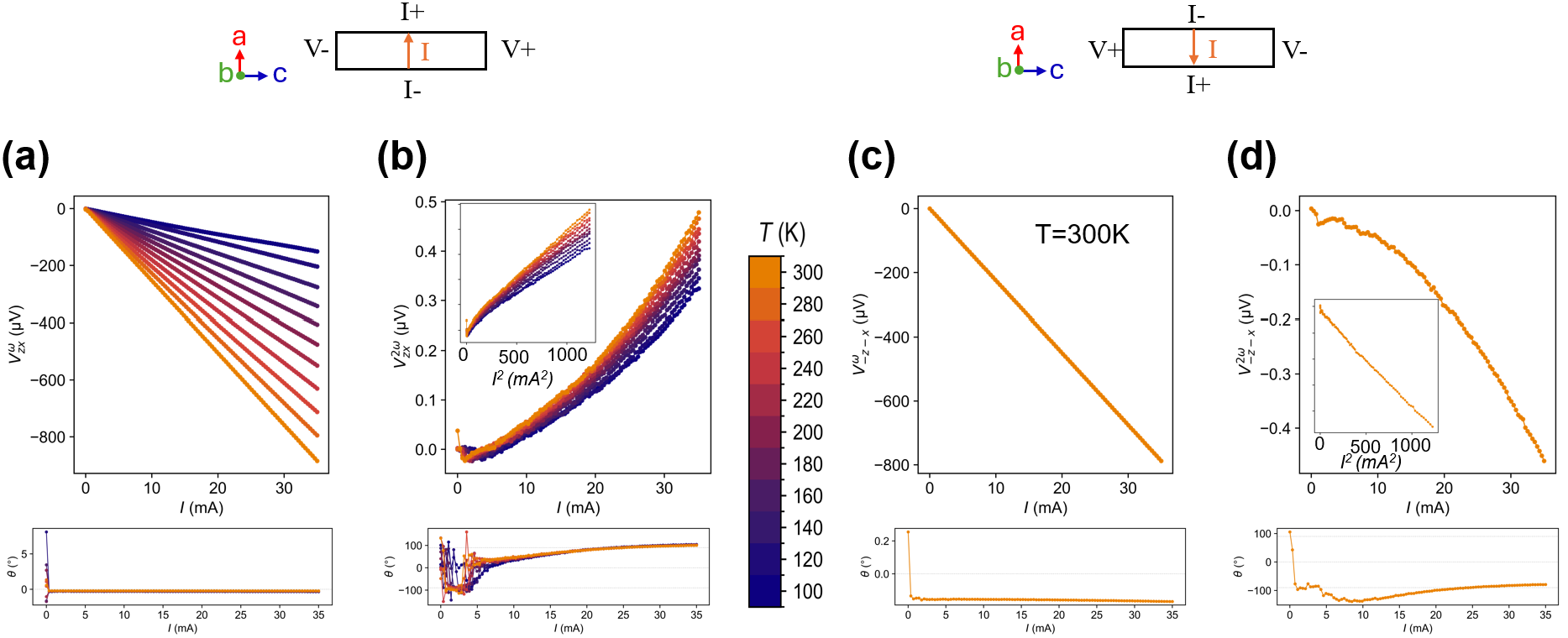}
    \caption{\textbf{Temperature-dependent nonlinear Hall conductivity of Au$_2$Pb$_{0.914}$P$_2$ in the $zx$ plane.} \textbf{(top)} schematic of two device geometries measured where current is applied along the $a$-axis and voltage measured along the polar $c$-axis. \textbf{(a)} First-order harmonic signal is linear with a phase angle very close to $0^\circ$. \textbf{(b)} Second-order harmonic signal taken simultaneously with the first-order harmonic. The signal is quadratic, as indicated by the linearity of $I^2-V$ shown in the inset, with a phase angle locking in at 90$^\circ$. \textbf{(c)} First-order harmonic signal after flipping both the current and voltage leads, converting the geometry of the measurement to V$_{-z-x}$. The signal is linear with a phase angle very close to $0^\circ$. Importantly, the slope has not changed signs as expected for a first-order signal. \textbf{(d)} Second-order response of V$_{-z-x}$. Flipping the contacts causes the second-order harmonic signal to flip, consistent with second-harmonic voltage signals.}
    \label{Vzx}
\end{figure}

Temperature-dependent measurements in the transverse geometry are shown in Figure  S\ref{Vzx}\textbf{(a)} and \textbf{(b)} for $V_{zx}$. The first-harmonic signal $V_{zx}^\omega$ (Figure S\ref{Vzx}\textbf{(a)}) displays a linear $I^\omega-V^\omega$ dependence whose slope decreases with decreased temperature. The corresponding second-harmonic signal $V_{zx}^{2\omega}$ grows quadratically with current across all temperatures, as confirmed by the linear $V_{zx}^{2\omega}$ versus $(I^{\omega})^2$ dependence shown in the inset, as well as its phase angle locking in to 90$^\circ$ above a few milliamperes. To verify that this second-order harmonic is real, as opposed to a measurement artifact, we then simultaneously flipped both the voltage and current contacts, converting the geometry of the measurement to V$_{-z-x}$ (Figure S\ref{Vzx}\textbf{(c)} and \textbf{(d)}). Under this full contact swap, the first-harmonic voltage is unchanged in sign (Figure S\ref{Vzx}\textbf{(c)}), as required for a linear resistive response, while the second-harmonic voltage reverses sign (Figure S\ref{Vzx}\textbf{(d)}), as expected for a polar nonlinear response. The $(I^{\omega})^2$ linearity and $\approx-90^\circ$ phase of V$_{-z-x}^{2\omega}$ are fully consistent with the sign-reversed counterpart of V$_{zx}^{2\omega}$. 

\begin{figure}
\renewcommand{\figurename}{Figure S}
    \centering
    \includegraphics[width=1\linewidth]{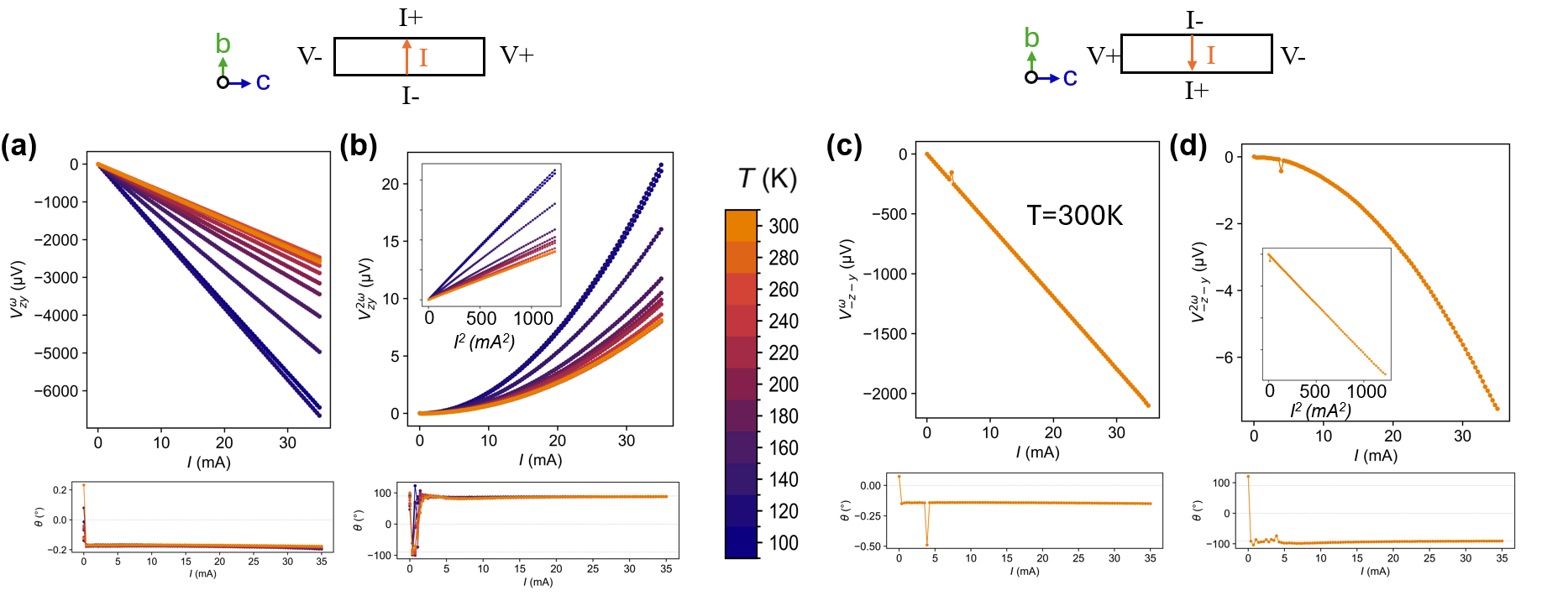}
    \caption{\textbf{Temperature-dependent nonlinear Hall conductivity of Au$_2$Pb$_{0.914}$P$_2$ in the $zy$ plane.} \textbf{(top)} schematic of two device geometries measured where current is applied along the $b$-axis and voltage measured along the polar $c$-axis. \textbf{(a)} First-order harmonic signal is linear with a phase angle very close to $0^\circ$. \textbf{(b)} Second-order harmonic signal taken simultaneously with the first-order harmonic. The signal is quadratic, as indicated by the linearity of $I^2-V$ shown in the inset, with a phase angle locking in at 90$^\circ$. \textbf{(c)} First-order harmonic signal after flipping both the current and voltage leads, converting the geometry of the measurement to V$_{-z-y}$. The signal is linear with a phase angle very close to $0^\circ$. Importantly, the slope has not changed signs as expected for a first-order signal. \textbf{(d)} Second-order response of V$_{-z-y}$. Flipping the contacts causes the second-order harmonic signal to flip, consistent with second-harmonic voltage signals.}
    \label{Vzy}
\end{figure}

Similar measurements were taken in the other symmetry-allowed transverse geometry, shown in Figure S\ref{Vzy} for $V_{zy}$. As shown in Figure S\ref{Vzy} \textbf{(a)}, the first-harmonic signal also displays a linear $I^\omega-V^\omega$ dependence. However, the slope now decreases with decreased temperature. The corresponding second-harmonic signal $V_{zy}^{2\omega}$ (Figure S\ref{Vzy} \textbf{(b)}) grows quadratically with current across all temperatures, as confirmed by the linear $V_{zy}^{2\omega}$ versus $(I^{\omega})^2$ dependence shown in the inset, as well as its phase angle locking in to 90$^\circ$ immediately. As before, we also verified that this second-order harmonic is real by simultaneously flipping both the voltage and current contacts, converting the geometry of the measurement to V$_{-z-y}$ (Figure S\ref{Vzx}\textbf{(c)} and \textbf{(d)}). Under this full contact swap the first-harmonic voltage is unchanged in sign (Figure S\ref{Vzy}\textbf{(c)}), as required for a linear resistive response, while the second-harmonic voltage reverses sign (Figure S\ref{Vzy}\textbf{(d)}), as expected for a polar nonlinear response.

\begin{figure}
\renewcommand{\figurename}{Figure S}
    \centering
    \includegraphics[width=0.5\linewidth]{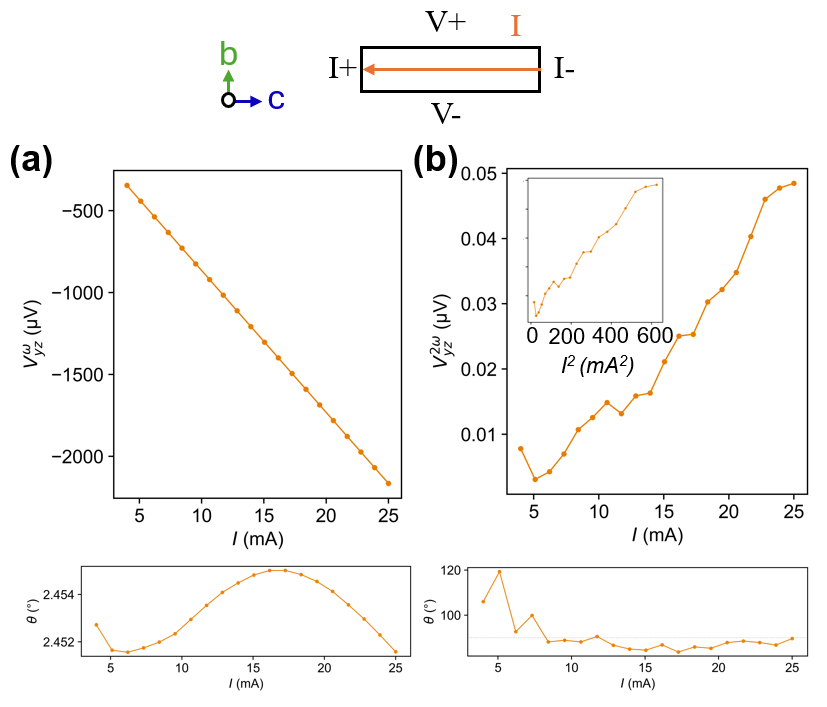}
    \caption{\textbf{Symmetry-disallowed nonlinear Hall conductivity of Au$_2$Pb$_{0.914}$P$_2$ in the $zy$ plane.} \textbf{(top)} schematic of the device geometry measured where current is applied along the polar $c$-axis and voltage measured along the polar $b$-axis. \textbf{(a)} First-order harmonic signal is linear with a phase angle close to $0^\circ$. \textbf{(b)} Second-order harmonic signal taken simultaneously with the first-order harmonic. The signal is roughly two orders of magnitude smaller than that observed in the allowed V$_zy$ which is attributed to slight contact misalignment. Signal is loosely quadratic with a phase angle locking in close to 90$^\circ$.}
    \label{Vyz}
\end{figure}

To qualitatively rule out Joule heating as the source of this second-order signal, as well as confirm the polar $mm2$ point group assignment, we also measured this crystal in a symmetry-disallowed geometry, shown in Figure S\ref{Vyz} for $V_{yz}$. Here, current flows along the polar $c$-axis and the voltage is sensed across the $b$-axis. For the $mm2$ point group, the second-order response in this geometry is identically zero by symmetry. As shown in Figure S\ref{Vyz}\textbf{(a)}, the first-order signal has a linear $I^\omega-V^\omega$ dependence with a similar magnitude as $V_{zy}$. However, the second-order response $V_{yz}^{2\omega}$ (Figure S\ref{Vyz}\textbf{(b)}) is approximately two orders of magnitude smaller than $V_{zy}^{2\omega}$ measured on the same crystal face. This residual signal is attributed to a slight misalignment of the voltage contacts away from a perfect $b$-axis, rather than to an intrinsic signal. Together, this contact-swap antisymmetry and geometric null constitute a robust symmetry argument for an intrinsic polar nonlinear conductivity in Au$_2$Pb$_{0.914}$P$_2$.

\begin{figure}
\renewcommand{\figurename}{Figure S}
    \centering
    \includegraphics[width=0.5\linewidth]{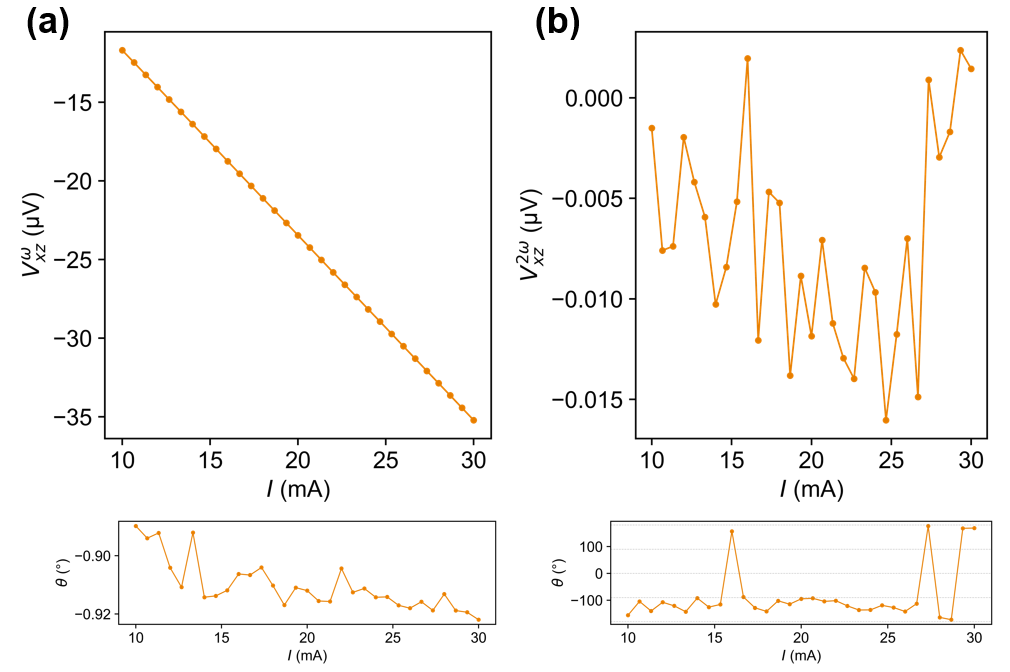}
        \caption{\textbf{Absence of nonlinear Hall conductivity of the parent compound, \ce{Au2PbP2}.} \textbf{(a)} First-order harmonic signal is linear with a phase angle close to $0^\circ$. \textbf{(b)} Second-order harmonic signal taken simultaneously with the first-order harmonic. As expected for a centrosymmetric compound, no second-order harmonic signal is observed.}
        \label{ParentVxz}
\end{figure}

Finally, we also attempted second-harmonic transport on crystals of our parent compound, \ce{Au2PbP2}. As the parent is in the centrosymmetric space group C\textit{mcm}, there should be a complete absence of second-order signal. In Figure S\ref{ParentVxz}\textbf{(a)}, the first-order harmonic signal is strong and linear in the current range of 10 to 30mA. However, in Figure S\ref{ParentVxz}\textbf{(b)}, a clear lack of any second-harmonic signal is observed, as expected.

\FloatBarrier

\section{Superconductivity Measurements}
\subsection{Electronic Transport}

Temperature- and field-dependent resistivity measurements are shown below in Figures S\ref{RT50} and S\ref{MR50}. After normalization to the normal state (taken as T=1.75~K and $\mu_0$H=300 mT, respectively), the exact transition, defined as a 50\% drop from normal state resistivity, was established. 

\begin{figure}
\renewcommand{\figurename}{Figure S}
    \centering
    \includegraphics[width=0.5\linewidth]{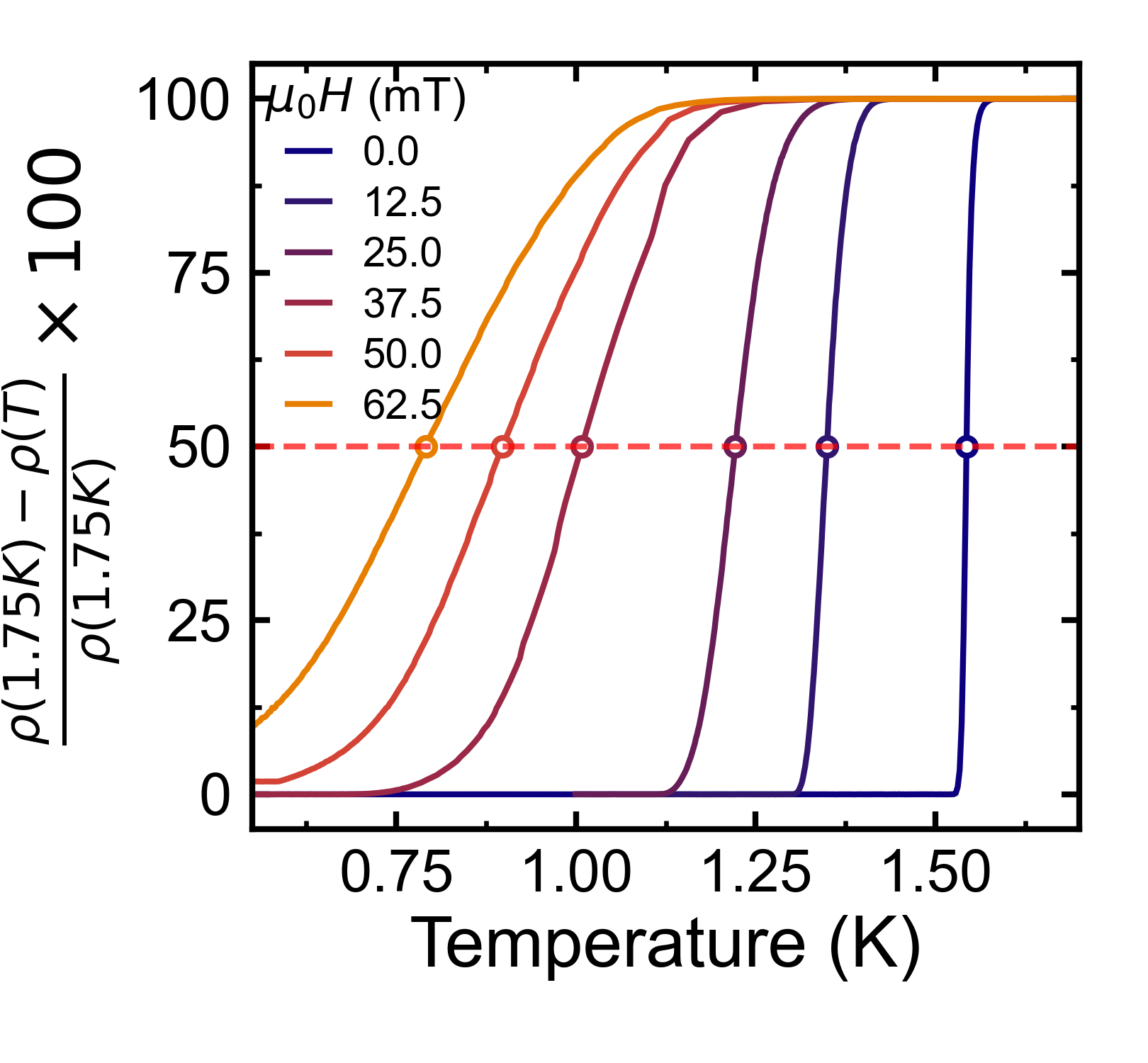}
        \caption{\textbf{Temperature-dependent resistivity data for Au$_2$Pb$_{0.914}$P$_2$.} Data is normalized to normal state resistivity just above T$_c$ at 1.75~K.}
        \label{RT50}
\end{figure}

\begin{figure}
\renewcommand{\figurename}{Figure S}
    \centering
    \includegraphics[width=0.5\linewidth]{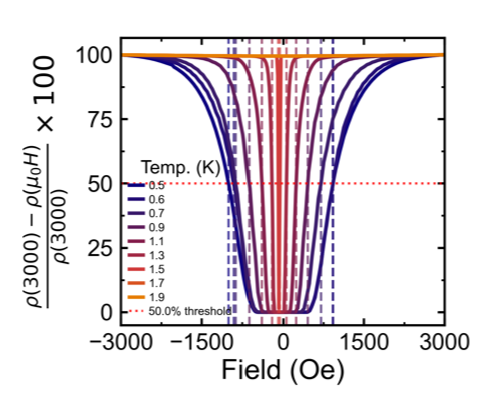}
        \caption{\textbf{Field-dependent resistivity isotherms for Au$_2$Pb$_{0.914}$P$_2$.} Data is normalized to normal state resistivity at $\mu_0$H=300~mT at 1.75~K.}
        \label{MR50}
\end{figure}

We then used the transition temperature and fields to create a phase diagram of the upper critical field, shown in Figure 6\textbf{(b)} of the main text. As an initial fit to find the upper critical field, we attempted a simplified Werthamer-Helfand-Hohenberg (WHH) fit. This simplified fit takes the form $\mu_0H_{c2}(0) = -0.693*T_c*\frac{d\mu_0H_{c2}}{dT}|_{_{T = T_c}}$. Here, $T_c$ is the critical temperature of 1.52 K, and 0.693 is the expected relationship in the dirty limit, which will be confirmed later, for WHH fits. The line representing the slope close to $T_c$, $\frac{d\mu_0H_{c2}}{dT}|_{_{T = T_c}}$, is shown as a gray dotted line and is found to be -56.67 mT/K, resulting in an initial upper critical field of $\mu_0H_{c2}(0)$ = 71.5 mT. We then constructed a line of the form $H_{c2}(T) = H_{c2}(0) * [1 - (\frac{T}{T_c})^2]$ to illustrate the phase boundary of an isotropic superconducting gap, given by the red dashed curve. In comparison, at 0.5 K, our $\rho$-$\mu_0H$ data show an upper critical field of 97.9 mT, suggesting a deviation in the upper critical field predicted by this equation. The deviation in the WHH fit appears from a positive curvature for the experimental data at temperatures close to T$_c$, contrasted to the typical dome-like shape predicted by models obeying WHH fits. This phenomenon has been discussed in noncentrosymmetric superconductors such as LaNiC$_2$\cite{Chen_2013, Hirose_2012} and MgB$_2$ \cite{Buzea_2001}, a positive curvature close to T$_c$ is an indication of multi-gap behavior due to interband coupling at higher temperatures. 

Thus, we also fit our data to a phenomenological fit based on a two-band Ginzburg-Landau (GL) model,\cite{Chen_2013, Buzea_2001} taking the form of $H_{c2}(T) = H_{c2}(0) * \frac{1 - (\frac{T}{T_c})^2} {1 + \alpha*(\frac{T}{T_c})^2}$. In this equation, $\alpha$ is a freely fit parameter that is refined to accurately describe the curvature within our data. This fit yields $\mu_0H_{c2}(0)$ = 140.2 mT and $\alpha$=2.21, shown as the orange curve in Figure 6\textbf{(b)} of the main text. The positive value of $\alpha$ is consistent with the upward curvature observed near $T_c$ and is characteristic of a significant disparity in quasiparticle diffusivities between two bands.

To obtain a microscopically interpretable description, we additionally fit the data to the full two-band dirty-limit theory of Gurevich \cite{Gurevich_2003} which generalizes the WHH formalism to a two-band superconductor. In this model, $\mu_0H_{c2}(T)$ is determined implicitly in equation \ref{GModel} 

\begin{equation}
a_0\left[\ln t + U(h)\right]\left[\ln t + U(\eta h)\right] + a_2\left[\ln t + U(\eta h)\right] + a_1\left[\ln t + U(h)\right] = 0 
\label{GModel}
\end{equation}

Where t=$\frac{T}{T_c}$, $U(x) = \psi(\frac{1}{2}+x)-\psi(\frac{1}{2})$ is the regularized digamma function, and $h=\frac{\mu_0H_{c2}D_1}{2\Phi_0T}$. The ratio $\eta=\frac{D_2}{D_1}$ is the intraband diffusivity ratio between the two bands, and the coefficients $a_0$, $a_1$, $a_2$, are determined by the intraband BCS coupling constants $\lambda_{11}$ and $\lambda_{22}$ with interband coupling $\lambda_{12}$=$\lambda_{21}$. \cite{Gurevich_2003} The free parameters $D_1$, $\eta$, $\lambda_{11}$, and $\lambda_{22}$ were detemined with interband coupling fixed at the weak-coupling value $\lambda_{12}$=$\lambda_{21}$=0.04.

\begin{figure}
\renewcommand{\figurename}{Figure S}
    \centering
    \includegraphics[width=0.5\linewidth]{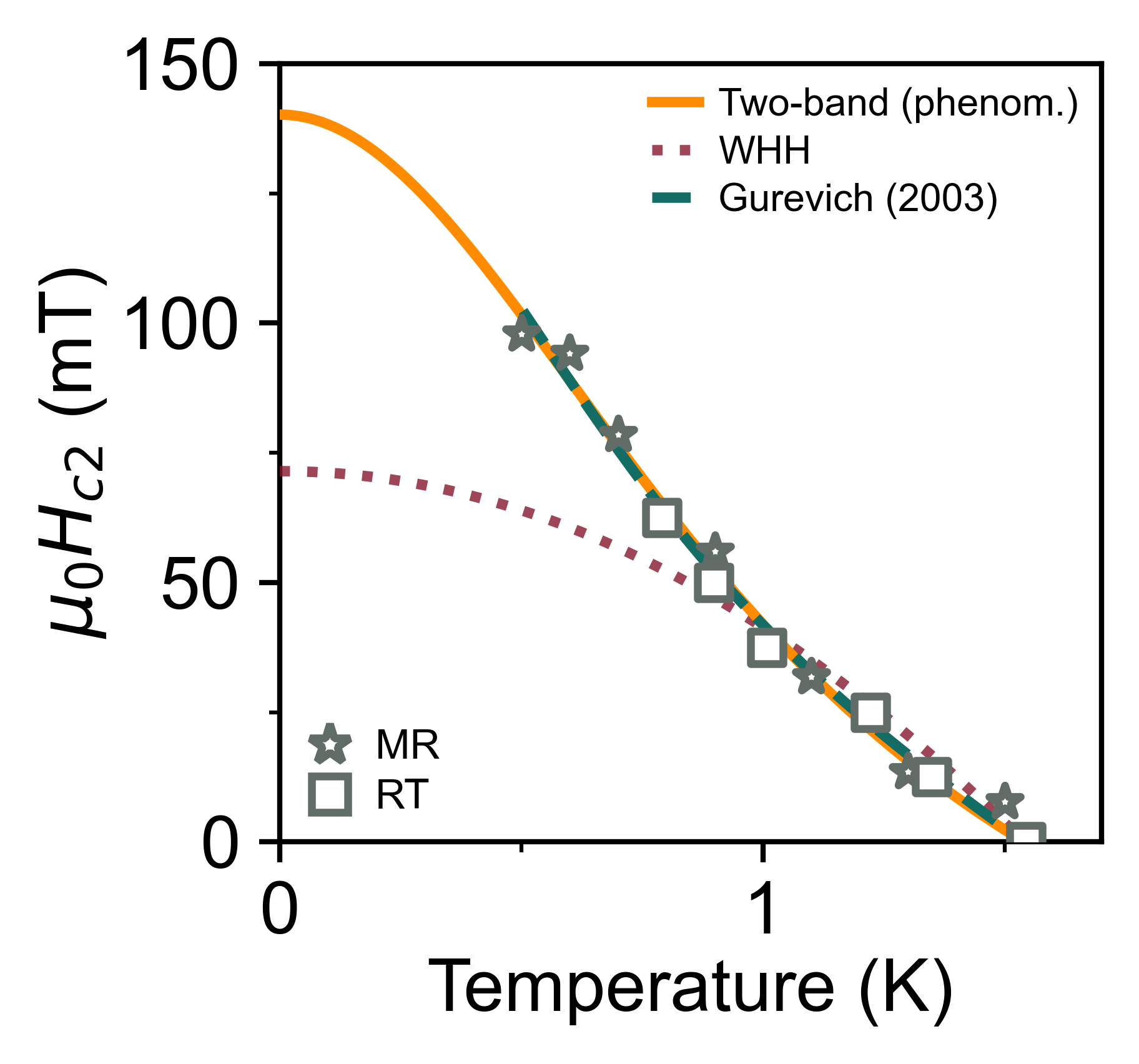}
        \caption{\textbf{Comparison of models used in determining the upper critical field of Au$_2$Pb$_{0.914}$P$_2$.} Stars indicate transitions from magnetoresistivity (MR) measurements, and squares indicate transitions from temperature-dependent resistivity (RT). The red dotted line indicates the simplified WHH fit, the yellow solid line is a phenomenological two-band fit which closely follows the full two-band dirty-limit theory of Gurevich, shown by a green dashed line.}
        \label{GFit}
\end{figure}

The Gurevich fit, shown as a green dashed line in Figure S\ref{GFit}, closely follows our phenomenological fit. The Gurevich fit yields an upper critical field of $\mu_0H_{c2}(0)$ = 136.7 mT, with fitted intraband diffusivities of $D_1$ = 1.81 $\times$ 10$^{-14}$ m$^2$/s and $D_2$ = 3.58 $\times$ 10$^{-15}$ m$^2$/s leading to a ratio $\eta \approx0.20$, indicating that one band is approximately five times more disordered than the other. 

\begin{figure}
\renewcommand{\figurename}{Figure S}
    \centering
    \includegraphics[width=0.5\linewidth]{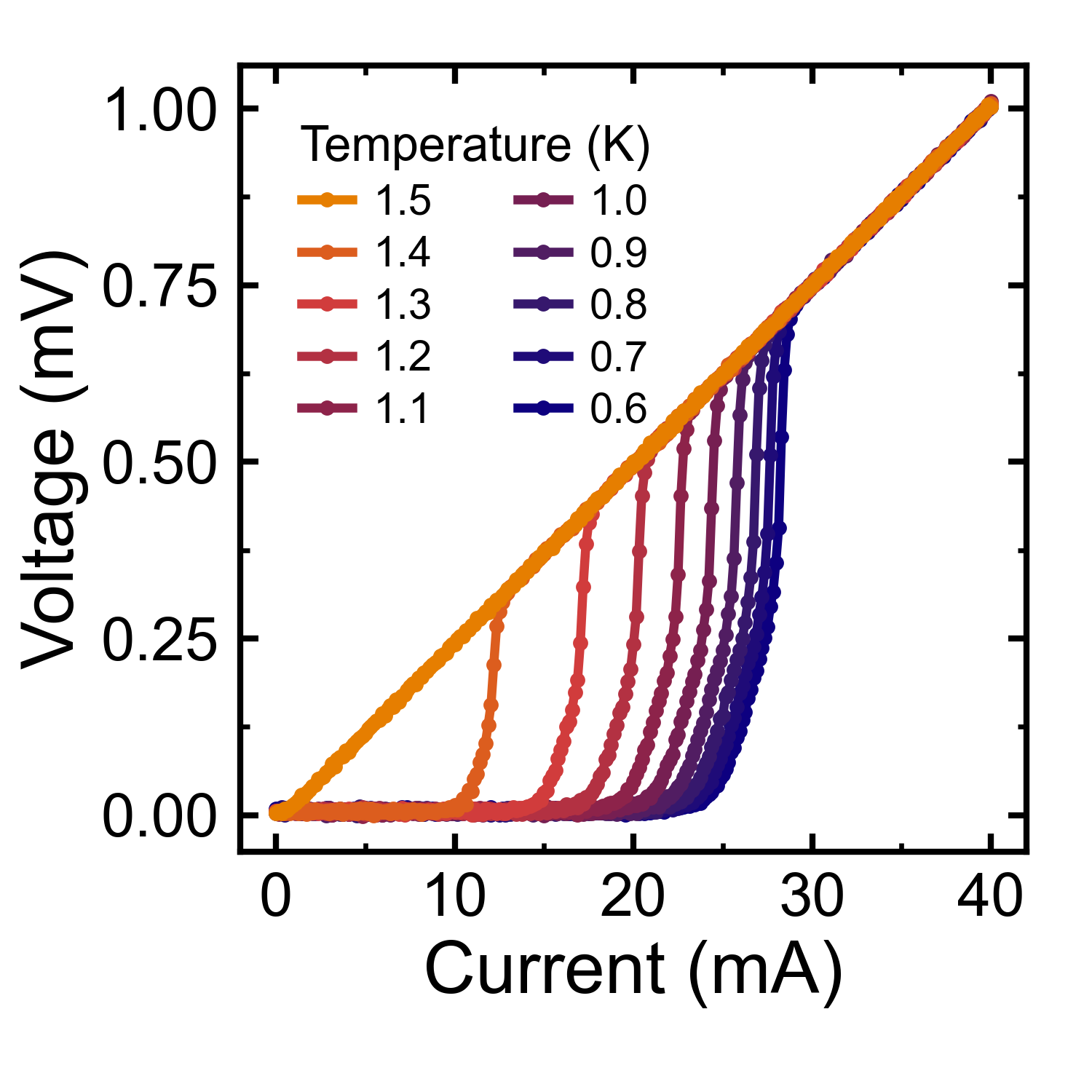}
        \caption{\textbf{Temperature-dependent $I-V$ characteristics in the superconducting state of Au$_2$Pb$_{0.914}$P$_2$.} The critical current at 0.6K is approximately 21.72 mA.}
        \label{IVTemp}
\end{figure}

\begin{figure}
\renewcommand{\figurename}{Figure S}
    \centering
    \includegraphics[width=0.5\linewidth]{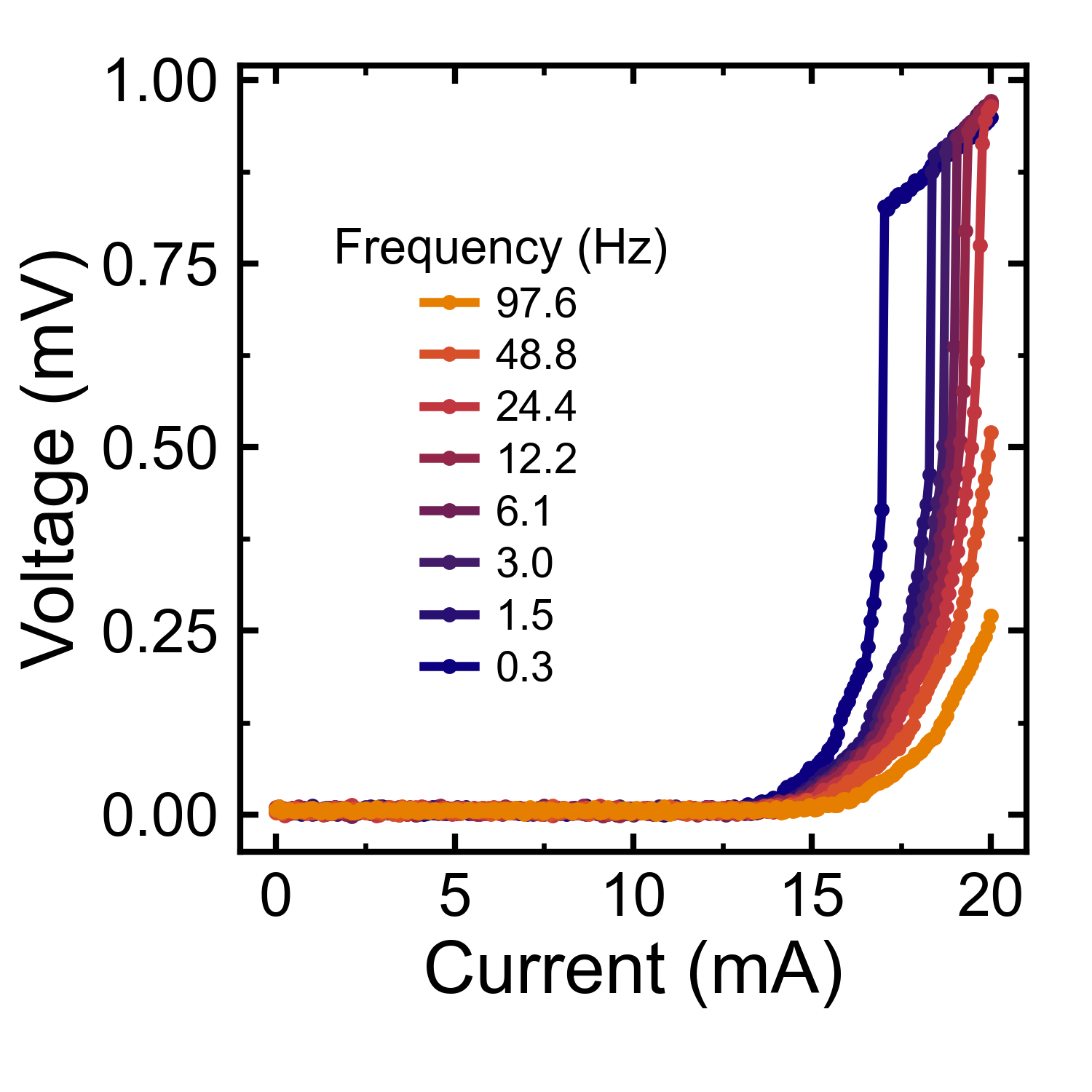}
        \caption{\textbf{Field-dependent $I-V$ characteristics in the superconducting state of Au$_2$Pb$_{0.914}$P$_2$.} The critical current at 97.66 Hz is approximately 25.78 mA.}
        \label{IVFreq}
\end{figure}

Current-voltage ($I$-$V$) characteristics, measured at a fixed AC frequency of 48.8 Hz (Figure S\ref{IVTemp}), reveal a well-defined critical current ($I_c$) of approximately 21.72 mA at 0.6 K. This critical current exhibits systematic temperature dependence, decreasing progressively with increasing temperature until the supercurrent vanishes entirely above $T_c$. Frequency-dependent measurements (Figure S\ref{IVFreq}) demonstrate that the critical current also varies with the AC excitation frequency. At 0.6 K, $I_c$ reaches a maximum value of approximately 25.78 mA at an excitation frequency of 97.66 Hz, while decreasing to approximately 21.88 mA at the lowest measured frequency of 0.30 Hz. For all frequencies investigated, the $I$-$V$ characteristics display pronounced nonlinearity above $I_c$, consistent with the onset of Joule heating in the normal state.

\FloatBarrier

\subsection{Heat Capacity}

Normal state behavior was fit from experimental data with an applied field of 5T. In the temperature range of T = 0.7-2.9~K, C$_p$/T vs T$^2$ and C$_p$ vs T were fit simultaneously to fit using equations \ref{Cp_T} and \ref{CpT_T2}. 

\begin{equation}
C_p(5T) = \gamma_N T + \beta T^3   
\label{Cp_T}
\end{equation}

\begin{equation}
C_p(5T)/T = \gamma_N + \beta T^2   
\label{CpT_T2}
\end{equation}

\begin{figure}[H]
\renewcommand{\figurename}{Figure S}
    \centering
    \resizebox{6.5in}{!}{\includegraphics{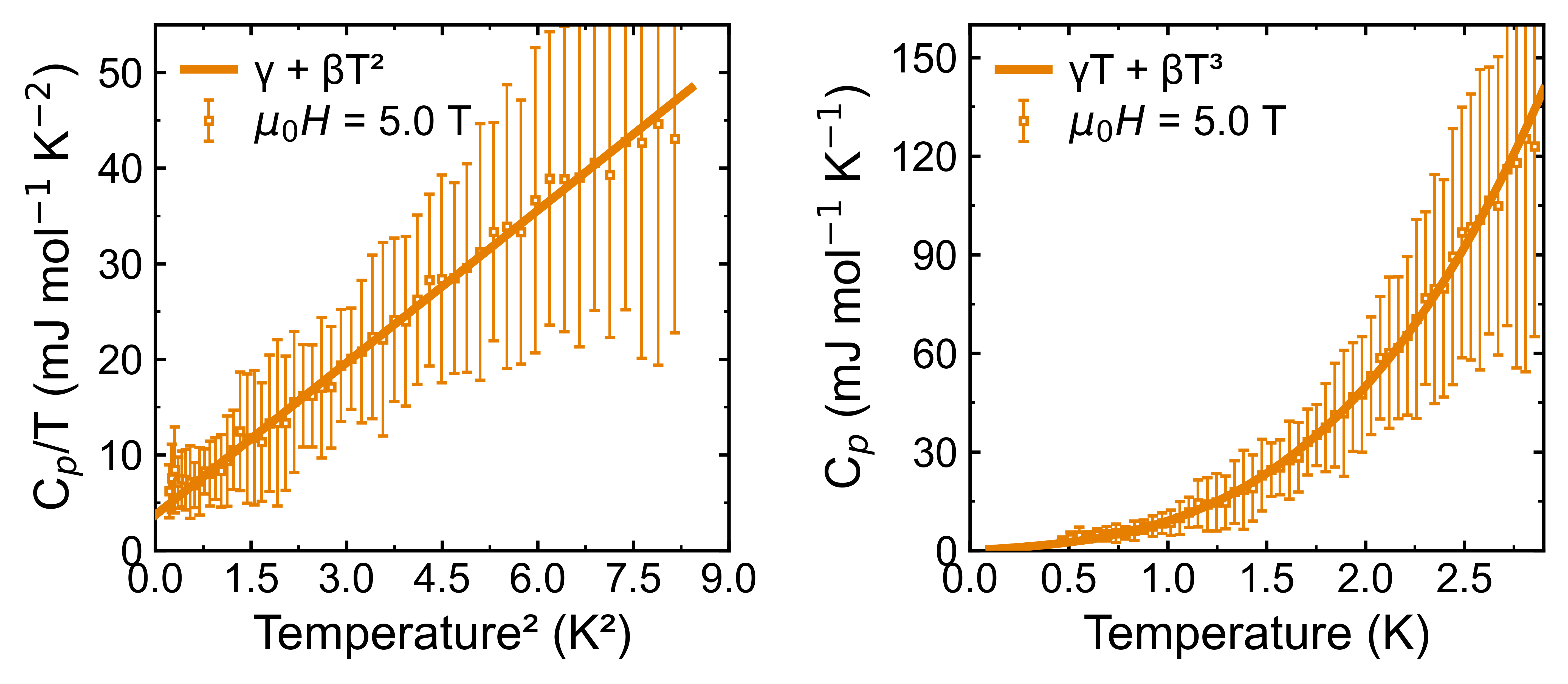}}
    \caption{\textbf{Simultaneous normal state fits of heat capacity data at $\mu_0(H)$ = 5~T.} Solid lines represent fits to the data (small squares) resulting in values $\gamma_N$ = 3.719 $\pm$ 0.925 mJ mol$^{-1}$K$^{-2}$ and $\beta$ = 5.310 $\pm$ 0.427 mJ mol$^{-1}$K$^{-4}$.}
\label{NormalStateFits}
\end{figure}

Where C$_p$ is the total sample heat capacity, the normal state Sommerfeld coefficient $\gamma_N$ = 3.719 mJ mol$^{-1}$K$^{-2}$ and phonon contribution $\beta$ = 5.310 mJ mol$^{-1}$K$^{-4}$ (Figures S\ref{NormalStateFits}a and b). Fits were successful with R$^2$'s of 0.9954 and 0.9922, respectively. 

Next, the fit parameters were used to subtract the phonon contribution term, $\beta$T$^3$ from the zero field data. Here,

\begin{equation}
C_{el}(0T) = C_p - \beta T^3
\end{equation}

\begin{figure}[H]
\renewcommand{\figurename}{Figure S}
    \centering
    \resizebox{3in}{!}{\includegraphics{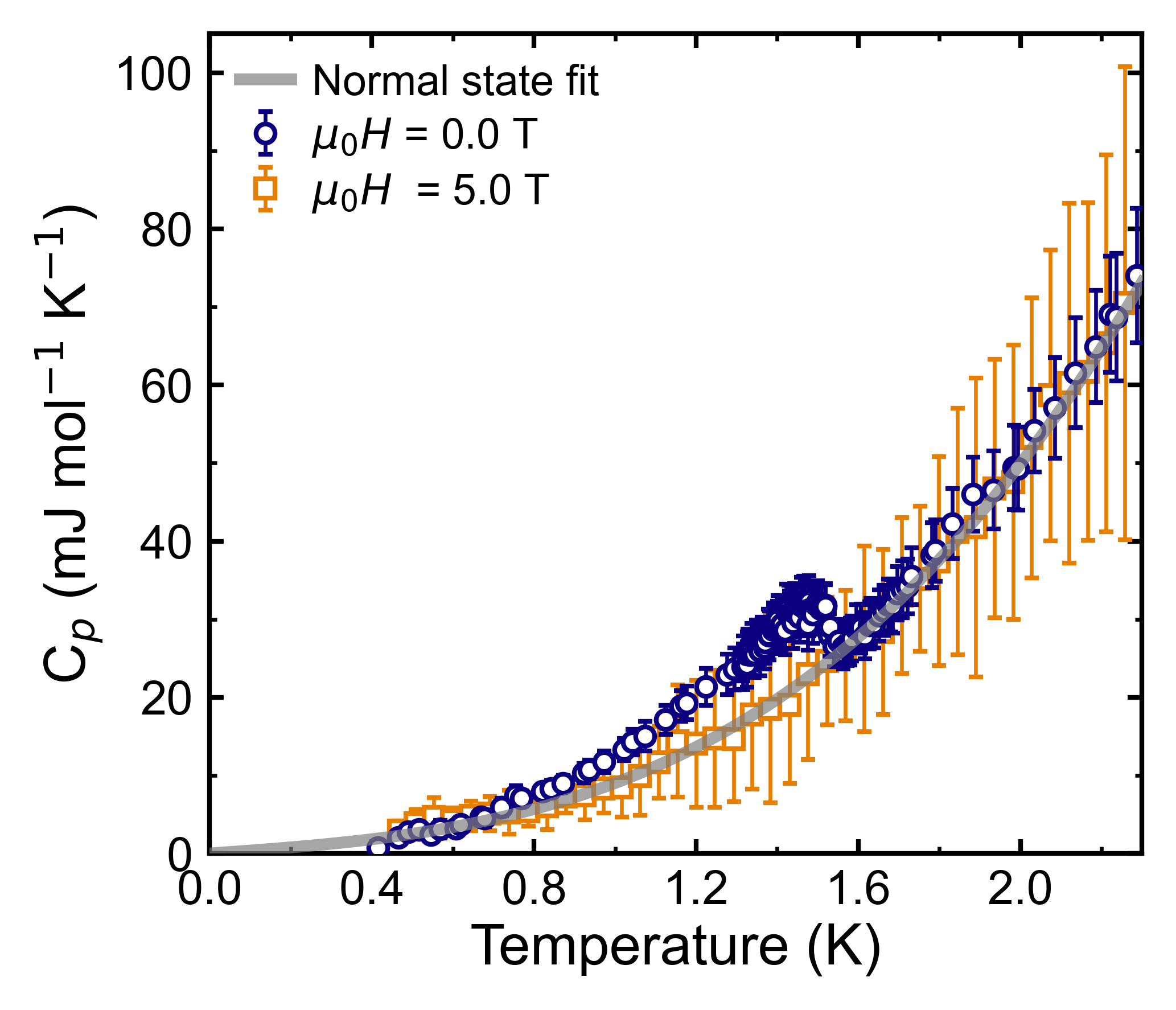}}
    \caption{\textbf{Superconducting state data and normal state fit of heat capacity data at 5T.} Solid lines represent fits to the normal state data (small squares) given parameters from the above simultaneous fit.}
\label{heat_capacity_fixed_parameters_plot}
\end{figure}

\begin{figure}[H]
\renewcommand{\figurename}{Figure S}
    \centering
    \resizebox{3in}{!}{\includegraphics{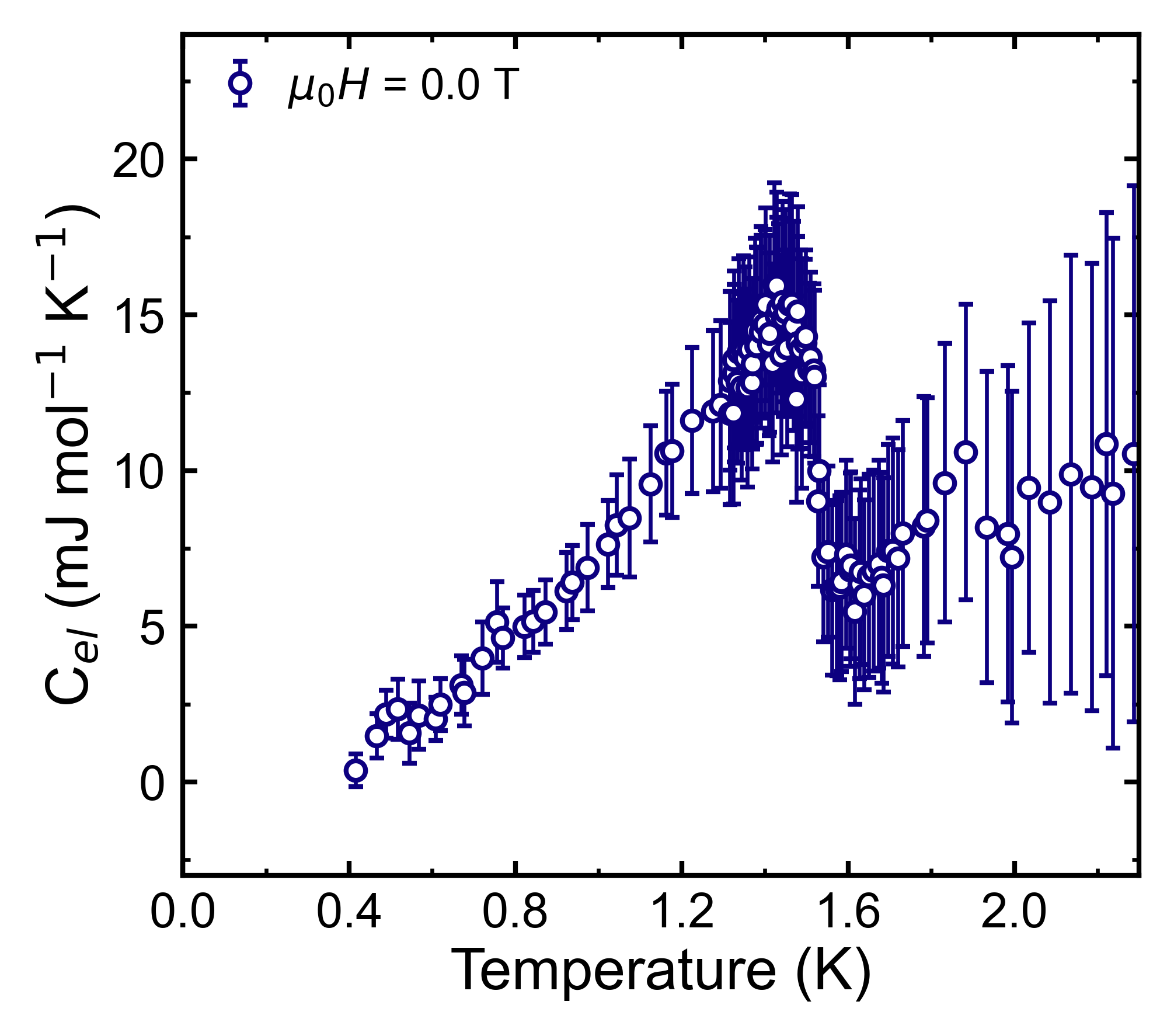}}
    \caption{\textbf{Phonon subtracted zero field heat capacity data.} Electronic component of the superconducting transition seen from heat capacity}
\label{electronic_heat_capacity_plot}
\end{figure}

Where C$_{el}$ represents the electronic component of the heat capacity, plotted in Figures S\ref{heat_capacity_fixed_parameters_plot} and S\ref{electronic_heat_capacity_plot}. 

With our fit $\beta$, we can then evaluate the Debye temperature ($\Theta_D$), given by equation \ref{Debye Temp}.~\cite{Carnicom_2018}

\begin{equation}
\Theta_D = \left(\frac{12\pi^4}{5 \beta}nR\right)^{1/3}
\label{Debye Temp}
\end{equation}

Here, $n$ = 4.914 (Au$_2$Pb$_{0.914}$P$_2$) and $R$ is the gas constant 8.314 J mol$^{-1}$ K$^{-1}$, resulting in a Debye temperature of 121.6 $\pm$ 3.3~K.

We can then estimate the electron-phonon coupling constant, $\lambda_{ep}$, using the inverted McMillan equation (equation \ref{Mcmillan}).

\begin{equation}
\lambda_{ep} = \frac{1.04 + \mu^*ln(\frac{\Theta_D}{1.45T_c})}{(1-0.62\mu^*)ln(\frac{\Theta_D}{1.45T_c})-1.04}
\label{Mcmillan}
\end{equation}

Where the Coulomb pseudopotential, $\mu*$, is assumed to be 0.13, resulting in $\lambda_{ep}$ = 0.589.~\cite{Carnicom_2018}

We can then also estimate the electronic states at our Fermi energy (N(E$_F$)) using our fit Sommerfeld coefficient using equation \ref{DensityofStates}.~\cite{Carnicom_2018}

\begin{equation}
N(E_F) = \frac{3\gamma_N}{\pi^2 k_B^2 (1+\lambda_{ep})}
\label{DensityofStates}
\end{equation}

Where k$_B$ is the Boltzmann constant 0.08617 meV/K, resulting in N(E$_F$) = 0.99 $\pm$ 0.25 states eV$^{-1}$ f.u.$^{-1}$, which matches closely with our DFT calculated density of states of 1.5 eV$^{-1}$ f.u.$^{-1}$.

\begin{figure}[H]
\renewcommand{\figurename}{Figure S}
    \centering
    \resizebox{3in}{!}{\includegraphics{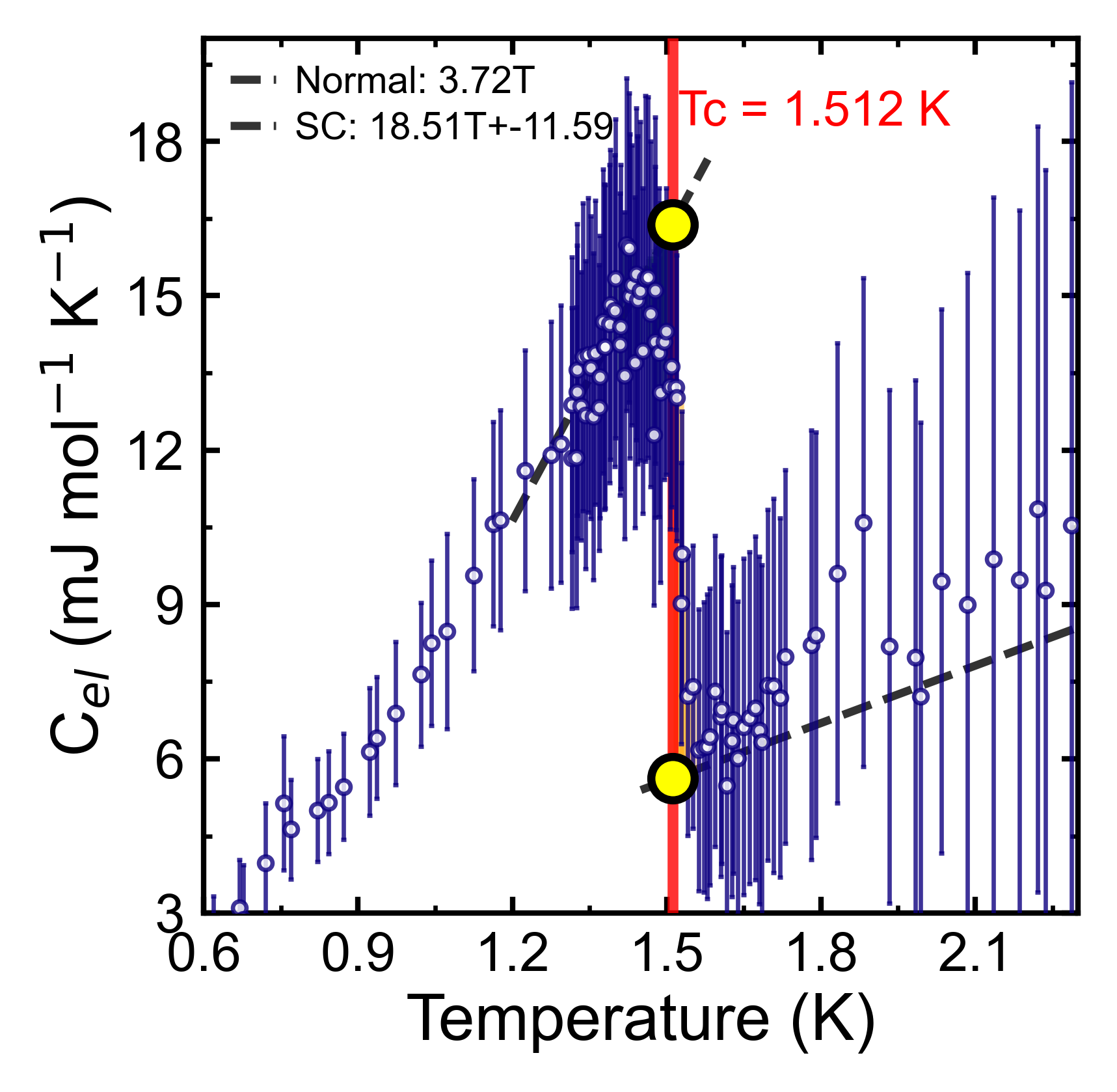}}
    \caption{\textbf{Equal area construction for determining thermodynamic T$_c$.} Plot shows two lines of best fit above and below a jump (red line) fit to the thermodynamic critical temperature. Intercept of the lines of best fit with T$_c$ (yellow points) are used to determine $\Delta$C.}
\label{equalarea}
\end{figure}

Evaluating the specific heat jump using an equal area construction, shown in Figure S\ref{equalarea} results in a thermodynamic critical temperature of 1.512~K. Additionally, the heat capacity jump ($\Delta C_{el}$) can be estimated by the difference intercepts of the normal and superconducting states' linear regression, resulting in $\Delta C_{el}$ = 10.77 $\pm$ 2.73 mJ mol$^{-1}$ K$^{-1}$. This can then be used to calculate the relative jump of the superconducting transition $\Delta C_{el}$/($\gamma_NT_c$) = 1.915 $\pm$ 0.486, falling just within error of the expected BCS weak coupling limit of 1.43.

The electronic contribution to the heat capacity can then be fit to various models.

First, we attempt semi-quantitative fits utilizing the $\alpha$-model.\cite{Padamsee1973, Johnston_2013, Muhlschlegel1959} This model utilizes the self-consistent temperature gap dependence from BCS-theory. This temperature gap dependence is then applied to various values of alpha, defined by equation \ref{alpha}.

\begin{equation}
\alpha = exp\left(-\frac{\Delta(0)}{k_BT}\right)
\label{alpha}
\end{equation}

\begin{figure}[H]
\renewcommand{\figurename}{Figure S}
    \centering
    \resizebox{3in}{!}{\includegraphics{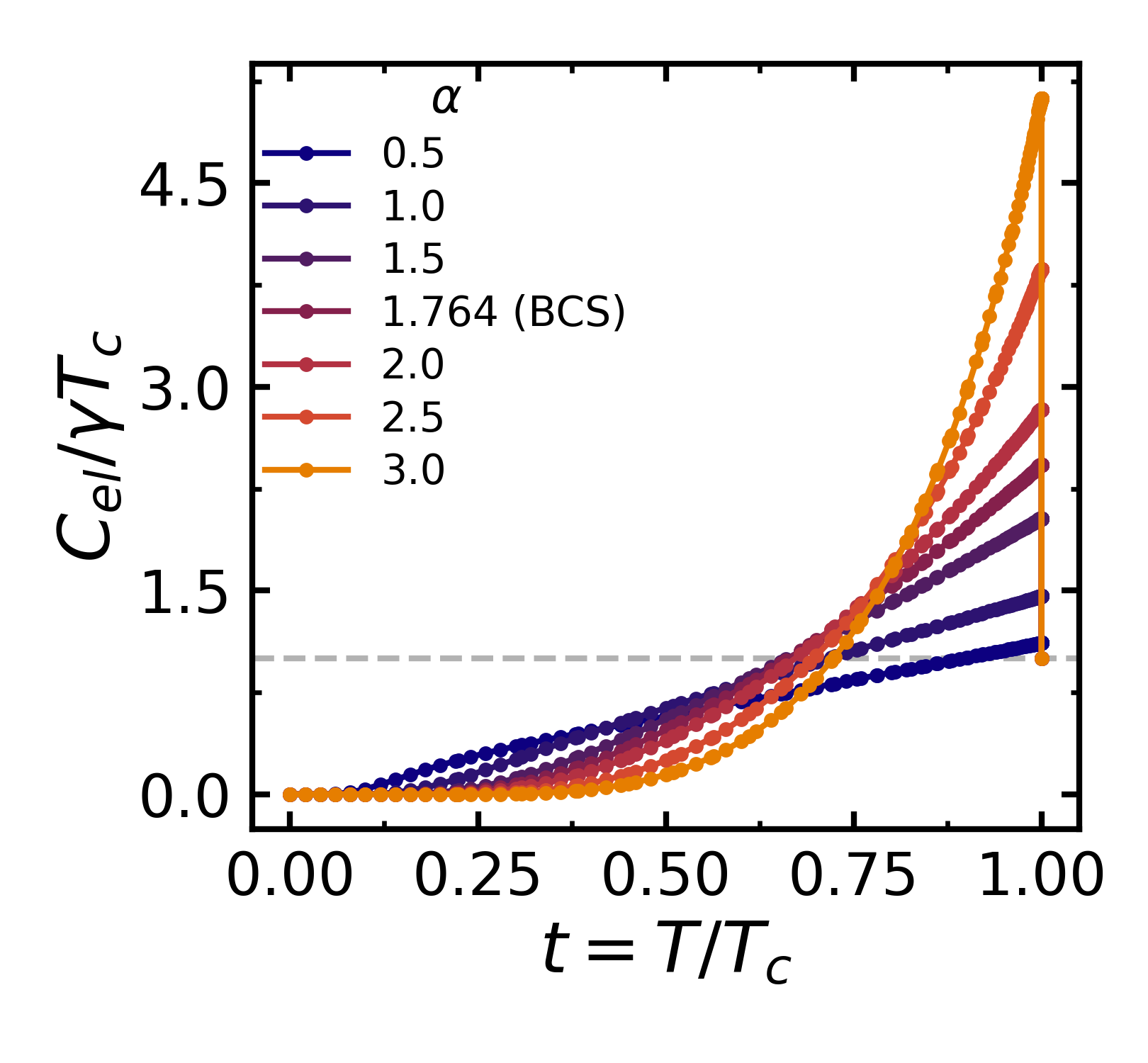}}
    \caption{\textbf{Simulated behavior of C$_{el}$ for various values of alpha.} These simulated plots are then used to fit the experimental data.}
\label{AlphaModelPlot}
\end{figure}

Generation of simulated data for C$_{el}$ for various values of $\alpha$ ranging from $\alpha$ = 0.201 to $\alpha$ = 2.999 with a step size of 0.001. This fitting analysis benefits from the fact that we can now perform fits with both strong and weak coupling (i.e., when alpha is greater or less than the BCS value of $\alpha_{BCS}$ = 1.764, respectively). A selected handful of these computed values are shown in Figure S \ref{AlphaModelPlot}.

In addition, we also attempted fits to a multi-gap scenario, taking the form of equation \ref{multiband fit}.

\begin{equation}
C_{el}=f*C_{el}(\alpha_1)+(1-f)*C_{el}(\alpha_2)
\label{multiband fit}
\end{equation}

Here, f is a fitting parameter representing the superconducting phase fraction, ranging from 0 to 1. 

We then attempted fits to our experimental data in four different ways. (1) A single $\alpha$ model, which assumes a phase pure, single, and isotropic superconducting gap. (2) A multi-gap model utilizing two separate $\alpha$ models, which assumes two separate isotropic superconducting gaps.(3) A single $\alpha$ model with impurity phase, which assumes option 1 mixed with some temperature electronic contribution from an impurity phase ($\gamma_{res}$T). And (4) A multi-gap model with impurity phase, assuming two separate isotropic superconducting gaps mixed with some contribution from an impurity phase. These fits to the experimental data are shown below in Figure S~\ref{AlphaModelComp}.

\begin{figure}[H]
\renewcommand{\figurename}{Figure S}
    \centering
    \resizebox{6in}{!}{\includegraphics{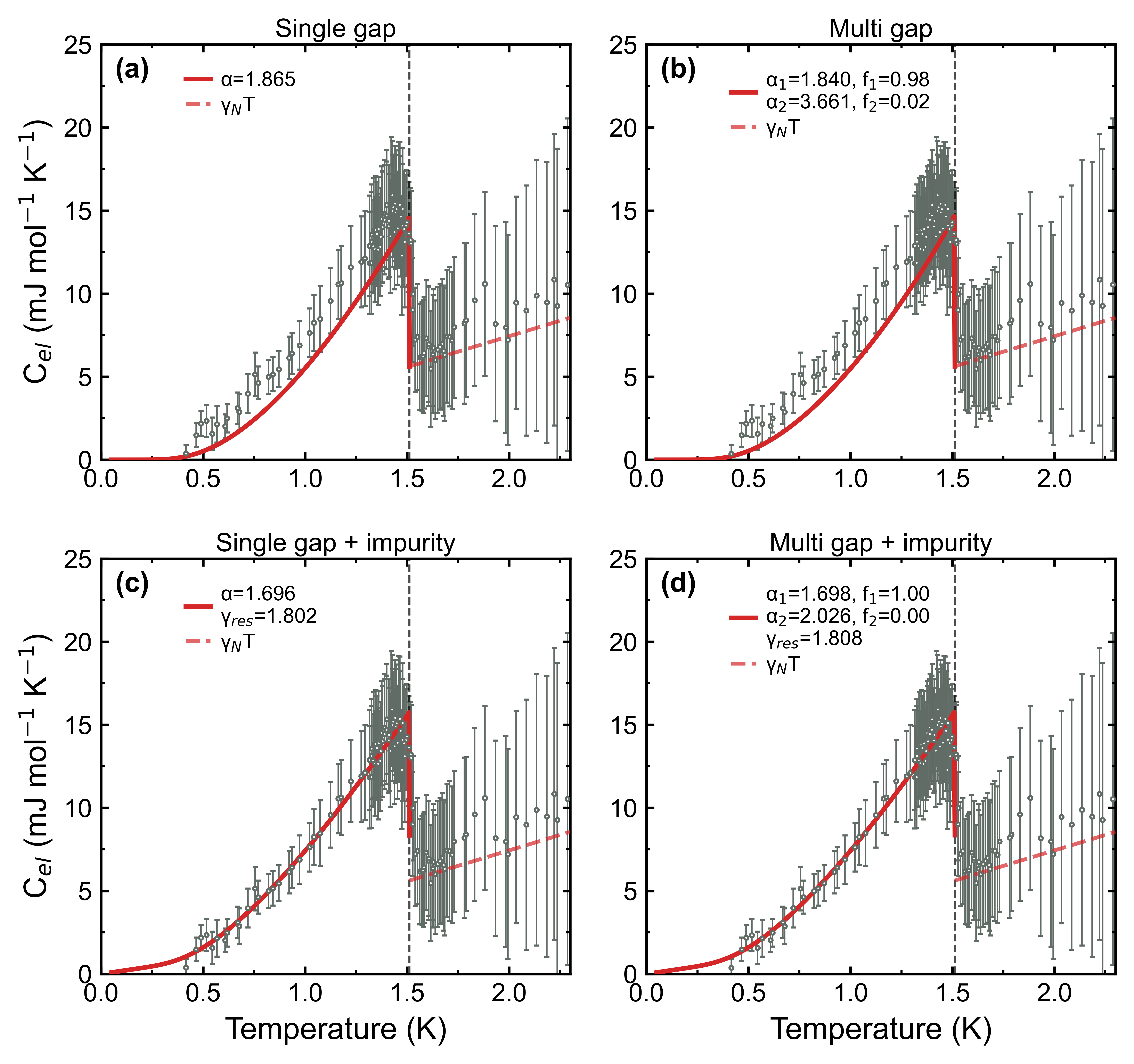}}
    \caption{\textbf{Various $\alpha$ model fits to the experimental heat capacity data.} \textbf{(a)} A single $\alpha$ model, assuming 100\% phase fraction of a isotropic gap. \textbf{(b)} A multi-gap model, assuming 100\% phase fraction of two isotropic gaps of different magnitude. \textbf{(c)} A single gap model with impurity phase, assuming some non-superconducting phase fraction defined by $\gamma_{res}$/$\gamma_N$. And \textbf{(d)} A multi-gap model with an impurity phase.}
\label{AlphaModelComp}
\end{figure}

We see from these fits that the multi-gap models closely approximate those of the best single-gap models: both multi-gap fits converge to 100\% phase fraction of a given alpha value. Also, in both single- and multi-gap fits, adding an impurity phase much better describes the experimental data. 
Therefore, for the $\alpha$ model, the best fit we can achieve is with two phases. First, a single, isotropic superconducting gap of $\alpha$=1.830, which is slightly higher than the expected BCS value of 1.764. Second, a non-superconducting impurity phase, such as Au, that has a phase fraction of $\gamma_{res}$/$\gamma_N$ = 0.484 (48.4\%). 

However, this brings up another issue with this model. If we consider the superconducting contribution to $\gamma_N$ (\textit{i.e.} $\gamma_{SC}=\gamma_N$-$\gamma_{res}$ = 1.918 mJ mol$^{-1}$ K$^{-2}$), we then also have to adjust the observed jump ratio at T$_c$. This inversely scales with the previously calculated heat capacity jump, recalculated now as C$_{el}$/($\gamma_{SC}T_c$) = 3.714. This value greatly exceeds the expected BCS value of 1.43. Put differently, none of these $\alpha$ model fits is good at explaining the entire experimentally observed data. Furthermore, while noncentrosymmetric superconductors may show predominantly s-wave character, many do not.~\cite{Smidman_2017} We therefore proceed with looking at other possible models focused on anisotropic gap structures.

\begin{figure}[H]
\renewcommand{\figurename}{Figure S}
    \centering
    \resizebox{3in}{!}{\includegraphics{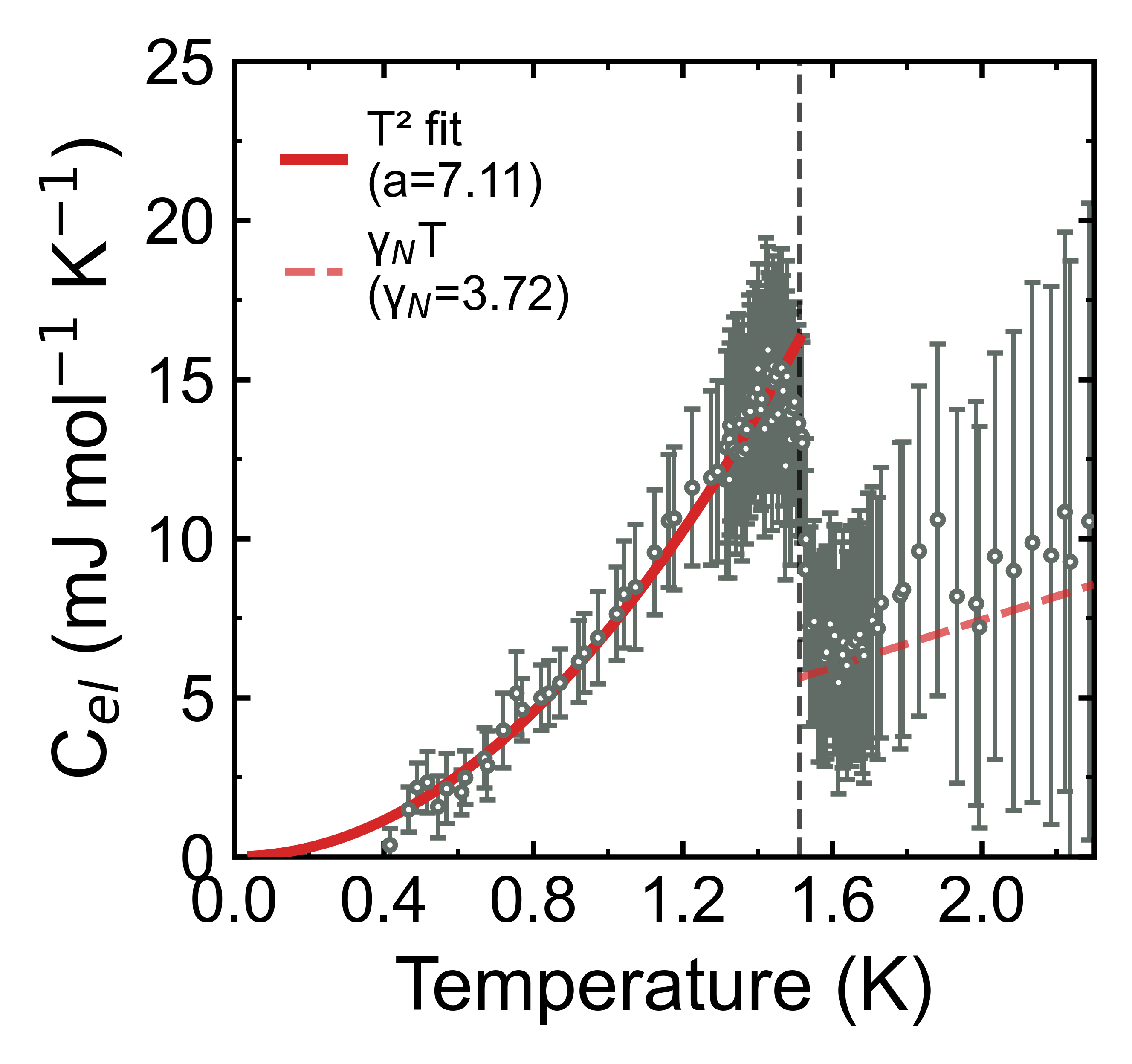}}
    \caption{\textbf{T$^2$ temperature dependence of C$_{el}$}.}
\label{T2FitPlot}
\end{figure}

We found that a simple quadratic fit, given in equation \ref{T2 fit}, provides an excellent agreement with our experimental data, shown in Figure S\ref{T2FitPlot} 

\begin{equation}
C_{el} = A_{sq} T^2
\label{T2 fit}
\end{equation}

This dependence on T$^2$ suggests possible nodal lines in the superconducting gap structure. Such nodal gaps can be rationalized as possible through a mix of strong spin-orbit coupling effects, as well as broken inversion symmetry.

Finally, we also conducted heat capacity measurements on our parent sample down to 0.22~K. Shown below in Figure S\ref{Cp_Parent}, we observe no superconducting transition down to 0.22~K, indicating that our topotactic deintercalation either has increased the critical temperature of our superconducting transition, or the allowance of the superconducting transition is enabled by mixed singlet--triplet pairing.

\begin{figure}[H]
\renewcommand{\figurename}{Figure S}
    \centering
    \resizebox{\textwidth}{!}{\includegraphics{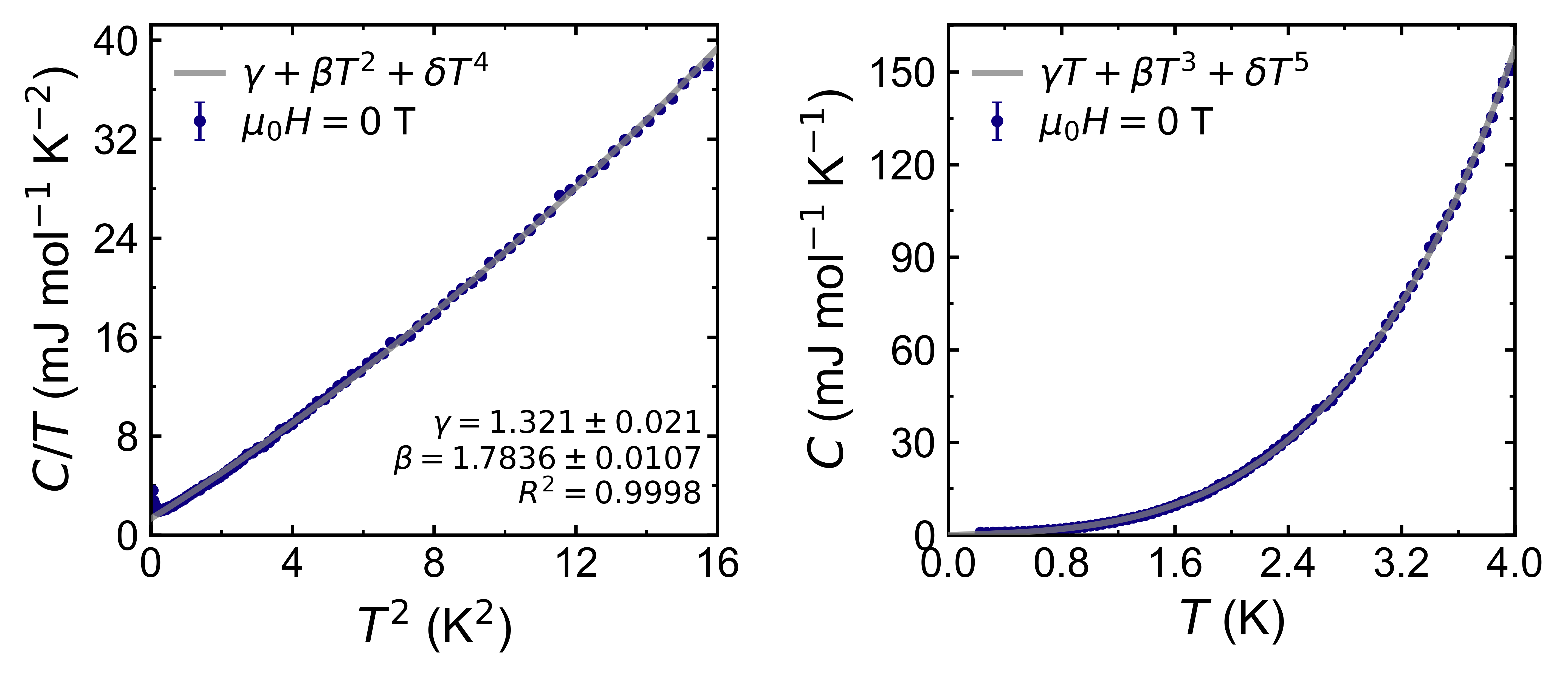}}
    \caption{\textbf{Zero field heat capacity of the parent compound, \ce{Au2PbP2}.} No superconducting jump is observed in this temperature range.}
\label{Cp_Parent}
\end{figure}

\FloatBarrier

The GL coherence length can then be calculated using  
\begin{equation}
\mu_0H_{c2}(0) = \frac{\Phi_0}{2\pi\xi_{GL}^2}
\label{GLCohrenceLength}
\end{equation}

where $\Phi_0$ is the quantum flux h/2e. Using the upper critical field of 140.2 mT from our WHH fit, we calculate $\xi_{GL}$ = 57.1 nm. The mean free path is then estimated using equation \ref{MFP}. 
\begin{equation}
l = 2.372*10^{-14}\frac{(\frac{m^*}{m_e})^2V_M^2}{N(E_F)^2\rho}
\label{MFP}
\end{equation}

Here, m$^*$/m$_e$ is assumed to be 1, $N(E_F)$ is the calculated density of states at the Fermi energy calculated to 1.25 states eV$^{-1}$, $\rho$ is the resistivity at the normal state measure to be 0.1 m$\Omega$ cm, and V$_M$ is the molar volume (62.02 cm$^3$ mol$^{-1}$). This leads to an estimated mean free path of 5.6 nm. The ratio of the mean free path to the GL coherence length results in $l$/$\xi_{GL}$ = 0.098, supporting the conclusion that we are in the dirty limit.

\subsection{Magnetic Susceptibility}

Demagnetization for magnetic measurements was calculated by estimating the sample as a rectangular prism where the demagnetization constant N is given by equation \ref{Demag}.\cite{Prozorov_2018, Lygouras_2025}

\begin{equation}
N \approx \frac{4AB}{4AB + 3C(A + B)}
\label{Demag}
\end{equation}

Here, A and B are the dimensions of the sample perpendicular to the applied magnetic field, while C is the dimension parallel to the applied magnetic field.

Direct Current (DC) Magnetometry was carried out in a 9T Quantum Design Magnetic Properties Measurement System with $^3$He attachment. For both M-T and M-H measurements, demagnetization corrections were needed. Importantly, the data presented in this paper used a sample geometry with the applied field oriented along the \textbf{c}-axis. That is, for these demagnetization corrections, A = 0.16 mm, B = 0.12 mm, and C = 0.97 mm. This results in a small demagnetization correction of N = 0.0861.

To get the corrected susceptibility, equation \ref{ChiDemag} was used.

\begin{equation}
\chi_{v,true} = \frac{\chi_{v,mea}}{1 - N * \chi_{v,mea}} 
\label{ChiDemag}
\end{equation}

The applied field for M-H was also corrected using equation \ref{MHDemag}.

\begin{equation}
H_{int} = H_{app} - N * M_v
\label{MHDemag}
\end{equation} 

Alternating Current (AC) Magnetic Susceptibility measurements were carried out in a 9T Quantum Design Physical Properties Measurement System attached with a Dilution Refrigerator. Measurements were taken with an applied AC excitation current of 1 Oe, with a AC frequency of 400 Hz. Each data point averages over 1 second (400 measurements). An applied magnetic field was then swept from 0 to 100 Oe at a rate of 0.4 Oe/Sec. These measurements were taken over a temperature range of 0.3 K to 1.5 K, although meaningful H$_{c1}$(T) could be determined only up to 1.2 K. 

\begin{figure}[H]
\renewcommand{\figurename}{Figure S}
    \centering
    \resizebox{6.5in}{!}{\includegraphics{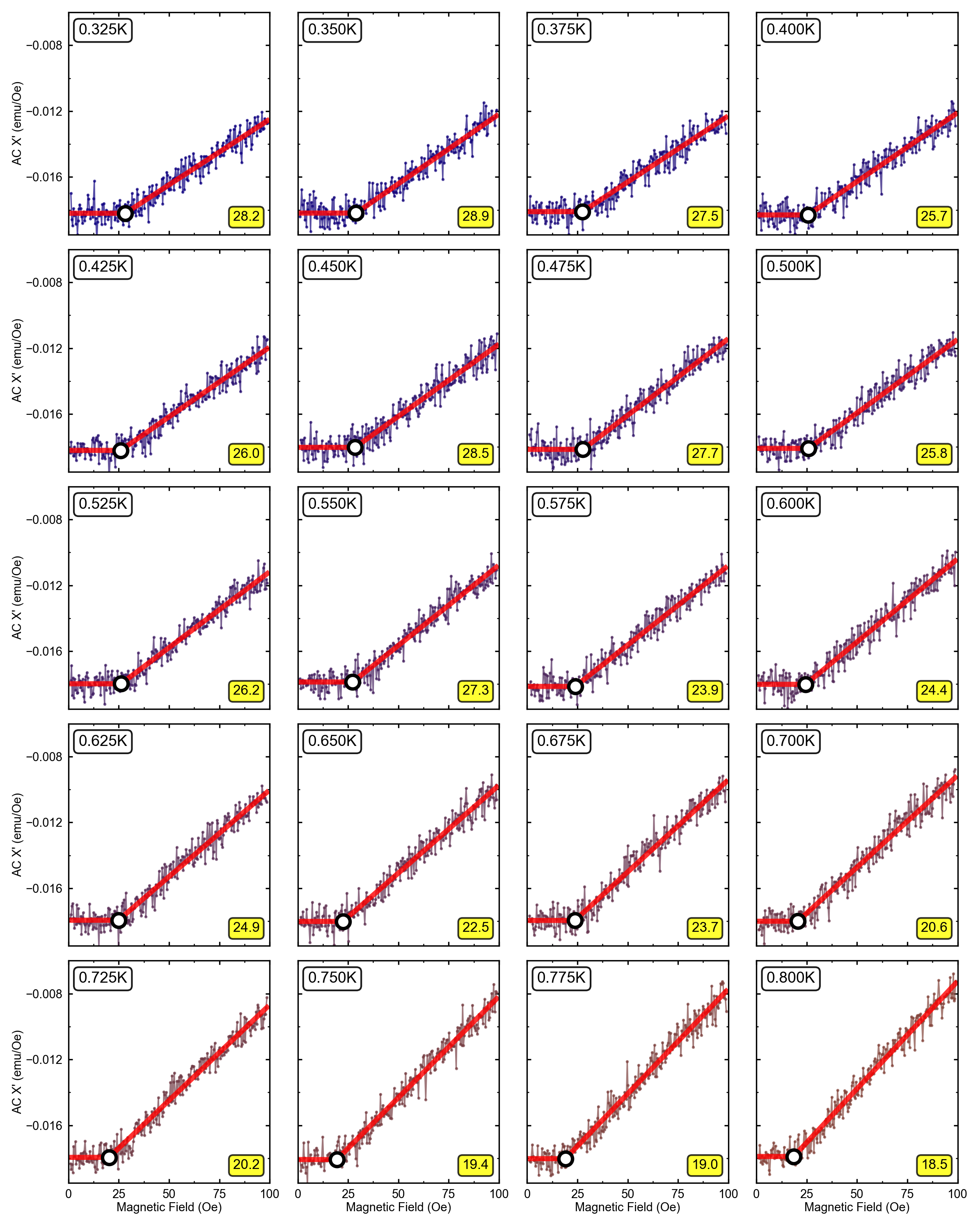}}
    \caption{\textbf{Hockey stick models for determining $\mu_0H_{c1}$}. Hockey stick models are fit to a breakout point to determine where the Meissner state is destroyed.}
    \label{HockeyStick1}
\end{figure}

\begin{figure}[H]
\renewcommand{\figurename}{Figure S}
    \centering
    \resizebox{6.5in}{!}{\includegraphics{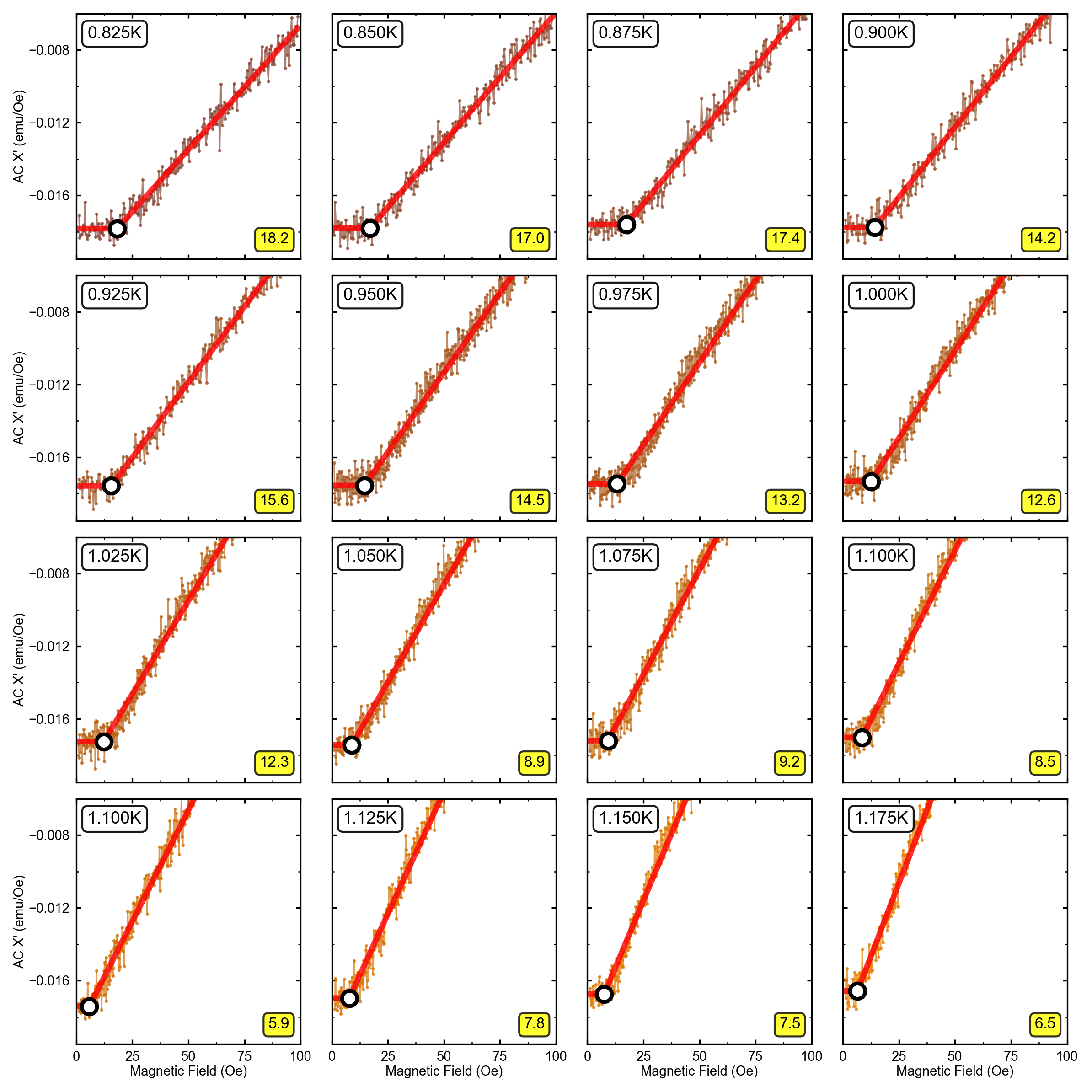}}
    \caption{\textbf{Hockey stick models for determining $\mu_0H_{c1}$}. Hockey stick models are fit to a breakout point to determine where the Meissner state is destroyed.}
    \label{HockeyStick2}
\end{figure}

The raw data, shown in Figures S\ref{HockeyStick1} and S\ref{HockeyStick2}, shows a constant negative value for AC X', reflecting a perfect Meissner state, before a critical field ($\mu_0H_{c1}^*$, white dot) begins to penetrate the sample via vortices. To determine $\mu_0H_{c1}^*$, the raw data is fit to a 'Hockey stick' model --- a piecewise linear function described by 
\begin{equation}
X'(\mu_0H) = X'(0); \text{ when } \mu_0H \leq \mu_0H_{c1}^*   
\end{equation}
and
\begin{equation}
X'(\mu_0H) = X'(0)+m(\mu_0H-\mu_0H_{c1}^*); \text{ when } \mu_0H > \mu_0H_{c1}^*   
\end{equation}

Where m is a fitted parameter representing the non-zero slope above $\mu_0H_{c1}^*$.

\begin{figure}[H]
\renewcommand{\figurename}{Figure S}
    \centering
    \resizebox{6.5in}{!}{\includegraphics{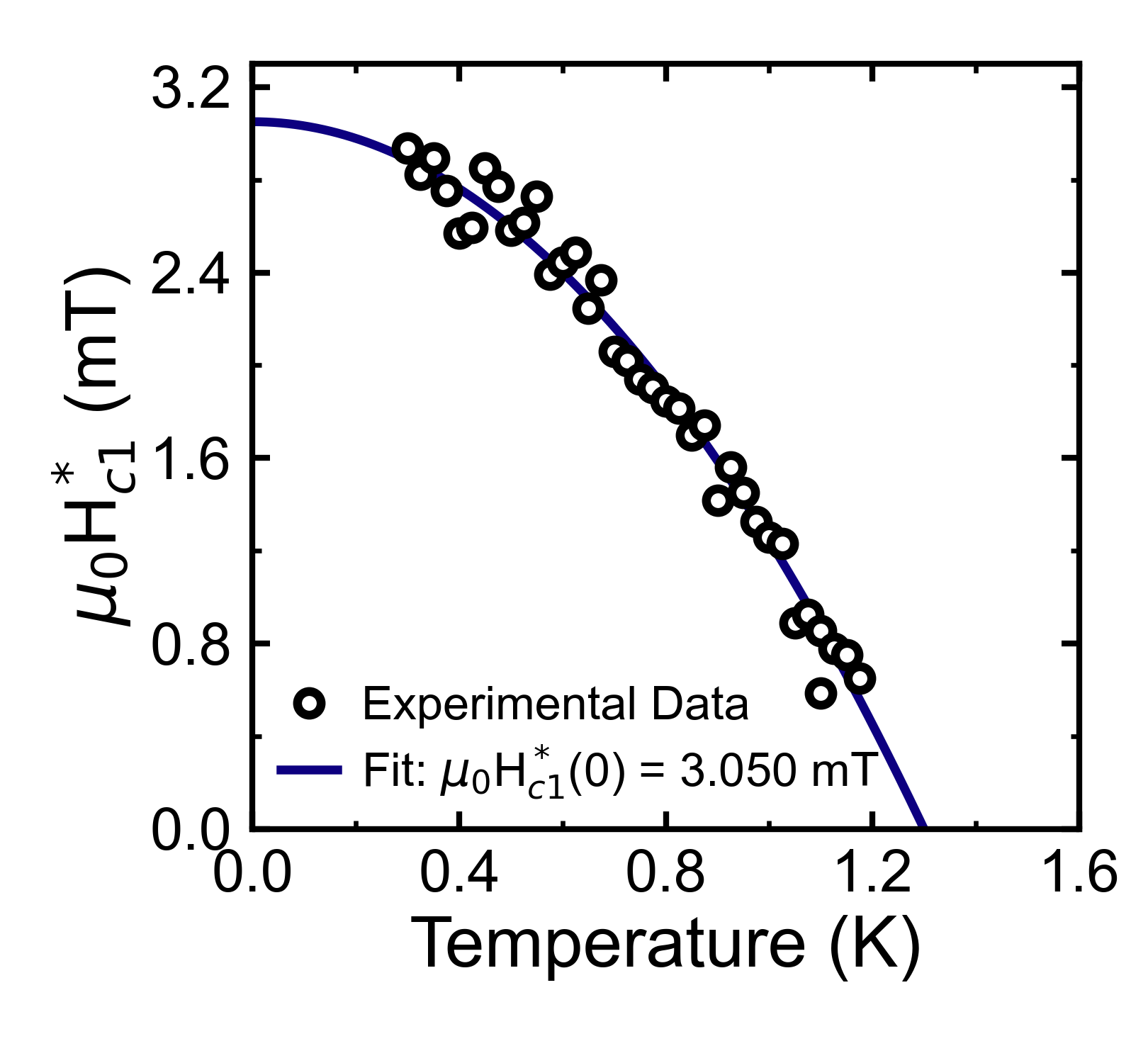}}
    \caption{\textbf{Fit breakout points for $\mu_0H_{c1}^*(T)$} versus temperature for temperatures between 0.3 K and 1.2 K. A clear trend can be seen, which is fit to the equation above.}
    \label{Critical_Field_Fit}
\end{figure}

The fitted breakout points are then plotted in Figure \ref{Critical_Field_Fit}. These data points can be well fit to equation:

\begin{equation}
\mu_0H_{c1}^*(T) = \mu_0H_{c1}^*(0)[1-(\frac{T}{T_c})^2]   
\end{equation}

During this fitting procedure, T$_c$ was adjusted from 1.521 K to 1.3 K. We reason that this is a reasonable value because the AC susceptibility data do not show any low field constant values below 1.2 K, either due to demagnetization effects or small paramagnetic impurities from GE varnish affecting the total resolution of these experiments at temperatures closer to T$_c$. Therefore, we expect both $\mu_0H_{c1}^*(0)$ and T$_c$ to be larger than reported here, though we assert that the general temperature dependence remains the same. 

Fitting this data results in a very good R$^2$ of 0.9791 and a critical $\mu_0H_{c1}^*(0)$ = 30.5 Oe.

With our fitted value of $\mu_0H_{c1}^*(0)$ = 3.05 mT, we may then begin to map how our relative penetration depth changes as a function of temperature. 

According to Ginzburg-Landau theory, the expression for the lower critical field is given as:

\begin{equation}
\mu_0H_{c1}(T) = \frac{\Phi_0}{4\pi\lambda^2(T)}  ln(\frac{\lambda(T)}{\xi(T)})
\end{equation}

Assuming in this data range that the $ln(\lambda(T)/\xi(T))$ term changes slowly, this can be written proportionally:

\begin{equation}
\frac{\lambda(T)}{\lambda(0)} = \frac{H_{c1}(0)}{H_{c1}(T)}
\end{equation}

The relative change in penetration depth can be related to the superfluid density where

\begin{equation}
\frac{\lambda(T)}{\lambda(0)} = \sqrt{\frac{n_s(0)}{n_s(T)}}
\label{superfluid_pendepth}
\end{equation}

where $n_s$ is the superfluid density. Gorter-Casimir then developed an empirical fit that closely resembles the BCS relationship:

\begin{equation}
\frac{n_s(T)}{n_s(0)} = 1 - 2\int_{0}^{\infty}\frac{\partial f}{\partial E}*(\frac{\Delta^2}{E^2}) d\xi \approx 1 - c(\frac{T}{Tc})^4
\label{superfluid_temp}
\end{equation}

where c is a fitting parameter. Putting together equations \ref{superfluid_pendepth} and \ref{superfluid_temp} gives us the relationship

\begin{equation}
\frac{\lambda(T)}{\lambda(0)} = \frac{1}{\sqrt{1 - c(\frac{T}{Tc})^4}}
\label{BCS_Pen_Fit}
\end{equation}

As a comparison, for a system with line nodes in the dirty limit, equation \ref{superfluid_temp} is modified to a second-order power law, derived from the approximately linear density of states near the nodal surface: 

\begin{equation}
\frac{n_s(T)}{n_s(0)} \approx 1 - c(\frac{T}{T_c})^2
\label{superfluid_temp_T2}
\end{equation}

resulting in equation \ref{T2_Pen_Fit}.

\begin{equation}
\frac{\lambda(T)}{\lambda(0)} = \frac{1}{\sqrt{1 - c(\frac{T}{Tc})^2}}
\label{T2_Pen_Fit}
\end{equation}

\begin{figure}[H]
\renewcommand{\figurename}{Figure S}
    \centering
    \resizebox{6.5in}{!}{\includegraphics{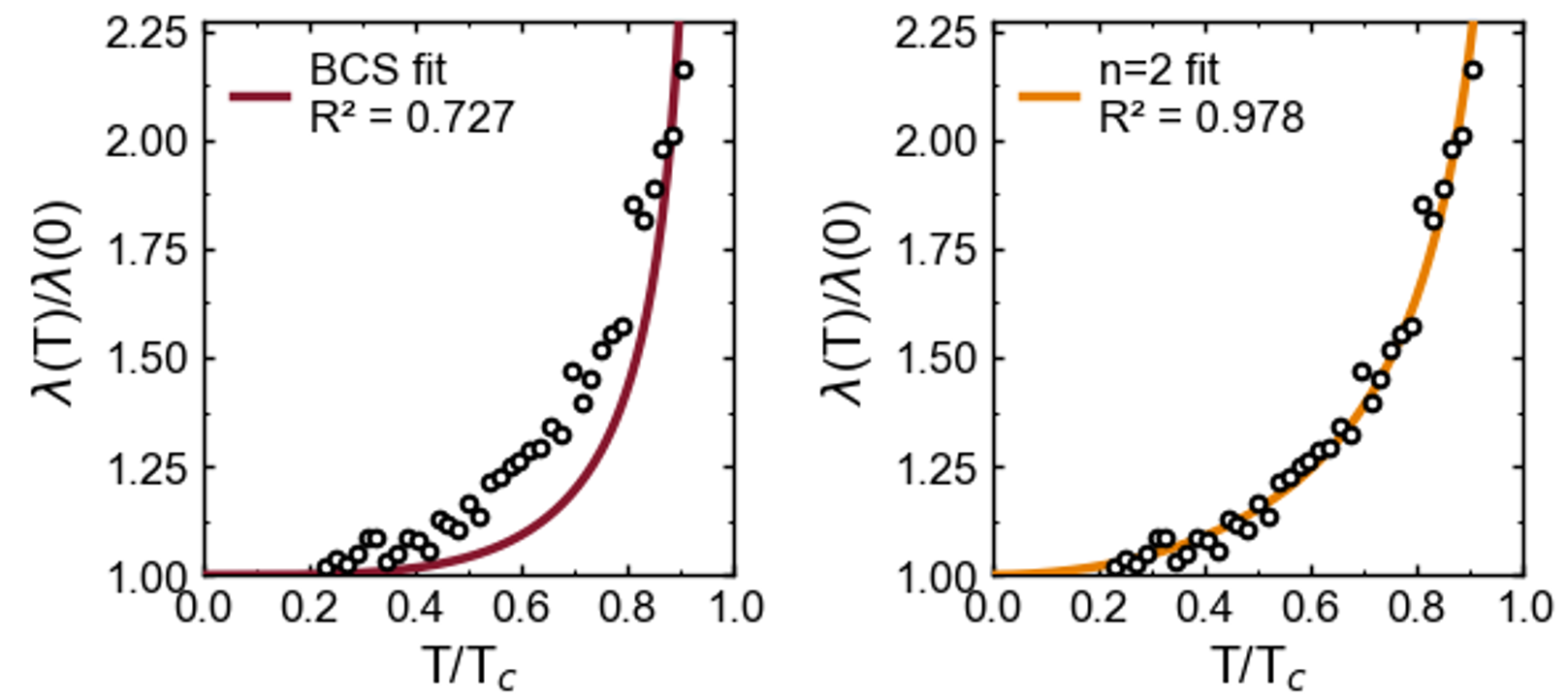}}
    \caption{\textbf{Comparison of relative penetration depth fits}. A. A BCS fit using equation \ref{BCS_Pen_Fit}, showing a departure from BCS behavior. B. A T$^2$ power law fit using equation \ref{T2_Pen_Fit}, with a close adherence to quadratic behavior.}
    \label{PenDepth_Fits}
\end{figure}

Comparing these two models, as shown in Figure S\ref{PenDepth_Fits}, we see strong deviations from the temperature-dependent BCS behavior (equation \ref{BCS_Pen_Fit}), while the n = 2 power law (equation \ref{T2_Pen_Fit}) fit is a much stronger fit.

\begin{table}[H]
\renewcommand{\tablename}{Table S}
\caption{\textbf{Superconducting parameters} for Au$_2$Pb$_{0.914}$P$_2$ based on strong coupling BCS model with no impurity phase ($\alpha = 1.865$).}
 \centering
 \begin{tabular}{l c c}
\hline
Parameter & Value & Units \\ 
\hline 
Transition temperature ($T_c$) & 1.521 & K \\
Sommerfeld coefficient ($\gamma$) & 3.719 $\pm$ 0.925 & mJ/mol/K$^2$ \\
Phonon coefficient ($\beta$) & 5.310 $\pm$ 0.427 & mJ/mol/K$^4$ \\
Debye temperature ($\theta_D$) & 121.6 $\pm$ 3.3 & K \\
Density of states $N(E_F)$ & 0.99 $\pm$ 0.25 & states/eV/f.u. \\
Specific heat jump ($\Delta C$) & 10.77 $\pm$ 2.73 & mJ/mol/K \\
BCS ratio ($\Delta C/\gamma T_c$) & 1.92 $\pm$ 0.680 & dimensionless \\
Gap energy ($\Delta_0$) & 0.243 & meV \\
Gap ratio ($2\Delta_0/k_BT_c$) & 3.730 & dimensionless \\
Electron-phonon coupling ($\lambda_{ep}$) & 0.589 & dimensionless \\
Lower critical field $\mu_0H_{c1}^*(0)$ & 3.05 & mT \\
Thermodynamic critical field $\mu_0H_c$ & 15.1 & mT \\
Upper critical field $\mu_0H_{c2}(0)$ & 140.2 & mT \\
Pauli paramagnetic limit & 2.81 & T \\
GL coherence length $\xi_{GL}(0)$ & 48.45 & nm \\
Penetration depth $\lambda_{GL}(0)$ & 318.8 & nm \\
GL parameter $\kappa$ & 6.58 & dimensionless \\
Mean free path ($l$) & 9.27 $\pm$ 4.61 & nm \\
$l/\xi$ ratio & 0.191 $\pm$ 0.095 & dimensionless \\
\hline
\label{Superconductingcharacterization}
\end{tabular}
\end{table}

\begin{table}[H]
\renewcommand{\tablename}{Table S}
\caption{\textbf{Superconducting parameters} for Au$_2$Pb$_{0.914}$P$_2$ based on a single gamp with impurity model ($\alpha = 1.696$, $\gamma_{res}$ = 1.802).}
 \centering
 \begin{tabular}{l c c}
\hline
Parameter & Value & Units \\ 
\hline 
Transition temperature ($T_c$) & 1.521 & K \\
Sommerfeld coefficient ($\gamma$) & 3.719 $\pm$ 0.925 & mJ/mol/K$^2$ \\
Phonon coefficient ($\beta$) & 5.310 $\pm$ 0.427 & mJ/mol/K$^4$ \\
Debye temperature ($\theta_D$) & 121.6 $\pm$ 3.3 & K \\
Density of states $N(E_F)$ & 0.51 $\pm$ 0.13 & states/eV/f.u. \\
Specific heat jump ($\Delta C$) & 10.77 $\pm$ 2.73 & mJ/mol/K \\
BCS ratio ($\Delta C/\gamma T_c$) & 3.716 $\pm$ 2.025 & dimensionless \\
Gap energy ($\Delta_0$) & 0.222 & meV \\
Gap ratio ($2\Delta_0/k_BT_c$) & 3.392 & dimensionless \\
Electron-phonon coupling ($\lambda_{ep}$) & 0.589 & dimensionless \\
Lower critical field $\mu_0H_{c1}^*(0)$ & 3.05 & mT \\
Thermodynamic critical field $\mu_0H_c$ & 15.1 & mT \\
Upper critical field $\mu_0H_{c2}(0)$ & 140.2 & mT \\
Pauli paramagnetic limit & 2.81 & T \\
GL coherence length $\xi_{GL}(0)$ & 48.45 & nm \\
Penetration depth $\lambda_{GL}(0)$ & 318.8 & nm \\
GL parameter $\kappa$ & 6.58 & dimensionless \\
Mean free path ($l$) & 34.82 $\pm$ 10.4 & nm \\
$l/\xi$ ratio & 0.719 $\pm$ 0.214 & dimensionless \\
\hline
\label{Superconductingcharacterization2}
\end{tabular}
\end{table}

\end{document}